%% file: var6.tex
\documentstyle[preprint,aps,prl]{revtex}
\begin{document}
\tightenlines

\def\4he{$^4$He}
\def\pc{\protect\cite}
\def\hpsi{\hat\psi}
\def\tpsi{\tilde\psi}
\def\br{{\bf r}}
\def\bk{{\bf k}}
\def\bu{{\bf u}}
\def\bw{{\bf w}}
\def\brt{\br,t}
\def\bbrt{(\brt)}
\def\cphio{\Phi_0}
\def\beq{\begin{equation}}
\def\eeq{\end{equation}}
\def\bea{\begin{eqnarray}}
\def\eea{\end{eqnarray}}
\def\bna{\bbox{\nabla}}
\def\bp{{\bf p}}
\def\bv{{\bf v}}
\def\tn{\tilde n}
\def\tp{\tilde p}
\def\be{\bbox{\eta}}

\draft

\title{
Dynamics of trapped Bose gases at finite temperatures}

\author{E. Zaremba}
\address{Department of Physics, Queen's University\\Kingston,
Ontario K7L 3N6, Canada} 

\author{T. Nikuni}
\address{Department of Physics, Tokyo Institute of Technology\\ Meguro,
Tokyo 152, Japan}

\author{and}
\address{}

\author{A. Griffin}
\address{Department of Physics, University of Toronto\\Toronto,
Ontario M5S 1A7, Canada}

\date{\today}

\maketitle

\begin{abstract}
Starting from an approximate microscopic model of a trapped
Bose-condensed gas at finite temperatures, we derive an equation of
motion for the condensate wavefunction and a quantum kinetic equation
for the distribution function for the excited atoms.  The kinetic 
equation is a generalization of our earlier work in that collisions
between the condensate and non-condensate ($C_{12}$) are now included, 
in addition to collisions between the excited atoms as described by the
Uehling-Uhlenbeck ($C_{22}$) collision integral. The continuity 
equation for the
local condensate density contains a source term $\Gamma_{12}$ which is
related to the $C_{12}$ collision term. If we assume that the $C_{22}$
collision rate is sufficiently rapid to ensure that the non-condensate
distribution function can be approximated by a local equilibrium Bose
distribution, the kinetic equation can be used to derive hydrodynamic
equations for the non-condensate. The $\Gamma_{12}$ source terms
appearing in these equations play a key role in describing the
equilibration of the local chemical potentials associated with the
condensate and non-condensate components. We give a detailed study of
these hydrodynamic equations and show how the Landau two-fluid equations
emerge in the frequency domain $\omega \tau_\mu \ll 1$, where $\tau_\mu$
is a characteristic relaxation time associated with
$C_{12}$  collisions. More generally, the
lack of complete local equilibrium between the condensate and
non-condensate is shown to give rise to a new relaxational mode which is
associated with the exchange of atoms between the two components. This
new mode provides an additional source of damping in the hydrodynamic
regime. Our equations are consistent with the generalized Kohn theorem
for the center of mass motion of the trapped gas even in the presence of
collisions. Finally, we formulate a variational solution of the
equations which provides a very convenient and physical way of
estimating normal mode frequencies. In particular, we use relatively
simple trial functions within this approach to work out some of the
monopole, dipole and quadrupole oscillations for an isotropic trap.
\end{abstract}
\vskip .25truein

\vskip 1.5 true cm
\section {Introduction}
\label{section1}
\vskip 0.2 true cm
At very low temperatures, the dynamics (collective oscillations) of a 
trapped Bose gas is described by the
time-dependent Gross-Pitaevskii (GP) equation\cite{dalfovo98} for the
macroscopic wave function $\Phi(\brt)$ associated with the Bose 
condensate. At higher temperatures, when an appreciable fraction of 
atoms is excited out of the condensate, the dynamics of the trapped 
gas becomes much more complicated since it now involves the coupled 
motion of the condensate and the non-condensate degrees of freedom.
In recent work\protect\cite{hutchinson97} the finite-temperature
excitations of the condensate were calculated within the 
Hartree-Fock-Popov (HFP) approximation. This treatment of the
excitations is equivalent to a description based on a generalized GP
equation in which the condensate density $n_c(\brt)$ is treated 
dynamically but the non-condensate density $\tilde{n}(\brt)$ is 
treated as static. That is, one ignores the effect of non-condensate
fluctuations on the dynamics of the condensate.
Several authors\protect\cite{minguzzi97,giorgini98} 
have recently extended this kind
of HFP theory by treating the non-condensate fluctuations 
$\delta\tilde{n} (\brt)$ using a time-dependent Hartree-Fock 
equation of motion (often referred to as the RPA).  This describes the 
collective modes in the collisionless mean-field regime.

In contrast, in the low frequency collision-dominated regime, one
expects a description in terms of hydrodynamic equations. In a recent
paper\protect\cite{zaremba98}, we discussed a 
method for obtaining two-fluid equations 
starting from a semi-classical kinetic equation for the distribution 
function $f(\bp,\br,t)$ of the excited atoms.  The final result was 
a closed set of two-fluid hydrodynamic equations in terms of the local 
densities and velocities of the condensate and non-condensate 
components. In this paper, we consider a generalized 
model and present a more complete discussion of the approximations 
which are involved. As we shall see, this new microscopic model,
though approximate, captures all the essential
physics of the collision-dominated hydrodynamic behavior involving the
condensate and non-condensate components. 

Deriving two-fluid equations within the context of a specific
microscopic model has several advantages:
\begin{enumerate}
\item It allows one to see in a very clear fashion how the condensate 
and non-condensate degrees of freedom arise and how they are coupled.
\item It gives a closed set of equations with all local thermodynamic
properties and relaxation times
determined in a self-consistent manner.  It thus ensures that the 
static and time-dependent properties are treated in a unified manner.
\item The equations we obtain can be used to derive the usual Landau 
equations (originally developed to describe superfluid 
$^4$He\cite{landau41,khalatnikov65}) in a
certain limit, 
and allow a clear identification of the Landau superfluid (normal 
fluid) density with the condensate (non-condensate) density.
\item The description of the dynamics can be extended beyond the regime 
in which the Landau phenomenological equations are valid, i.e., when the
condensate and non-condensate are not in local diffusive equilibrium.
\item An explicit microscopic formulation provides a starting point for
improved descriptions, for example, the inclusion of the effect 
of the off-diagonal (or anomalous) self-energies which are
important at very low temperatures~\cite{giorgini98}.
\end{enumerate}

In the ZGN paper\cite{zaremba98}, we only included collisions between
the excited (non-condensate) atoms. Moreover, we only
discussed the solutions of the kinetic equation for $f(\bp,\br,t)$ which
correspond to dynamic local equilibrium within the non-condensate. This
local equilibrium state is defined by the condition that the 
collision integral for the excited atoms (denoted by $C_{22}[f]$) 
vanishes identically. As a result, no explicit dependence on the
$C_{22}$ collision integral appeared in the ZGN hydrodynamic equations.
In the present 
paper, we make a major improvement by also considering the effect
of collisions between atoms in the condensate with those in
the non-condensate. This collision term (denoted by $C_{12}[f]$)
allows us to discuss how the condensate is coupled to the non-condensate
beyond mean-field effects, and how the condensate comes into dynamic
local equilibrium with the non-condensate. In this connection, we 
should note that the
different effects of the $C_{12}$ and $C_{22}$ collisions was first
discussed in a qualitative way by Eckern~\cite{eckern84}.
The generalized set of
hydrodynamic equations which arise when the effects of $C_{12}[f]$ are
included will be referred to as the ZGN$^\prime$ equations to
distinguish them from the set we derived previously~\cite{zaremba98}.

Even when collisions among non-condensate atoms are sufficiently rapid
to enforce the distribution function $f$ to take the form of a local
equilibrium Bose distribution with chemical potential $\tilde \mu$, it 
is possible that the non-condensate is not in complete local 
equilibrium with the condensate. In this situation, the chemical
potential $\tilde \mu$ of the non-condensate can be different from 
that of the condensate, $\mu_c$. Thus the variable $\mu_{\rm diff} 
\equiv \tilde \mu - \mu_c$ emerges as a key new local variable which is 
needed to represent the equilibration of the condensate and
non-condensate components. In the original ZGN analysis, where the
effects of $C_{12}$ were neglected, the difference between $\mu_c$ and
$\tilde \mu$ was implicitly built into the ZGN hydrodynamic equations,
but this fact was overlooked in our attempt 
to reduce these equations to the Landau
two-fluid hydrodynamic equations. To arrive at this latter description,
we assumed additional local equilibrium thermodynamic relations 
which implicitly forced $\mu_{\rm diff} = 0$ (i.e., $\tilde\mu=\mu_c$).
Thus the ZGN equations are in fact distinct from 
the Landau two-fluid equations and, in general, will lead to
different dynamical predictions. In the present paper, we show how
the more general ZGN$^\prime$ equations reduce to either the ZGN or 
Landau descriptions, in the appropriate limits. This more
complete description involves an additional equation of motion for
$\mu_{\rm diff}$, which is coupled to the condensate and non-condensate
velocity fields. The new relaxational mode which emerges involves the
transfer of atoms between the two components, and allows one 
to understand the conditions under which the Landau two-fluid equations 
are valid. 

The possibility of the condensate being out of equilibrium with the
non-condensate was studied many years ago, mainly with regard
to the behaviour of superfluid helium near the $\lambda$ point. In a
well-known paper, Landau and Khalatnikov\cite{landau54} calculated the 
absorption of sound near a second-order phase transition due to the
coupling to a relaxing order parameter. 
Pitaevskii\cite{pitaevskii59,khalatnikov65} subsequently generalized 
the phenomenological two-fluid equations for superfluid helium to take
into account the relaxation of the superfluid density near the $\lambda$
point. The same problem was looked at within the Mori formalism by
Miyake and Yamada\cite{miyake76}. All of these approaches are
phenomenological in nature since they introduce new dissipation 
parameters to describe the process of atomic exchange between the
condensate (superfluid) and non-condensate (normal fluid). In contrast,
our theory is based on a well-defined microscopic model of 
a Bose-condensed gas, and as a result, the relaxation times 
responsible for the dissipation can be calculated explicitly. 

We briefly reported on some of our main results in Ref.\cite{nikuni99} 
which discusses the ZGN$^\prime$ equations for the special case of a 
{\it uniform} Bose gas. Here we give a more complete derivation of the
ZGN$^\prime$ equations and a more detailed discussion of their
properties. Our work may be viewed as an extension to the case of a
trapped Bose-condensed gas of the pioneering work of Kirkpatrick and
Dorfman\cite{kirkpatrick} (referred to in the following as KD). 
However, these authors concentrated on the
derivation of the Landau two-fluid equations with the inclusion of
dissipation, and did not address the interesting possibility of the
condensate and non-condensate components being out of diffusive local
equilibrium ($\mu_c \ne \tilde \mu$). In this context, we should also
mention Bogoliubov's rigorous field-theoretic derivation of the Landau
two-fluid equations\cite{bogoliubov70}, which also implicitly assumed
that the superfluid and normal fluid were always in local thermodynamic
equilibrium.

In the last few years, several papers dealing with non-equilibrium
properties of trapped Bose gases have appeared. We mention specifically
the work of Gardiner {\it et al.}\cite{gardiner98}, Proukakis {\it et
al.}\cite{proukakis98}, Stoof\cite{stoof99} and Walser 
{\it et al.}\cite{walser98}.
Each of these papers presents a formal derivation of a quantum
kinetic equation using quite different theoretical methods.
The equations of motion we derive for the condensate and
non-condensate are closest in form to those of Stoof\cite{stoof99}
and Walser {\it et al.}~\cite{walser98}, although there are also many
formal similarities with the work of Proukakis {\it et
al.}\cite{proukakis98}.
However, in contrast to this other work, we focus on
deriving hydrodynamic equations, and on establishing the criteria
under which such equations are valid. In particular, we identify the
crucial role that collisions between condensate and non-condensate atoms
have in estalishing a state of local equilibrium. The paper of
Stoof\cite{stoof99} is also useful for its thorough and insightful
discussion of the dynamics of condensate formation and growth, a topic
which Gardiner {\it et al.}\cite{gardiner98} study in considerable
detail. In this regard, we note that our kinetic 
equation can also be used to investigate condensate growth and 
applications to this problem will be presented
elsewhere~\cite{zaremba99}.

A brief summary of our paper follows. In Section II, we derive the 
equation of motion for the condensate, and in Section III, we obtain
the quantum kinetic equation describing the non-condensate. These
equations are applicable to the coupled dynamics of the condensate and
non-condensate at finite temperatures, and have been derived taking
careful account of correlation functions involving three field
operators, which describe the transfer of atoms between the two 
components. Taking moments of our kinetic equation, we derive in 
Section IV the hydrodynamic 
equations for the non-condensate, based on the assumption of
local equilibrium within the excited-atom component. In Section V, we 
then analyze the full set of linearized ZGN$^\prime$ equations and show
that a relaxation time $\tau_\mu$ naturally arises which characterizes 
the equilibration of the condensate and non-condensate chemical 
potentials. We show (in the simple case of a uniform Bose gas) that the
ZGN equations give a good description in the limit $\omega\tau_\mu 
\gg 1$, while the Landau two-fluid hydrodynamic theory emerges in the 
opposite limit, $\omega\tau_\mu \ll 1$. In Section VI, we demonstrate 
that the generalized Gross-Pitaevskii equation for the condensate and 
the kinetic equation for the non-condensate are consistent with the
generalized Kohn theorem for the center of mass motion in a harmonic
confining potential. In Section VII, we reformulate the ZGN hydrodynamic
equations in terms of a variational functional which is then used to
calculate the frequencies (and associated condensate and non-condensate 
velocities) of the lowest-lying monopole $(l=0)$, dipole $(l=1)$ and
quadrupole $(l=2)$ collective modes. The results for the out-of-phase
dipole mode were 
first reported in ZGN\cite{zaremba98}. Finally, in Section VIII, we
assess what we have accomplished and briefly discuss some further
extensions of the theory.

\input{condensate.tex}
\input{kinetic.tex}
\input{hydro.tex}
\input{landau.tex}
\input{kohn.tex}
\input{varsol.tex}
\input{conclusions.tex}
\vfil \break
\appendix
\input append.a.tex
\input append.b.tex
\input append.c.tex

\vfil\break
\centerline{\bf Figure Captions}

\begin{itemize}
\item[Fig. 1:]
Squares of the first and second sound velocities (normalized by
the first sound velocity of the ideal gas at $T = T_{BEC}$) 
vs. $T/T_{BEC}$. The solid lines are for the Landau two-fluid
theory while the dashed lines are for the ZGN hydrodynamics. The curves
were generated for $gn/k_B T_{BEC} = 0.2$.

\item[Fig. 2:] Damping of the second sound mode in a uniform gas
due to the relaxation of
the condensate and non-condensate chemical potentials at a temperature
$T/T_{BEC} = 0.9$. The damping,
$\Gamma_2$, relative to the mode frequency, $\Omega_2$, is plotted 
as a function of $\Omega_2 \tau_\mu$.

\item[Fig. 3:] Density of atoms as a function of position in an 
isotropic parabolic trap containing 5000 $^{87}$Rb atoms.  
Parameters defining the trap and atomic interactions are given in 
the body of the paper. The solid line is the total density
and the dashed line is the non-condensate density. The parameter $d$ is
the oscillator length, $d=\sqrt{\hbar/m\omega_0}$.

\item[Fig. 4:] Monopole ($l=0$) and quadrupole ($l=2$) 
mode frequencies, in units of the trap frequency $\omega_0$, vs. 
temperature for 5000 $^{87}$Rb atoms in an isotropic parabolic trap. 
The `+' and `--' signs merely distinguish the two modes of a given
angular character. The lower part of the figure shows the condensate
fraction as a function of $T$. 

\item[Fig. 5:] Variational condensate ($A_c$) and non-condensate ($A_n$)
amplitudes for the monopole modes as a function of temperature. Panel
(a) is for the $0^+$ mode and panel (b) is for the $0^-$ mode. The
vertical line at $T \simeq 149$ nK denotes the BEC transition
temperature.

\item[Fig. 6:] As in Fig. 5, but for the quadrupole $2^+$ and
$2^-$ modes.

\end{itemize}

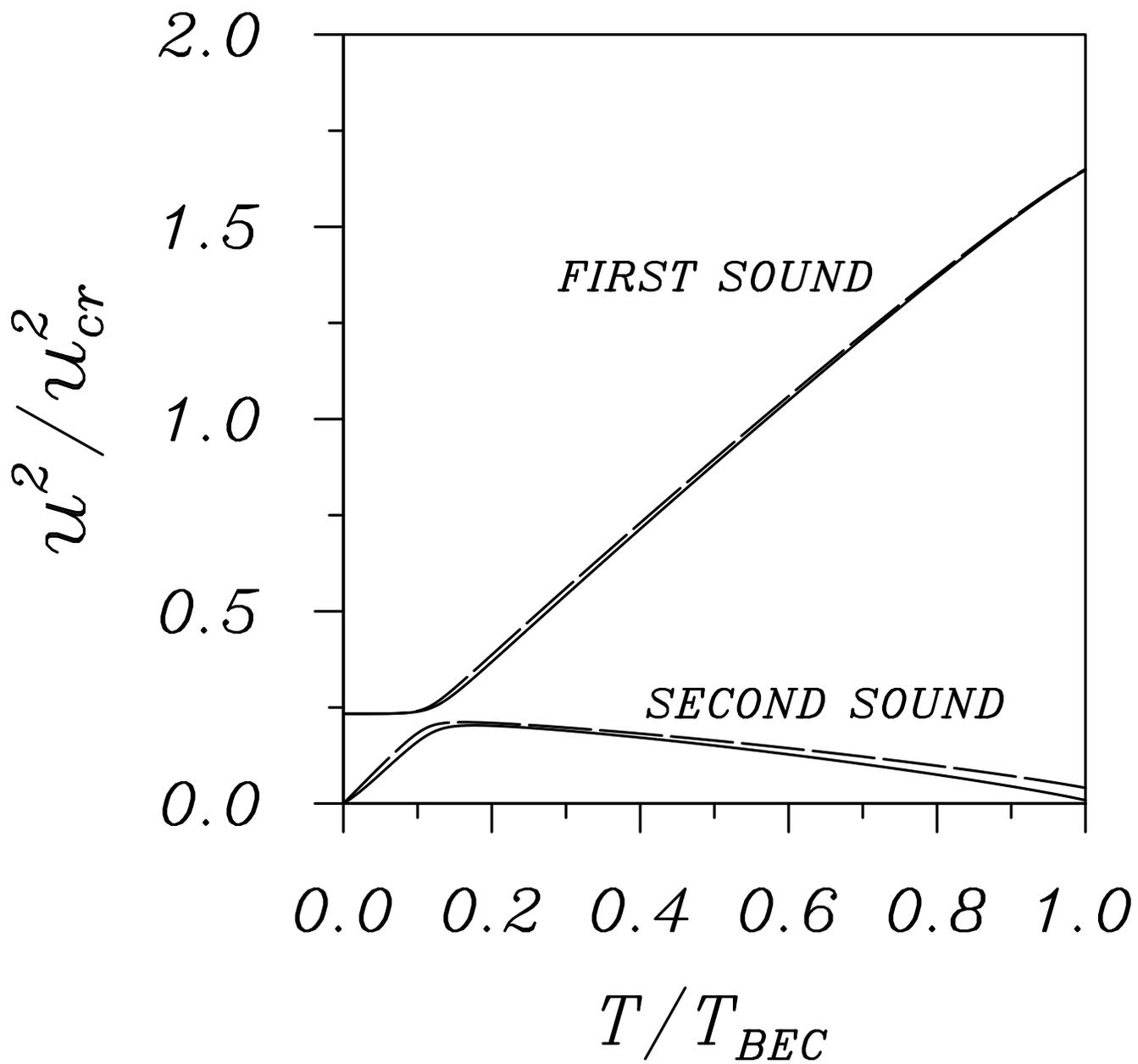
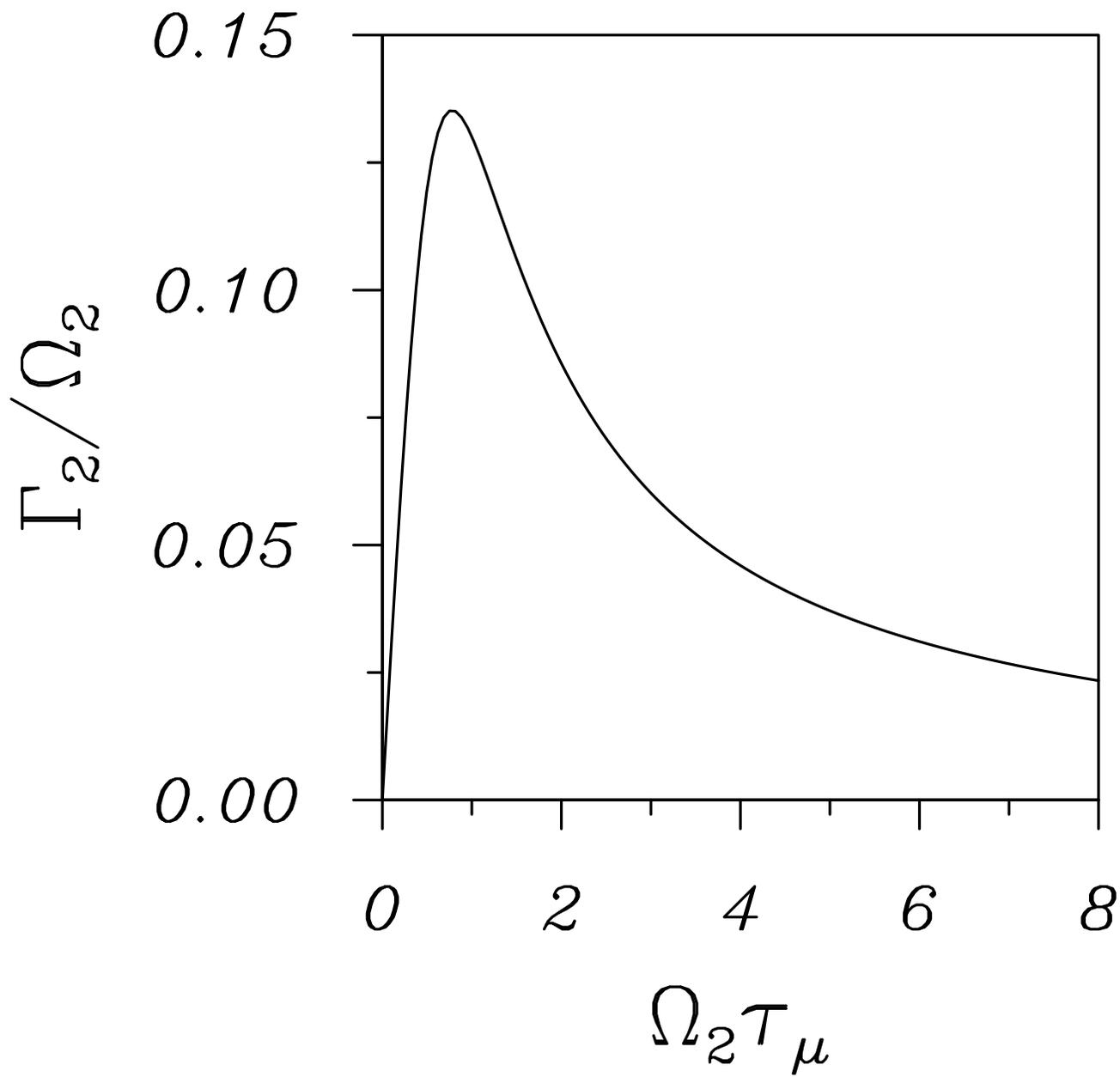
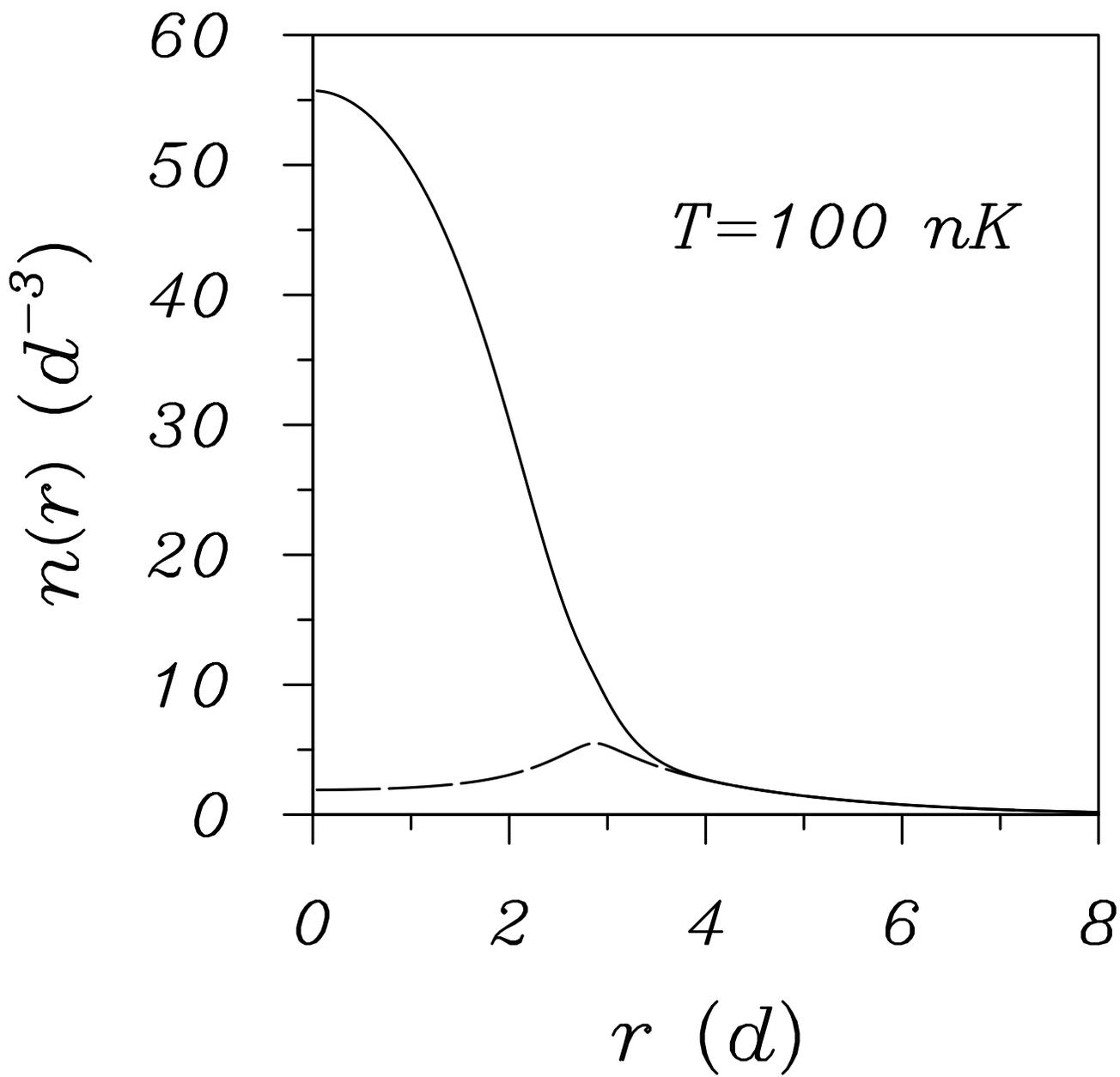
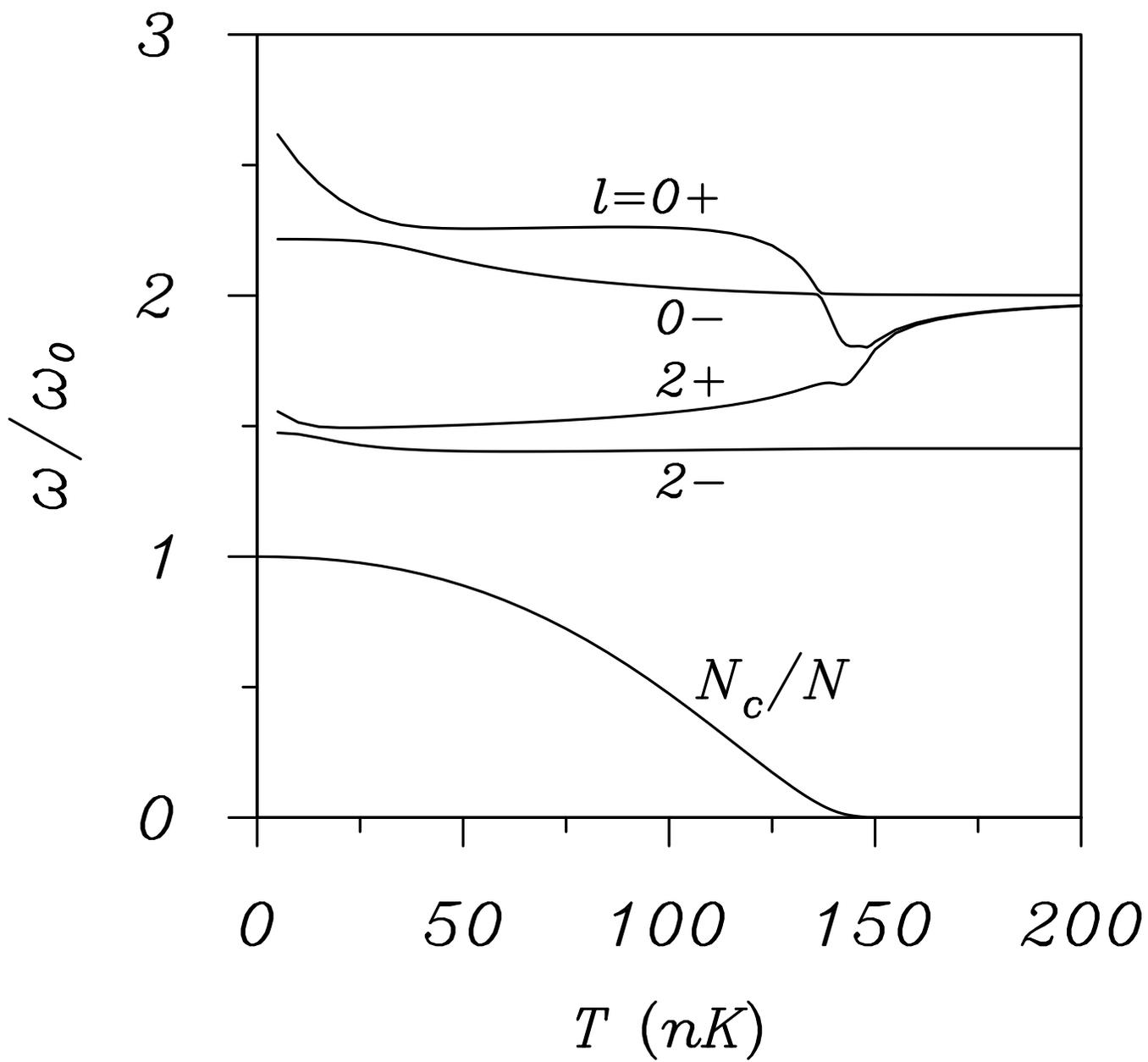
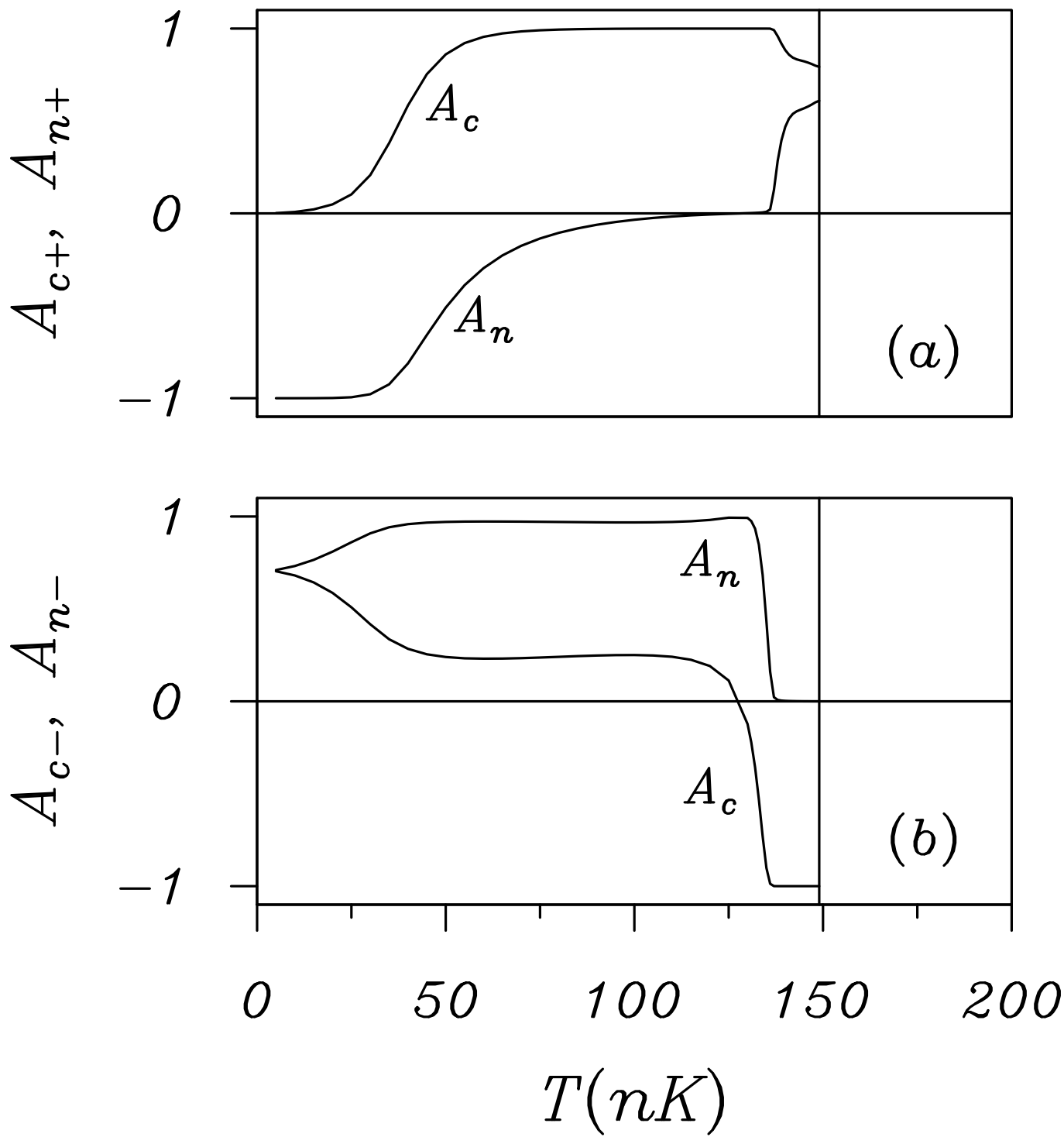
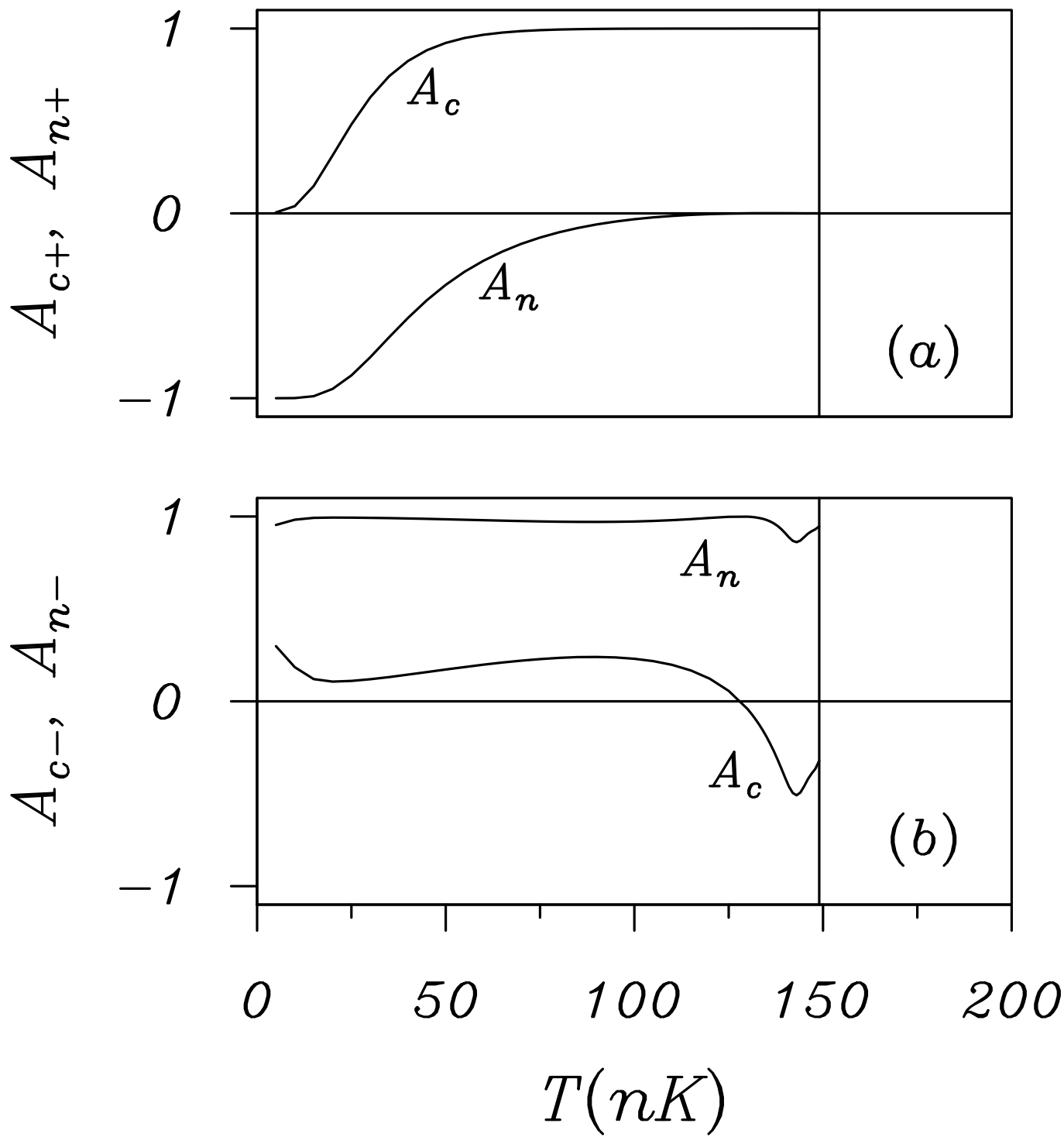
\end{document}

%% file: condensate.tex
\section {Hydrodynamic equations for the condensate atoms}
\label{section2}
We start with the usual Heisenberg equation of motion for the quantum
field operator,
\begin{equation}
i\hbar {\partial \hat \psi ({\bf r},t) \over \partial t} = 
[\hat \psi ({\bf r},t),\hat H] = \left ( -{\hbar^2 \nabla^2
\over 2m} + U_{ext}({\bf r}) \right ) \hat \psi ({\bf r},t) 
+ g \hat \psi^\dagger ({\bf r},t) \hat \psi ({\bf r},t) \hat 
\psi ({\bf r},t)\,,
\label{eq1}
\end{equation}
where $U_{ext}({\bf r})$ is the confining potential. We have here
assumed that the interaction potential can be represented as a
short-ranged pseudopotential of strength $g =
4\pi a \hbar^2/m$, where $a$ is the $s$-wave scattering length.
The equation for the condensate wave function is obtained by taking an
average of (\ref{eq1}) with respect to a broken symmetry 
nonequilibrium ensemble in
which the field operator takes a non-zero expectation value (denoted by
angular brackets),
\begin{equation}
\Phi({\bf r},t) = \langle \hat \psi ({\bf r},t) \rangle\,.
\label{eq2}
\end{equation}
Introducing the usual definition of the non-condensate field operator 
$\tilde \psi ({\bf r},t)$ according to
\begin{equation}
\hat \psi ({\bf r},t) = \Phi({\bf r},t) + \tilde \psi ({\bf
r},t)\,,
\label{eq3}
\end{equation}
with $\langle \tilde \psi ({\bf r},t) \rangle = 0$, the 
expectation value of (\ref{eq1}) thus yields
\begin{eqnarray}
i\hbar {\partial \Phi ({\bf r},t) \over \partial t} = \Bigg ( 
 -{\hbar^2 \nabla^2 \over 2m} &+& U_{ext}({\bf r}) 
+ g n_c({\bf r},t) + 2g \tilde n({\bf r},t) \Bigg ) \Phi
({\bf r},t) \nonumber \\
&+& g \tilde m({\bf r},t) \Phi^*({\bf r},t) 
+ g \langle
\tilde \psi^\dagger ({\bf r},t) \tilde \psi ({\bf r},t) \tilde \psi
({\bf r},t) \rangle\,,
\label{eq4}
\end{eqnarray}
where $n_c({\bf r},t) = | \Phi({\bf r},t) |^2$ is the
non-equilibrium density of atoms in the condensate and
$\tilde n({\bf r},t) = \langle \tilde \psi^\dagger ({\bf r},t)
\tilde \psi ({\bf r},t) \rangle$ is the non-equilibrium non-condensate
density. Eq.~(\ref{eq4}) also involves the off-diagonal non-condensate
density $\tilde m({\bf r},t) = \langle \tilde \psi ({\bf r},t)
\tilde \psi ({\bf r},t) \rangle$ and the three-field correlation
function $\langle \tilde \psi^\dagger \tilde \psi \tilde \psi
\rangle$, both of which in principle have non-zero expectation values 
because of the assumed Bose broken symmetry. We note that 
(\ref{eq4}) is formally exact within the pseudopotential 
approximation. A more detailed discussion starting from the bare pair
interaction is given in Ref.~\cite{proukakis98}.

It is useful at this point to discuss some standard approximations to
(\ref{eq4}). It reduces to the well-known Gross-Pitaevskii (GP) 
equation\cite{dalfovo98} if all the atoms are in the condensate 
(i.e., $\tilde n = 0$) and the anomalous correlations ($\tilde m$ and 
$\langle \tilde \psi^\dagger \tilde \psi \tilde \psi \rangle$) are 
ignored. This is a very good
approximation for $T \ll T_{BEC}$; at $T=0$, the non-condensate 
fraction in trapped atomic gases is estimated to be less than 
$1\%$\cite{javanainen96,hutchinson97}.
As a first step in dealing with finite temperatures, Hutchinson 
{\it et al.}\protect\cite{hutchinson97} (referred to in the 
following as HZG) calculated the excitations in the HFP 
approximation\cite{griffin96}.  The thermal occupation of these
quasiparticle excitations provides an expression for the equilibrium
non-condensate density $\tilde{n}_0({\bf r})$, while the  
condensate density $n_{c0}({\bf r})\equiv|\Phi_0({\bf r})|^2$ is
obtained from the solution of a generalized GP equation.
The self-consistent calculation of these two densities determines the
equilibrium properties of the trapped gas within the Popov approximation
and the excitation energies are themselves identified with the 
collective excitations of the condensate. In the context of 
Eq.~(\ref{eq4}), the HZG approximation\cite{hutchinson97} 
consists of neglecting
$\tilde m$ and $\langle\tilde\psi^{\dagger}\tilde\psi\tilde\psi
\rangle$ and solving the linearized version of the
resulting equation with the additional approximation
$\tilde{n}({\bf r}, t)\simeq\tilde{n}_0({\bf r})$. In
physical terms, this latter approximation corresponds to 
the condensate atoms moving in the {\it static} Hartree-Fock field 
due to the non-condensate atoms. That is, only the dynamics of the
Hartree mean field of the condensate atoms is treated self-consistently,
while the {\it collective} dynamics of the non-condensate atoms is 
completely ignored.

Returning to the exact equation for the macroscopic wavefunction in
(\ref{eq4}), it is useful to introduce  phase and amplitude 
variables, $\Phi({\bf r}, t)=|\Phi({\bf r}, t)|e^{i\theta({\bf r}, t)}$.
After a little algebra, one finds that (\ref{eq4}) is equivalent to
\begin{mathletters}
\bea
&&{\partial n_c \over \partial t} + \bbox{\nabla}\cdot(n_c{\bf v}_c)= 
\frac{2g}{\hbar}{\rm Im}[(\Phi^*)^2 \tilde m + \Phi^*
\langle\tilde\psi^{\dagger}\tilde\psi\tilde\psi\rangle],
\label{eq5a}
\\
&&\hbar\frac{\partial\theta}{\partial t}=-\mu_c-\frac{\hbar^2}{2m}
(\bbox{\nabla}\theta)^2=-(\mu_c+\frac{1}{2}mv_c^2).
\label{eq5b}
\eea
\label{eq5}
\end{mathletters}

\noindent
Here the new local variables are the condensate density
\begin{equation}
n_c({\bf r}, t)\equiv  |\Phi({\bf r}, t)|^2, 
\label{eq6}
\end{equation}
and the condensate velocity
\begin{equation}
{\bf v}_c({\bf r}, t) \equiv  \frac{\hbar}{m}\bbox{\nabla}
\theta({\bf r}, t),
\label{eq7}
\end{equation}
while the local condensate chemical potential is defined by
\bea
\mu_c({\bf r}, t) \equiv
 -{\hbar^2{\bbox\nabla}^2\sqrt{n_c({\bf r}, t)}\over
2m\sqrt{n_c({\bf r}, t)}} &+& U_{ext}({\bf r}) +gn_c({\bf r}, t) +
2g\tilde{n}({\bf r,} t)\nonumber \\
&+& \frac{g}{\hbar n_c}{\rm Re}[(\Phi^*)^2 \tilde m + \Phi^*
\langle \tilde\psi^{\dagger}\tilde\psi\tilde\psi\rangle].
\label{eq8}
\eea
The condensate velocity in our theory is clearly identifiable 
with the superfluid velocity in the usual discussions of
superfluid hydrodynamics. It will be seen later when we discuss the
kinetic equation for the excited atoms that, in our model, the
condensate density $n_c$ can be identified with the superfluid density
as defined in the Landau theory~\cite{landau41,khalatnikov65}.

One may give a physical interpretation of the generalized condensate 
equation in (\ref{eq5b}). In static thermal equilibrium, we have
${\bf v}_c=0$ and all variables are time-independent. In this case, the
RHS of (\ref{eq5a}) must vanish and (\ref{eq5b}) has the solution 
$\theta(t)=-\mu_{c0}t/\hbar$.  
Thus it is clear that in thermal equilibrium, a condensate atom has an 
energy $\varepsilon_{c0}$ equal to $\mu_{c0}$, the equilibrium chemical
potential. We now suppose that we are in a regime where the variables 
$\mu_c$ and $\bv_c$ are slowly-varying in space and time. In this
situation, we have 
\bea
\theta(\br,t)&\simeq& \theta(\br_0,t_0)+{\partial \theta \over \partial
t}(t-t_0) +\bbox{\nabla}\theta \cdot (\br-\br_0) + \dots \nonumber \\
&\equiv& \theta(\br_0,t_0)-\frac{1}{\hbar} \varepsilon_c(\br_0,t_0)
(t-t_0) +\frac{1}{\hbar} m\bv_c(\br_0,t_0)\cdot(\br-\br_0)+\dots\,,
\label{eq9}
\eea
where $\varepsilon_c(\br_0,t_0) \equiv \mu_c(\br_0,t_0)+
\frac{1}{2}mv_c^2(\br_0,t_0)$. With this phase variation of the
condensate wavefunction, it is natural to interpret $\varepsilon_c$ as
the local energy of a condensate atom and $\bv_c$ as its local velocity.
This interpretation is consistent with the energy and momentum 
conservation laws we derive later when we consider collisions between 
condensate and non-condensate atoms (see, in particular, the discussion
at the end of Section~\ref{section3} and in Appendix A).

The condensate equations summarized by (\ref{eq5}-\ref{eq8}) are
formally exact, but are not a closed set of equations since they depend
on the additional non-condensate variables $\tilde n$, $\tilde m$ and  
$\langle \tilde \psi^\dagger \tilde \psi \tilde \psi \rangle$.
As in Ref.\cite{zaremba98}, we restrict ourselves in the present paper 
to the Popov approximation\cite{griffin96} which corresponds to 
neglecting the anomalous pair correlation $\tilde m$ in (\ref{eq5a}) and
(\ref{eq8}). Some justification for this approximation is given in
Appendix A.
However a crucial difference from ZGN is that we retain 
the three-field correlation function $\langle \tilde \psi^\dagger 
\tilde \psi \tilde \psi \rangle$ in (\ref{eq5a}) and (\ref{eq8}) which
we evaluate in an approximation consistent with setting $\tilde m$ to
zero. We should remark that the importance of $\langle \tilde 
\psi^\dagger \tilde \psi \tilde \psi \rangle$ when dealing with
collisions involving condensate atoms has also been emphasized in
Ref.~\cite{proukakis98}.

%% file: kinetic.tex
\section {Derivation of a quantum Boltzmann equation}
\label{section3}
In this section, we turn to  the dynamics of the non-condensate. 
The physical properties of interest are in principle defined by the 
following equation of
motion obtained from (\ref{eq1}) and (\ref{eq4}),
\bea
i\hbar {\partial \tilde \psi \over \partial t} = \Bigg ( -{\hbar^2
\nabla^2 \over 2m} &+& U_{ext} + 2gn \Bigg ) \tilde \psi - 2g\tilde n
\tilde \psi + g\Phi^2 \tilde \psi^\dagger \nonumber \\ &+&g\Phi^*(\tilde
\psi \tilde \psi - \tilde m) + 2g\Phi (\tilde \psi^\dagger \tilde \psi
-\tilde n) + g(\tilde \psi^\dagger \tilde \psi \tilde \psi -
\langle \tilde \psi^\dagger \tilde \psi \tilde \psi \rangle )\,,
\label{eq10}
\eea
where $n=n_c + \tilde n$ is the total density. This equation preserves 
$\langle \tilde \psi \rangle = 0$ as a function of time. It will allow
us to derive a kinetic equation for the excited atoms.
Following Kirkpatrick and Dorfman\cite{kirkpatrick}, it
is convenient to define the time evolution of $\tilde \psi$ by
\beq
\tilde \psi (\brt) = \hat S^\dagger (t,t_0) \tilde \psi (\brt_0) 
\hat S(t,t_0)\,,
\label{eq11}
\eeq
where the unitary operator $\hat S(t,t_0)$ evolves according to
the equation of motion
\beq
i\hbar {d\hat S(t,t_0) \over dt} = \hat H_{\rm eff}(t) \hat S(t,t_0)\,,
\label{eq12}
\eeq
with $\hat S(t_0,t_0) = 1$. Here, $t_0$ is the time at which the 
initial nonequilibrium density matrix $\hat \rho(t_0)$ is specified.
The effective Hamiltonian in (\ref{eq12}) is given by
\bea
\hat H_{\rm eff}(t) &=& \hat H_0(t) + \hat H'(t)\nonumber \\
\hat H'(t) &=& \hat H_1^{'}(t)+\hat H_2^{'}(t) + \hat H_3^{'}(t)
+ \hat H_4^{'}(t)\,,
\label{eq13}
\eea
where the various contributions are defined as 
\bea
\hat H_0(t) &=& \int d\br \, \tilde \psi^\dagger \Bigg ( -{\hbar^2
\nabla^2 \over 2m} + U(\brt) \Bigg ) \tilde \psi \nonumber \\
\hat H_1^{'}(t) &=& \int d\br \left ( L_1(\brt) \tilde \psi^\dagger 
+ L_1^*(\brt) \tilde \psi \right ) \nonumber \\ \hat H_2^{'}(t) 
&=& {g \over 2}\int d\br \left ( \Phi^2(\brt) \tilde \psi^\dagger
\tilde \psi^\dagger  + \Phi^{*2}(\brt) \tilde \psi \tilde \psi 
\right )  \nonumber \\ \hat H_3^{'}(t) 
&=& g\int d\br \left ( \Phi^*(\brt) \tilde \psi^\dagger
\tilde \psi \tilde \psi + \Phi(\brt) \tilde \psi^\dagger \tilde 
\psi^\dagger \tilde \psi \right )\nonumber \\
\hat H_4^{'}(t) &=& {g\over 2}\int d\br \, \tilde \psi^\dagger \tilde 
\psi^\dagger \tilde \psi \tilde \psi - 2g\int d\br \, \tilde n(\brt)
\tilde \psi^\dagger \tilde \psi \nonumber \\
U(\brt) &=& U_{ext}(\br) + 2gn(\brt) = U_{ext}(\br) + 
2g[ \tilde n(\brt) + n_c(\brt)] \nonumber \\
L_1(\brt) &=& -g \left ( 2\tilde n(\brt) \Phi(\brt) + \tilde m(\brt)
\Phi^*(\brt) + \langle \tilde \psi^\dagger \tilde \psi \tilde \psi
\rangle \right )\,.
\label{eq14}
\eea
It is understood that in all these expressions the arguments of the 
field operators are $(\brt_0)$; the dependence of the various terms of
the Hamiltonian on $t$ arises through the mean-field 
expectation values. It can be shown that
Eqs.~(\ref{eq11})-(\ref{eq14}), together with the equal time
commutator $[\tilde \psi(\brt_0),\tilde \psi^\dagger(\br',t_0)]=$
$\nobreak{\delta(\br-\br')}$,
reproduce the equation of motion for $\tilde \psi$
in (\ref{eq10}), as well as that for $\tilde \psi^\dagger$.
In writing (\ref{eq13}) we consider $\hat H'(t)$ to be a perturbation to
the zeroth-order Hamiltonian $\hat H_0(t)$. Noting that $U(\brt)$ is the
total Hartree-Fock (HF) mean field, $\hat H_0(t)$ defines
excitations of the system at the level of the time-dependent HF 
approximation. Other choices of the zeroth-order Hamiltonian are also
possible. For example, $\hat H_0(t)$ could be combined with $\hat
H_2'(t)$ to define a dynamic Bogoliubov Hamiltonian which would then
lead to local Bogoliubov quasiparticle excitations. 
This possible extension~\cite{kirkpatrick}, which would be appropriate 
when considering very low temperatures, will not be considered in 
this paper.

We now consider an arbitrary operator $\hat O(t)$ which is made up of
some combination of non-condensate field operators $\tilde \psi (\brt)$ 
and $\tilde \psi^\dagger (\brt)$, for example, the local density
operator $\tilde
n(\brt) = \tilde \psi^\dagger (\brt)\tilde \psi(\brt)$. By making 
use of (\ref{eq11}), the expectation value of $\hat O(t)$ 
with respect to the initial density matrix $\hat \rho(t_0)$ can 
be expressed as
\bea
\langle \hat O(t) \rangle \equiv \langle \hat O \rangle_t 
&=& {\rm Tr} \hat \rho(t_0) \hat O(t)\nonumber
\\ &=& {\rm Tr} \tilde \rho(t,t_0) \hat O(t_0)\,,
\label{eq15}
\eea
where $\tilde \rho(t,t_0) \equiv \hat S(t,t_0) \hat \rho(t_0) \hat
S^\dagger (t,t_0)$ satisfies the following equation
\beq
i\hbar {d\tilde \rho(t,t_0) \over dt} = [\hat H_{\rm eff}(t),\tilde
\rho(t,t_0)]\,.
\label{eq16}
\eeq
We refer to Appendix A for further discussion of this equation of
motion.

Our ultimate objective is to obtain a quantum kinetic equation for 
the non-condensate atoms. For this purpose, we define the Wigner 
{\it operator}
\beq
\hat f(\bp,\br,t_0) \equiv \int d\br'\,e^{i\bp \cdot \br'/\hbar}\tilde 
\psi^\dagger (\br + {\br' \over 2},t_0) 
\tilde \psi (\br -{\br'\over 2},t_0)\,.
\label{eq17}
\eeq
Its expectation value according to Eq.~(\ref{eq15}) 
then yields the Wigner distribution {\it function}
\beq
f(\bp,\br,t)= {\rm Tr} \tilde \rho(t,t_0) \hat f(\bp,\br,t_0)\,.
\label{eq18}
\eeq
Knowledge of this function allows one to calculate various 
nonequilibrium expectation values, such as the non-condensate density
\beq
\tilde n(\brt) = \int {d\bp \over (2\pi\hbar)^3} f(\bp,\br,t)\,.
\label{eq19}
\eeq

Using (\ref{eq16}), the equation of motion for $f$ is
\bea
{\partial f(\bp,\br,t) \over \partial t} &=& {1\over i\hbar}
{\rm Tr} \tilde \rho(t,t_0) [\hat f(\bp,\br,t_0),\hat H_{\rm eff}(t) ] 
\nonumber \\ &=& {1\over i\hbar} {\rm Tr} \tilde \rho(t,t_0) 
[\hat f(\bp,\br,t_0),\hat H_0(t) ] 
+ {1\over i\hbar} {\rm Tr} \tilde \rho(t,t_0) 
[\hat f(\bp,\br,t_0),\hat H'(t) ]\,.
\label{eq20}
\eea
The first term on the right hand side of (\ref{eq20}) defines the
free-streaming operator in the kinetic equation. With the assumption
that $U(\br,t)$ varies slowly in space, we then have
\beq
{\partial f(\bp,\br,t) \over \partial t} + {\bp \over m} \cdot
\bbox{\nabla} f(\bp,\br,t)- \bbox{\nabla} U \cdot
\bbox{\nabla}_{\bp} f(\bp,\br,t)  = {1\over i\hbar}
{\rm Tr} \tilde \rho(t,t_0) [\hat f(\bp,\br,t_0),\hat H'(t) ] \,.
\label{eq21}
\eeq
The right hand side of this equation clearly represents the effect of
collisions between the atoms. The reduction of this term to the form of
a binary collision integral is a lengthy exercise (some details are
given in Appendix A). However, the final result has a
physically transparent form. The collision integral is the sum of two
contributions:  
\beq
\left . {\partial f\over\partial t}\right|_{coll}=C_{12}[f]+C_{22}[f]\,,
\label{eq22}
\eeq
where
\begin{mathletters}
\bea
C_{22}[f] &=& {2g^2\over (2\pi)^5 \hbar^7}
 \int d{\bf p}_2\int d{\bf p}_3
\int d{\bf p}_4\delta ({\bf p}+{\bf p}_2 
-{\bf p}_3 -{\bf p}_4)\nonumber\\
&&\times \delta(\tilde \varepsilon_{p}+\tilde \varepsilon_{p_2}
-\tilde \varepsilon_{p_3}-\tilde \varepsilon_{p_4})
\left[(1+f)(1+f_2)f_3f_4-ff_2(1+f_3)(1+f_4)\right],
\label{eq23b}
\eea
\bea
C_{12}[f]&=&{2 g^2 n_c \over (2\pi)^2\hbar^4} \int d{\bf p}_1 \int
d{\bf p}_2 \int d{\bf p}_3
\delta(m{\bf v}_c+{\bf p}_1-{\bf p}_2-{\bf p}_3) \cr
&&\times
\delta(\varepsilon_c+\tilde \varepsilon_{p_1}
-\tilde \varepsilon_{p_2}-\tilde \varepsilon_{p_3}) 
[\delta({\bf p}-{\bf p}_1)-\delta({\bf p}-{\bf p}_2)
-\delta({\bf p}-{\bf p}_3)] \cr
&&\times [(1+f_1)f_2f_3-f_1(1+f_2)(1+f_3)],
\label{eq23a}
\eea
\label{eq23}
\end{mathletters}

\noindent
with $f \equiv f(\bp,\br, t),\, f_i\equiv f(\bp_i,\br,t)$.
We also note that the variables $n_c$, $\bv_c$, $\varepsilon_c$ 
and $\tilde \varepsilon_p$ are
all functions of $\br$ and $t$.
As a result of Bose statistics, the creation of an atom in a state 
$i$ is associated with the statistical factor $(1+f_i)$.
Clearly $C_{22}$ describes two-body collisions between excited atoms 
(2 atoms $\rightleftharpoons$ 2 atoms) while $C_{12}$ describes 
collisions which involve one condensate atom (1 atoms 
$\rightleftharpoons$ 2 atoms). The momentum and energy delta functions 
in the $C_{12}$ collision term take into account the fact that 
a condensate atom locally has energy $\varepsilon_c(\brt)$ defined in
(\ref{eq9}), and momentum $m{\bf v}_c$. On the other hand, the excited
atoms locally have the HF energy
\beq
\tilde \varepsilon_p({\bf r}, t) ={p^2\over 2m} + U({\bf r}, t)\,.
\label{eq24}
\eeq
We note that the HF mean-field defined in (\ref{eq14}) involves the 
condensate 
density $n_c({\bf r}, t)\equiv |\Phi({\bf r}, t)|^2,$  which must
be determined self-consistently using the condensate equation of 
motion in (\ref{eq4}). 

To complete our derivation of a closed set of equations for both the
condensate and non-condensate, we must finally deal with the {\it
off-diagonal} HF self-energy $\tilde m$ and the term $\langle
\tilde \psi^\dagger \tilde \psi \tilde \psi \rangle$ which appear in
(\ref{eq5}) and (\ref{eq8}). We show in Appendix A that 
$\tilde m$ can be consistently neglected, and that
\begin{eqnarray}
\langle \tilde\psi^{\dagger}\tilde\psi\tilde\psi\rangle
&=&-{ig\Phi\over (2\pi)^5\hbar^6}
\int d{\bf p}_1\int d{\bf p}_2
\int d{\bf p}_3
\delta(m{\bf v}_c+{\bf p}_1-{\bf p}_2-{\bf p}_3) \cr
&&\times \delta(\varepsilon_c+\tilde \varepsilon_{p_1}-
\tilde \varepsilon_{p_2}- \tilde \varepsilon_{p_3})
[f_1(1+f_2)(1+f_3)-(1+f_1)f_2f_3].
\label{eq25}
\end{eqnarray}
We do not display the real part of $\langle \tilde \psi^\dagger 
\tilde \psi \tilde \psi \rangle$ which contributes to the local 
condensate chemical potential defined in (\ref{eq8}). This contribution
is of order $g^2$ and will be neglected since our approximations do not
adequately treat all terms of this order (see Appendix A for further
discussion).  However, the imaginary part of $\langle 
\tilde \psi^\dagger \tilde \psi \tilde \psi \rangle$ shown in
(\ref{eq25}) contributes an important source
term, $-\Gamma_{12}[f]$, to the right hand side of the continuity 
equation in (\ref{eq5a}). Comparing (\ref{eq25}) with the expression for
$C_{12}[f]$ in (\ref{eq23a}), this source term is given by
\beq
-\Gamma_{12}[f] \equiv {2g\over \hbar} {\rm Im} [\Phi^*\langle 
\tilde\psi^{\dagger}\tilde\psi\tilde\psi\rangle ] =
- \int {d\bp \over (2\pi\hbar)^3} C_{12}[f(\bp,\br,t)]\,.
\label{eq26}
\eeq
The latter form in (\ref{eq26}) emphasizes the fact that the $C_{12}$
collisions
can locally change the relative number of condensate and non-condensate
atoms and that separate number conservation laws no longer hold
for each of these components. As we shall see, the close connection
between $\langle \tilde \psi^\dagger \tilde \psi \tilde \psi \rangle$ in
(\ref{eq25}) and the $C_{12}$ collision integral in (\ref{eq23a})
is essential for the conservation of the {\it total} number of 
particles. This indicates that the approximations leading to our final 
equations for the condensate and non-condensate are internally
consistent. 

To summarize the results of this section, as well as that of 
Section \ref{section2},
we write out the final form of the equations we have derived for
the dynamics of a trapped Bose-condensed 
gas at finite temperatures. For the condensate, we have the equations
\begin{mathletters}
\bea
{\partial n_c \over \partial t} + \bbox{\nabla}\cdot(n_c{\bf v}_c)&=& 
-\Gamma_{12}[f] \label{eq27a} \\
m\left({\partial\over\partial t}+{\bf v}_c\cdot 
\bbox{\nabla}\right) \bv_c&=&-\bbox{\nabla}\mu_c\,,
\label{eq27b}
\eea
\label{eq27}
\end{mathletters}
where the condensate chemical potential is given by
\beq
\mu_c({\bf r}, t) =
 -{\hbar^2{\bbox\nabla}^2\sqrt{n_c({\bf r}, t)}\over
2m\sqrt{n_c({\bf r}, t)}} + U_{ext}({\bf r}) +gn_c({\bf r}, t) +
2g\tilde{n}({\bf r,} t)\,.
\label{eq28}
\eeq
As we have discussed in detail, these
equations are approximations to the exact equations given in
(\ref{eq5}) and (\ref{eq8}).
For the non-condensate we have the kinetic equation
\beq
{\partial f(\bp,\br,t) \over \partial t} + {\bp \over m} \cdot
\bbox{\nabla} f(\bp,\br,t)- \bbox{\nabla} U \cdot
\bbox{\nabla}_{\bp} f(\bp,\br,t)  = C_{12}[f] + C_{22}[f]\,,
\label{eq29}
\eeq
with the two collision terms given explicitly by the expressions in 
(\ref{eq23}). These equations are
coupled through the HF mean fields and must therefore be solved
self-consistently. An equation equivalent to (\ref{eq27}) was derived by
Proukakis {\it et al.}~\cite{proukakis98}, while our kinetic equation in
(\ref{eq29}) is implicitly contained in the work of
Stoof~\cite{stoof99}.

It is apparent that the equations in (\ref{eq27}) are already in the
form of ``hydrodynamic" equations for the condensate variables. In
Section \ref{section4}, we use (\ref{eq29}) to derive analogous 
hydrodynamic equations for the non-condensate atoms.
Although these equations have been derived for a trapped gas at finite
temperatures, we note that they go over smoothly to the correct
$T = 0$ theory, since as $f(\bp,\br,t)$ and hence $\tilde n$ become
negligible, the 
condensate equations (\ref{eq27}) and (\ref{eq28}) reduce to the 
standard Gross-Pitaevskii dynamics of a pure condensate at $T = 0$ 
\cite{dalfovo98,stringari96}. We therefore expect our equations to
provide a reasonable description of the dynamics over the full range of
temperatures from the strongly degenerate limit well below $T_{BEC}$ 
through to the classical high-temperature limit.

Before leaving this Section, we compare these results with some earlier
theoretical studies. 
For temperatures above the Bose-Einstein transition $(T>T_{\rm BEC})$ 
where $C_{12}$ vanishes, the kinetic equation (\ref{eq29}) reduces to 
the well-known Uehling-Uhlenbeck equation~\cite{uehling33}, as derived
in Ch.~6 of Kadanoff and Baym~\cite{kadanoff62}, for example, and 
studied numerically by several authors\cite{lopez98}. Below
$T_{BEC}$, (\ref{eq29}) is analogous to the kinetic equation 
derived by KD \cite{kirkpatrick} for a homogeneous weakly-interacting 
Bose-condensed gas at finite temperatures. The main difference is that
these authors choose to work with {\it excitations} which are defined 
within the local rest frame of the superfluid (condensate), and thus 
the excitation energies are measured relative to that of the 
condensate atoms. In contrast, our kinetic equation involves {\it atoms}
moving in a self-consistent HF field. It is perhaps useful to make this
comparison more precise. Our $C_{12}[f]$ collision integral involves 
the energy-momentum conservation factors 
\begin{displaymath}
\delta(\varepsilon_c+\tilde \varepsilon_1 -\tilde \varepsilon_2 -\tilde
\varepsilon_3) \delta(m\bv_c+\bp_1-\bp_2-\bp_3)
\end{displaymath}
where $\tilde \varepsilon_i = p_i^2/2m + 2g(\tilde n + n_c)$ 
(assuming $U_{ext} = 0$ for simplicity) and 
$\varepsilon_c = \mu_c + {1\over 2}m v_c^2$ is the local energy of an 
atom in the condensate. On the other hand, in the work of KD,
$C_{12}[f]$ contains the energy-momentum conservation factors
\begin{displaymath}
\delta(E_1-E_2-E_3)\delta(\bp_1'-\bp_2'-\bp_3')\,,
\end{displaymath}
where $E_i$ is the quasiparticle (Bogoliubov)
excitation energy and $\bp_i' =
\bp_i-m\bv_c$ is the quasiparticle momentum in the local rest frame.
In the ``high-temperature limit" (large momenta), these excitation
energies reduce to $E_i \simeq {p'_i}^2/2m + gn_c = \tilde 
\varepsilon_i -
\mu_c - \bv\cdot\bp_i+{1\over 2}mv_c^2$ where $\mu_c = g(n_c + 2\tilde
n)$ is the local condensate chemical potential as given by 
(\ref{eq28}).  One can easily verify
that the two energy-momentum conservation factors defined above are
completely equivalent in this high-temperature limit, and thus the two 
seemingly different formulations are essentially the same.

It should be noted that the high-temperature limit in the case of a 
trapped Bose gas is not as restrictive as it might seem. The work 
reported in Ref.~\cite{giorgini97} indicates that the differences 
arising from using the HF {\it vs.} Bogoliubov excitation spectrum 
in calculating thermodynamic
properties are very small down to quite low temperatures, in contrast to
the situation for a uniform Bose gas. A more detailed explanation for
the dominance of single-particle excitations in determining the
statistical behaviour of trapped Bose gases can be found in
Ref.~\cite{dalfovo98}.

%% file: hydro.tex
\section {Non-condensate hydrodynamic equations}
\label{section4}
\subsection{General moment equations}
\vskip 0.2 true cm

Following the standard procedure in kinetic theory, we proceed to take 
moments of the kinetic equation in (\ref{eq29}).
It follows immediately from their definitions that
\begin{eqnarray}
\int d{\bf p} \ {\bf p}C_{22}&=&0, \qquad \int d{\bf p} \ 
({\bf p}-m{\bf v}_c)C_{12}=0 \cr
\int d{\bf p} \ \tilde \varepsilon_pC_{22}&=&0, \qquad \int d{\bf p}
 \ (\tilde \varepsilon_p-\varepsilon_c)C_{12}=0.
\label{eq30}
\end{eqnarray}
This shows that both the $C_{12}$ and $C_{22}$ collision processes
conserve energy and momentum.
In the case of $1\rightleftharpoons 2$ collisions, the momentum of
the non-condensate atoms is conserved in the local rest frame of the
condensate; in addition, the energy conservation condition takes into
account explicitly the different mean-field energies of condensate and
non-condensate atoms. In addition to the exact relations in
(\ref{eq30}), we have 
\begin{equation}
\int d{\bf p} \ C_{22}=0\,,
\label{eq31}
\end{equation}
which implies that the number of excited atoms is conserved in 
$C_{22}$ collisions. In contrast, as already mentioned after 
(\ref{eq26}), 
$C_{12}$ does {\it not} conserve the number of excited atoms and thus
we find
\begin{equation}
\Gamma_{12}[f] \equiv \int {d{\bf p}\over (2\pi\hbar)^3}\ C_{12} \neq 0.
\label{eq32}
\end{equation}

Taking (\ref{eq30})--(\ref{eq32}) into account and, after a certain 
amount of rearrangement, the exact moment equations derived from 
(\ref{eq29}) can be written in the form ($\mu$ and $\nu$ denote 
Cartesian components):

\begin{mathletters}
\beq
{\partial{\tilde n}\over\partial t}+\bbox{\nabla}\cdot 
(\tilde{n}{\bf v}_n) = \Gamma_{12}[f]\,,
\label{eq33a}
\eeq
\beq
m{\tilde n}\left({\partial\over\partial t}+{\bf v}_n\cdot 
\bbox{\nabla}\right) v_{n\mu}=-{\partial P_{\mu\nu}\over\partial x_\nu}
-{\tilde n}{\partial U\over\partial x_\mu}
-m(v_{n\mu}-v_{c\mu})\Gamma_{12}[f]\,,
\label{eq33b}
\eeq
\beq
{\partial\tilde\epsilon\over\partial t} +
\nabla\cdot(\tilde\epsilon{\bf v}_n) = -\bbox{\nabla}\cdot{\bf Q}
-D_{\mu\nu} P_{\mu\nu}+
\left[\frac{1}{2}m(\bv_n-\bv_c)^2+\mu_c-U\right]\Gamma_{12}[f].
\label{eq33c}
\eeq
\label{eq33}
\end{mathletters}

\noindent
The non-condensate density is defined in (\ref{eq19}) while
the non-condensate local velocity is defined by
\beq
{\tilde n}({\bf r},t){\bf v}_n({\bf r}, t)\equiv\int{d{\bf p}
\over(2\pi\hbar)^3} {{\bf p}\over m} f(\bp, \br, t)\,.
\label{eq34}
\eeq
In addition, we have

\begin{mathletters}
\begin{eqnarray}
P_{\mu\nu}({\bf r}, t)&\equiv& m
\int{d{\bf p}\over(2\pi\hbar)^3}\left({p_\mu\over m} - v_{n\mu}\right)
\left({p_\nu\over m} - v_{n\nu}\right)
f(\bp, \br, t), \label{eq35a} \\
{\bf Q}({\bf r}, t)&\equiv& \int{d{\bf p}\over(2\pi\hbar)^3}{1\over 2m} 
({\bf p}-m{\bf v}_n)^2\left({{\bf p}\over m}-{\bf v}_n\right)
f(\bp, \br, t),\label{eq35b}  \\
\tilde \epsilon({\bf r}, t) &\equiv&\int{d{\bf p}\over(2\pi\hbar)^3}
{1\over 2m} ({\bf p}-m{\bf v}_n)^2
f(\bp, \br, t)\label{eq35c} \,.
\end{eqnarray}
\label{eq35}
\end{mathletters}
\noindent
The symmetric rate-of-strain tensor appearing in (\ref{eq33c})
is defined by
\beq
D_{\mu\nu}({\bf r}, t) \equiv {1 \over 2} \left({\partial v_{n \mu}
\over \partial x_\nu} + {\partial v_{n \nu} \over \partial x_\mu}
\right).
\label{eq36}
\eeq
It is clear from the addition of (\ref{eq27a}) and (\ref{eq33a}) that 
the {\it total} number of atoms is conserved, as it must be.

The set of equations in (\ref{eq33})--(\ref{eq35}) are formally exact 
but they are obviously not closed since the number of local variables
($\tilde n, n_c, {\bf v}_n, {\bf v}_c$ as well as the stress tensor 
$P_{\mu\nu}$, kinetic energy density $\tilde \epsilon$, heat current 
$\bf Q$, and $\Gamma_{12}$) exceeds the total number (nine) of coupled 
scalar equations. To proceed we must specify the conditions under
which the dynamics of the trapped gas is to be determined. The range of
possibilities include: (i) the collisionless regime in which $C_{12} =
C_{22} = 0$. Here, the collisionless kinetic equation for the 
non-condensate must be
solved explicitly in order to determine the various physical quantities
of interest; (ii) an intermediate regime in which collisions between
excited atoms are sufficiently rapid to establish a state of local 
equilibrium within the non-condensate component, {\it but} in which 
collisions between the condensate and non-condensate, as described be
$C_{12}$, can be neglected. We shall refer to this as a state of
{\it partial local equilibrium} which might be expected to arise near
$T_{BEC}$ when the density of condensate atoms is small (note the
proportionality of $C_{12}$ to $n_c$); and (iii) a regime in which the
collision rates are so high that a condition of {\it complete local
equilibrium} is established. As will be discussed in detail later, 
regime (iii) is the one conventionally dealt with using Landau two-fluid
hydrodynamics in uniform Bose superfluids~\cite{khalatnikov65}. Of
course, the most general situation will not fall neatly into any of the
categories listed so far and will require a detailed solution of the
kinetic equation in (\ref{eq29}) together with the
quantum hydrodynamic equations for the condensate.

We shall not deal with the collisionless regime (i) in the present
paper, but note that some work along these lines has recently 
appeared~\cite{bijlsma98}. Instead, we
shall focus on the other two regimes indicated above, in which the
$C_{12}$ and $C_{22}$ collisions play a crucial role. Of particular
interest is the transition from the partial to complete local
equilibrium conditions. An analysis of this regime will clarify the way
in which the Landau theory emerges within the context of our explicit
microscopic model. This turns out to be a rather subtle problem, which 
was not fully appreciated in our earlier work\cite{zaremba98,griffin97},
nor addressed in the analogous work of KD~\cite{kirkpatrick}.

\subsection{Partial local equilibrium}

We now consider the regime in which Eqs.~(\ref{eq33})--(\ref{eq35})
reduce to a closed set of equations for the non-condensate variables.
When the collision rate among the excited atoms is high, the collision 
integral $C_{22}$ drives the distribution function $f(\bp,\br,t)$ 
towards the ``local equilibrium" Bose distribution
\begin{equation}
\tilde f(\bp,\br,t)=
\frac{1}{e^{\beta[\frac{1}{2m}({\bf p}-m{\bf v}_n)^2+U-\tilde \mu]}-1},
\label{eq37}
\end{equation}
where now the temperature parameter $\beta$, local fluid velocity 
${\bf v}_n$, chemical potential $\tilde \mu$ and mean field $U$ are all
functions of ${\br}$ and $t$. It is important to appreciate that the
local chemical potential $\tilde \mu$ which appears here is distinct 
from the local chemical potential $\mu_c$ defined in (\ref{eq28}) for 
the condensate.  How these 
two chemical potentials acquire a common value in the limit of complete
local equilibrium requires a careful analysis of the effect of $C_{12}$
(see Section \ref{section5}). For the moment, we shall
simply note the consequences that (\ref{eq37}) has for the collision
integrals.

First, one may immediately verify that the local equilibrium form of 
$\tilde f$ in (\ref{eq37}) guarantees that
\begin{equation}
C_{22}[\tilde f]=0\,.
\label{eq38}
\end{equation}
Indeed, it is this condition that defines the ``local equilibrium"
solution.
The result is independent of the value of $\tilde \mu$ and only
makes use of the energy and momentum conservation factors in 
the definition of $C_{22}$ in (\ref{eq23b}), and the key identity for 
the Bose distribution
\begin{equation}
1+f(x)=\frac{e^x}{e^x-1}=-f(-x)\,.
\label{eq39}
\end{equation}
In contrast, one finds from (\ref{eq23a}) that $C_{12}[\tilde f]\neq0$, 
namely
\begin{eqnarray}
C_{12}[\tilde f]&=&{2 g^2 n_c\over (2\pi)^2\hbar^4} \left [
1 - e^{-\beta(\tilde \mu-\frac{1}{2}m({\bf v}_n-{\bf v}_c)^2-\mu_c)}
\right ] \cr && \times \int d{\bf p}_1
\int d{\bf p}_2 \int d{\bf p}_3
\delta(m{\bf v}_c+{\bf p}_1-{\bf p}_2-{\bf p}_3)
\delta(\varepsilon_c+\tilde \varepsilon_1
-\tilde \varepsilon_2-\tilde \varepsilon_3) \cr
&& \times [\delta({\bf p}-{\bf p}_1)-\delta({\bf p}-{\bf p}_2)
-\delta({\bf p}-{\bf p}_3)]
(1+\tilde f_1)\tilde f_2 \tilde f_3.
\label{eq40}
\end{eqnarray}
A factor similar to the first square bracket in (\ref{eq40}) appears in
the growth equation derived by Gardiner {\it et al.}~\cite{gardiner98}.
In their work, our time- and space-dependent $\tilde \mu$ is replaced 
by a constant chemical potential characterizing a non-condensate 
particle reservoir. On the other hand, they use a kinetic description
for low-lying non-condensate states which we describe in terms
of a local equilibrium distribution with local chemical
potential $\tilde \mu$. Thus, although different in detail, there is
considerable overlap in physical content. The important point is that
$C_{12}[\tilde f]$ is in general non-vanishing, reflecting
the fact that excited atoms in the non-condensate need not be in 
local diffusive equilibrium with the condensate atoms ($\tilde \mu \ne
\mu_c$).

Assuming that the collision rate among excited atoms is sufficiently
high, the local equilibrium distribution in (\ref{eq37}) provides the
appropriate zeroth-order solution to the kinetic equation in 
(\ref{eq29}). This is the same reasoning used in classical kinetic
theory~\cite{huang87}. Of
course, even in the absence of the $C_{12}$ collision term, $\tilde f$
is not an exact solution of (\ref{eq29}).  A systematic procedure 
based on the Chapman-Enskog method gives rise to corrections to 
$\tilde f$ which are necessary in order to describe thermal conduction 
and viscosity. A detailed discussion of these damping processes in 
a trapped Bose gas {\it above}
$T_{BEC}$ is given in Refs.~\cite{nikuni98} and  \cite{kavoulakis98}.
Such calculations can be extended to below $T_{BEC}$ and have been
carried out by KD~\cite{kirkpatrick} for a uniform Bose-condensed gas.
However, we ignore these corrections in the present analysis and
proceed with the assumption that $f \simeq \tilde f$ is a good
approximation (see Section~\ref{section8} for further remarks.)

Using
(\ref{eq37}) to evaluate the quantities in (\ref{eq35}), we find that 
the heat current ${\bf Q}(\brt) = 0$, and that
\beq
P_{\mu\nu} ({\bf r}, t) = \delta_{\mu\nu}{\tilde P}({\bf r}, t)
\equiv\delta_{\mu\nu}\int{d{\bf p}\over (2\pi\hbar)^3} {p^2\over 3m}
\left. \tilde f(\bp, \br, t)\right|_{{\bf v}_n=0}\,. 
\label{eq41}
\eeq
In addition, we note that the kinetic energy density 
(\ref{eq35c}) is given by
\beq
{\tilde \epsilon(\brt) ={3\over 2} \tilde P (\brt)\,,
\label{eq42}}
\eeq
which is the relation found for a uniform ideal gas. Since the momentum 
dependence of the integrand in (\ref{eq41}) is the same as for a static
equilibrium Bose distribution, the integrations can be carried 
out explicitly to give the expression,
\beq
{\tilde P}({\bf r}, t) = {1\over\beta\Lambda^3} g_{5/2}(z)\,,
\label{eq43}
\eeq
where the local thermodynamic variables $\beta$ and $z$
are again functions of $\br$ and
$t$. The Bose-Einstein functions are defined as $g_n(z) 
\equiv\sum^{\infty}_{l=1} z^l/l^n$, the local equilibrium fugacity is
\beq
z(\brt) \equiv e^{\beta(\brt)[\tilde \mu(\brt)- U(\brt)]},
\label{eq44}
\eeq
and the local thermal de Broglie wavelength is
\beq
\Lambda(\brt) \equiv\left({2\pi\hbar^2\over mk_B T(\brt)} \right)^{1/2}.
\label{eq45}
\eeq
Finally, we note that the non-condensate density associated with
(\ref{eq37}) is given by
\beq
\tilde n(\brt) = \int{d{\bf p}\over (2\pi\hbar)^3}
\left. \tilde f(\bp, \br, t)\right|_{{\bf v}_n=0}
= {1\over\Lambda^3} g_{3/2}(z)\,.
\label{eq46}
\eeq

To summarize, using the local equilibrium approximation $f \simeq 
\tilde f$, the hydrodynamic equations in (\ref{eq33}) simplify to
\begin{mathletters}
\beq
{\partial{\tilde n}\over\partial t}+\bbox{\nabla}\cdot 
(\tilde{n}{\bf v}_n) = \Gamma_{12}[\tilde f]\,,
\label{eq46'a}
\eeq
\beq
m{\tilde n}\left({\partial\over\partial t}+{\bf v}_n\cdot 
\bbox{\nabla}\right) \bv_n=-\bbox{\nabla} \tilde P
-{\tilde n}\bbox{\nabla} U
-m(\bv_n-\bv_c)\Gamma_{12}[\tilde f]\,,
\label{eq46'b}
\eeq
\beq
{\partial\tilde P\over\partial t} +
\nabla\cdot(\tilde P{\bf v}_n) = -{2\over 3}\tilde P \bbox{\nabla}
\cdot{\bv_n} + {2\over 3}
\left[\frac{1}{2}m(\bv_n-\bv_c)^2+\mu_c-U\right]\Gamma_{12}[\tilde f],
\label{eq46'c}
\eeq
\label{eq46'}
\end{mathletters}

\noindent
where 
\bea
\Gamma_{12}[\tilde f] &\equiv& \int {d\bp \over (2\pi\hbar)^3}
C_{12}[\tilde f] \nonumber \\
&=& -{2g^2n_c\over (2\pi)^5\hbar^7} \left[ 1-e^{-\beta(\tilde \mu -
\mu_c - {1\over 2}m(\bv_n-\bv_c)^2)} \right ] 
\int d\bp_1 \int d\bp_2 \int d\bp_3 \nonumber \\ 
&& \hskip .9truein \times \delta(m\bv_c + \bp_1 - \bp_2 - \bp_3)
\delta(\varepsilon_c + \tilde \varepsilon_1 - \tilde \varepsilon_2
-\tilde \varepsilon_3) (1+\tilde f_1)\tilde f_2 \tilde f_3\,.
\label{eq46'd}
\eea
It is clear that the $\Gamma_{12}[\tilde f]$ terms in (\ref{eq46'}) 
will play a crucial role in the solution of these hydrodynamic 
equations. In particular, these terms describe the
transfer of atoms between the condensate and non-condensate and
are responsible for bringing the two components into complete local 
equilibrium.

In this paper, we concentrate on solving (\ref{eq46'}) for small 
amplitude oscillations around static (i.e., absolute) equilibrium.
For the non-condensate atoms, absolute equilibrium is
specified by a Bose distribution at a uniform temperature $T_0$ and 
chemical potential $\tilde \mu_0$ (all {\it static} equilibrium 
quantities are indicated by a subscript 0).  The equilibrium density 
and pressure are then given by the expressions $\tilde n_0(\br) =
g_{3/2}(z_0)/\Lambda_0^3$ and $\tilde P_0(\br) =
g_{5/2}(z_0)/\beta_0\Lambda_0^3$, respectively. These are of
the same form as in Eqs.~(\ref{eq46}) and 
(\ref{eq43}), but with all thermodynamic quantities taking on
their equilibrium values. In particular, the equilibrium fugacity
is $z_0({\bf r}) =e^{\beta_0(\tilde \mu_0-U_0({\bf r}))}$ with
$U_0({\bf r})=U_{ext}({\bf r}) + 2g[\tilde{n}_0({\bf r})+
n_{c0}({\bf r})].$ It is clear that the equilibrium condensate density 
$n_{c0}(\br)$ is also required to complete the definition of the 
non-condensate equilibrium variables. This we obtain from the 
equilibrium condensate wavefunction which is a solution of the 
static equation
\beq
\left[-{\hbar^2\nabla^2\over 2m} + U_{ext}({\bf r}) 
+ gn_{c0}({\bf  r}) + 2g\tilde{n}_0({\bf r})\right]\Phi_0({\bf r}) =
\mu_{c0} \Phi_0({\bf r})\,.
\label{eq47}
\eeq
This is the HFP equation for the condensate used 
previously~\cite{hutchinson97,giorgini97}, but with the important
difference that $\tilde n_0$ is determined in the HF approximation as
described above.

The eigenvalue of (\ref{eq47}) can be written as
\beq
\mu_{c0} = -{\hbar^2\nabla^2\sqrt{n_{c0}({\bf r})}\over 2m\sqrt{n_{c0}
({\bf r})}} + U_{ext}({\bf r}) + gn_{c0}({\bf r}) +  
2g\tilde{n}_0({\bf r})\,,
\label{eq48}
\eeq
and is identified with the static equilibrium condensate chemical 
potential. To see the 
connection between $\tilde \mu_0$ and $\mu_{c0}$, we refer to
the expression for $C_{12}[\tilde f]$ in Eq.~(\ref{eq40}). 
With $\bv_n = \bv_c = 0$, we note that the factor in
the square brackets vanishes if $\tilde \mu_0 = \mu_{c0}$, which is 
precisely the expected condition for diffusive
equilibrium between the condensate and non-condensate. Thus the
constraint $\tilde \mu_0 = \mu_{c0}$
must be imposed in order to obtain the self-consistent solution
for $\tilde{n}_0(\bf{r})$ and $n_{c0}(\bf{r})$. Further discussion
regarding the static equilibrium properties of a trapped Bose gas is 
given in Appendix B.

\subsection{Linearized hydrodynamic equations}

To study the collective modes, we linearize the system of equations in
(\ref{eq46'}) around the static equilibrium solutions given above 
to obtain the non-condensate hydrodynamic equations
\begin{mathletters}
\bea
{\partial\delta \tilde{n}\over\partial t} &=& -\bbox{\nabla}\cdot
(\tilde{n}_0\delta{\bf v}_n)+\delta \Gamma_{12}, \label{eq49a}\\
m\tilde{n}_0{\partial\delta{\bf v}_n\over\partial t} &=& -\bbox{\nabla}
\delta\tilde{P}-\delta\tilde{n}\bbox{\nabla}U_0 - 
2g\tilde{n}_0\bbox{\nabla} (\delta\tilde{n}+\delta n_c), \label{eq49b}\\
{\partial\delta\tilde{P}\over\partial t} &=& -{5 \over 3}{\bbox\nabla}
\cdot(\tilde{P}_0\delta{\bf v}_n)+ {2 \over 3}\delta{\bf v}_n
\cdot{\bbox\nabla}\tilde{P}_0+{2\over 3} (\mu_{c0} - U_0) \delta
\Gamma_{12}\,.
\label{eq49c}
\eea
\label{eq49}
\end{mathletters}
\noindent
We observe that $\delta \Gamma_{12}$ does not appear explicitly in
(\ref{eq49b}) as it contributes a second order correction.
These non-condensate equations must be solved in conjunction with the
following linearized equations for the condensate [see (\ref{eq27})]
\begin{mathletters}
\bea
{\partial\delta n_c\over\partial t} &=& - \bbox{\nabla}\cdot
(n_{c0} \delta{\bf v}_c)-\delta \Gamma_{12}, \label{eq50a}\\
m{\partial\delta{\bf v}_c\over\partial t} 
&=& -\bbox{\nabla}\delta\mu_c\,,
\label{eq50b}
\eea
\label{eq50}
\end{mathletters}
\noindent
where 
\beq
\delta \mu_c(\brt) = {1\over 2 \Phi_0(\br)} \hat
{\cal L}_0(\br)\left [ {1\over \Phi_0(\br)}
\delta n_c(\brt) \right ] + g\delta n_c(\brt) 
+ 2g\delta \tilde n(\brt)\,.
\label{eq51}
\eeq
The operator $\hat{\cal L}_0(\br)$ appearing in this equation is defined
as
\beq
\hat{\cal L}_0(\br)
\equiv\Bigg [-{\hbar^2 \nabla^2\over 
2m} +U_{ext}({\bf r}) + 2g\tilde n_0({\bf r})
+gn_{c0}({\bf r})- \mu_{c0}\Bigg ]\,,
\label{eq52}
\eeq
and is essentially the HFP Hamiltonian in Eq.~(\ref{eq47}).
We must finally obtain an expression for $\delta \Gamma_{12}$ in
(\ref{eq49}) and (\ref{eq50}) which is the linearized form of the
expression in (\ref{eq46'd}). We have
\beq
\delta \Gamma_{12}[\tilde f]=
-{\beta_0 n_{c0} \over \tau_{12}} \delta\mu_{\rm diff}\,,
\label{eq83}
\eeq
where we have introduced the difference between the local chemical
potentials of the condensate and non-condensate,
\beq
\mu_{\rm diff} \equiv \tilde \mu - \mu_c,
\label{eq69}
\eeq
and the ``equilibrium" $C_{12}$ collision rate
\beq
{1\over \tau_{12}} \equiv {2 g^2 \over (2\pi)^5\hbar^7}
\int d{\bf p}_1 \int d{\bf p}_2 \int d{\bf p}_3 
\delta({\bf p}_1-{\bf p}_2-{\bf p}_3)\delta(\mu_{c0}+\tilde\varepsilon_1
-\tilde \varepsilon_2-\tilde \varepsilon_3) (1+f_1^0) f_2^0 f_3^0 \,.
\label{eq84}
\eeq
Here, $f^0_i$ is the static equilibrium Bose distribution function.

\subsection{ZGN hydrodynamics}

In the remaining part of this Section, we
suppose that on the time scale of the collective modes of interest,
collisions between condensate and non-condensate atoms can be ignored.
This amounts to setting $C_{12}[\tilde f]$ (and hence
$\Gamma_{12}[\tilde f]$) to zero, and gives the hydrodynamic
equations derived in ZGN\cite{zaremba98}. The linearized version of
these equations is obtained by setting $\delta \Gamma_{12}$ to zero in
(\ref{eq49}) and (\ref{eq50}). They are a closed set of equations 
for the variables $\delta n_c$, $\delta
\tilde n$, $\bv_c$, $\bv_n$ and $\delta \tilde P$ and may be solved to
obtain the self-sustained oscillations of the trapped gas at finite
temperatures. A powerful variational method for solving equations of 
this kind is presented in Section \ref{section7}, where we consider the
explicit solution of the ZGN equations for a trapped gas.

An approximation commonly used in the recent literature on trapped 
Bose gases involves the neglect of the non-local ``quantum pressure'' 
terms, i.e., the first terms on the RHS of (\ref{eq48}) and 
(\ref{eq51}).
This so-called Thomas-Fermi (TF) approximation~\cite{stringari96,baym96}
has been found to give quite good results when the number of atoms in
the trap is large. Within the TF approximation, the equilibrium 
condensate chemical potential in (\ref{eq48}) simplifies to
\beq
\mu_{c0} \simeq U_{ext}(\br) + gn_{c0} ({\bf r})
+ 2g\tilde{n}_0(\br) \,.
\label{eq53}
\eeq
At $T=0$, where $\tilde n_0(\br) = 0$, (\ref{eq53}) gives a condensate
density which goes smoothly to zero at points where $U_{ext}(\br) =
\mu_{c0}$. However at finite temperatures, the combined solution of
(\ref{eq53}) and the semiclassical expression for the non-condensate
density ($\tilde n_0 =  g_{3/2}(z_0)/\Lambda_0^3$)
gives rise to unphysical discontinuous behaviour of the
densities at the sharp boundary of the condensate~\cite{huse82}. This
behaviour is especially problematic when attempting to obtain solutions
of Eqs.~(\ref{eq49})--(\ref{eq51}) for the collective modes, since
elaborate boundary conditions must be formulated in order to join the 
fluctuating variables at the condensate boundary. These 
problems can be avoided by retaining the quantum pressure terms 
which ensure that the densities are smooth functions of position in all
regions of the trapped Bose-condensed gas. Although these gradient terms
might appear to complicate the analysis considerably, it
will be seen that they can be handled easily using the variational 
method of solution developed in Section~\ref{section7}.

For the purpose of comparison with results obtained in
Section~\ref{section5}, it
is useful to consider the ZGN equations in the limit of a {\it uniform}
Bose gas. In this case, the ZGN equations yield first and second sound 
modes\protect\cite{griffin97}.
Eliminating the variables $\delta n_c$, $\delta \tilde n$ and
$\delta \tilde P$ from Eqs.~(\ref{eq49})--(\ref{eq50}) (with $\delta
\Gamma_{12} \equiv 0$), 
one obtains two coupled equations\cite{griffin97}:
\begin{mathletters}
\bea
m{\partial^2 \delta \bv_c \over \partial t^2} &=& gn_{c0}
\bbox{\nabla}(\bbox{\nabla} \cdot \delta \bv_c) + 2g\tilde n_0
\bbox{\nabla}(\bbox{\nabla} \cdot \delta \bv_n)\label{eq54a} \\
m{\partial^2 \delta \bv_n \over \partial t^2} &=& 2gn_{c0}
\bbox{\nabla}(\bbox{\nabla} \cdot \delta \bv_c) + 
\left ( {5\over 3}{\tilde P_0 \over \tilde n_0} + 2g\tilde n_0 \right )
\bbox{\nabla}(\bbox{\nabla} \cdot \delta \bv_n)\label{eq54b}\,.
\eea
\label{eq54}
\end{mathletters}
\noindent
Introducing velocity potentials according to
$\delta \bv_c \equiv \bbox{\nabla} \phi_c$ and
$\delta \bv_n \equiv \bbox{\nabla} \phi_n$ and 
assuming a plane wave solution
of the form $\phi_{c,n}(\brt) = \phi_{c,n}e^{i(\bk\cdot\br-\omega t)}$,
one obtains
\begin{mathletters}
\bea
m\omega^2\phi_c &=& gn_{c0} k^2 \phi_c + 2g\tilde n_0 k^2 \phi_n
\label{eq55a} \\
m\omega^2\phi_n &=& 2gn_{c0} k^2 \phi_c + 
\left ( {5\over 3}{\tilde P_0 \over \tilde n_0} + 2g\tilde n_0 \right )
k^2 \phi_n \label{eq55b}\,.
\eea
\label{eq55}
\end{mathletters}
The nontrivial solution of these equations yields the normal mode
frequencies of the uniform gas. These modes are first and second 
sound waves with the dispersion relation $\omega = uk$, where the 
sound velocities $u$ are the solution of
\beq
u^4 - u^2\left ( {5\over 3} {\tilde P_0 \over m\tilde n_0} + {2g\tilde
n_0 \over m} + {gn_{c0} \over m} \right ) +{gn_{c0} \over m}
\left ( {5\over 3} {\tilde P_0 \over m\tilde n_0} -{2g\tilde n_0 \over
m} \right ) = 0\,.
\label{eq56}
\eeq
Writing this equation as $u^4 - Au^2 +B =0$, we see that the
coefficients $A$ and $B$ are completely determined by equilibrium 
properties of the gas. These coefficients will be compared in 
Section~\ref{section5} with the
corresponding coefficients obtained on the basis of the Landau two-fluid
equations. As we show there, the $A$ and $B$ coefficients are slightly 
different in the Landau regime and reflect the different conditions 
underlying the propagation of first and second sound waves in the 
two cases.

We note that the quartic equation in
(\ref{eq56}) yields four solutions while Eqs.~(\ref{eq49}) and
(\ref{eq50}) are a total of five equations and must therefore 
admit five distinct
longitudinal modes. The missing solution is, in fact, a zero frequency
mode with mode amplitudes $\delta \bv_c = \delta \bv_n = 0$, $\delta n_c
= -2\delta \tilde n$ and $\delta \tilde P = 2g\tilde n_0 \delta \tilde
n$. The physical interpretation of such a static mode is not immediately
obvious. However, since we require some 
results from Section~\ref{section5} in order to explain its meaning,
we shall defer further discussion until then.

%% file: landau.tex
\section {Complete local equilibrium: Landau two-fluid equations}
\label{section5}

In this Section, we show that within the TF approximation, the set of 
equations in (\ref{eq27}) and (\ref{eq46'}) can be used to derive 
linearized two-fluid equations which reduce in a certain limit to
the standard Landau form 
\cite{landau41}, as generalized to include an external trapping 
potential. This derivation is of interest for several reasons. First, 
it shows that our approximate microscopic model is consistent
with the Landau two-fluid equations when conditions of complete local 
equilibrium exist. Under these conditions, the Landau
equations are valid for dense fluids such as superfluid $^{4}$He as
well as for dilute Bose gases. These equations have been
derived using very general arguments\cite{bogoliubov70}, but this very 
generality tends to obscure the underlying physics which a 
microscopic model such as ours can reveal in a very clear manner.
For example, our model not only leads to the Landau equations, but at
the same time gives the static equilibrium thermodynamic functions 
required in the calculation of the normal modes. An
even more significant advantage of our microscopic model is that we can
explore the {\it transition} from partial to complete
local diffusive equilibrium. In this respect, our two-fluid
hydrodynamics based on Eqs.~(\ref{eq27}) and (\ref{eq46'}) provides a
more complete description of the possible dynamical behaviour in 
trapped Bose gases.

In a previous paper\protect\cite{zaremba98}, we indicated how the
linearized ZGN equations could be combined to give the Landau form of 
the two-fluid equations for a {\it uniform} gas. However, the arguments
presented there were incomplete.
Here, we present a more careful discussion
starting form Eqs.~(\ref{eq27}) and (\ref{eq46'}) and show how an
explicit consideration of the source term $\Gamma_{12}[f]$ associated
with the $C_{12}$ collisions leads to
generalized two-fluid equations that reduce to the Landau equations 
in the appropriate limit. In addition, we shall extend our previous
analysis to the case of an inhomogeneous system. In doing so, however,
we make use of the Thomas-Fermi approximation which effectively treats
the system as locally homogeneous. If this approximation is not made,
one cannot obtain the usual Landau equations since the
gradient term of the chemical potential in (\ref{eq28}) contributes to
the right hand side of (\ref{eq27b}). This nonlocal term is a quantum
correction which is always ignored in microscopic derivations of the 
Landau equations for uniform systems~\cite{bogoliubov70}.

For completeness, we first display the linearized Landau two-fluid 
hydrodynamic equations\protect\cite{landau41,khalatnikov65,nozieres90},
in the form they take
in the presence of an external potential~\cite{shenoy98}:

\begin{mathletters}
\beq
{\partial\delta n\over\partial t} + \bbox{\nabla}\cdot\delta{\bf j}=0 
\label{eq57a}
\eeq

\beq
m{\partial\delta{\bf j}\over\partial t} = -\bbox{\nabla}\delta P-\delta
 n\bbox{\nabla} U_{ext}
\label{eq57b}
\eeq

\beq
m{\partial\delta{\bf v}_s\over\partial t} = -\bbox{\nabla}\delta\mu 
\label{eq57c}
\eeq

\beq
{\partial\delta s\over\partial t} +{\bbox\nabla}\cdot(s_0\delta 
{\bf v}_n) = 0\,,
\label{eq57d}
\eeq
\label{eq57}
\end{mathletters}
where 

\bea
m\delta{\bf j}({\bf r}, t) & = &\rho_{s0}({\bf r})\delta{\bf v}_s
({\bf r}, t) + \rho_{n0}({\bf r})\delta{\bf v}_n
({\bf r}, t), \nonumber\\
m\delta n({\bf r}, t) & = &\delta\rho_s
({\bf r}, t)+ \delta\rho_n
({\bf r}, t).\label{eq58}
\eea
We have here distinguished the superfluid variables $\rho_s$ and $\bv_s$
from the corresponding variables for the condensate; a similar
distinction applies to the normal fluid variables as opposed to the
non-condensate variables. However, the
condensate and non-condensate variables can be identified with the more
usual two-fluid variables within the context of our model which treats
the excitations in the HF approximation. It should be noted that this
correspondence will no longer be valid if one were to go beyond this
approximation. For example, in the HFP approximation, there is a finite
depletion of the condensate even at zero temperature~\cite{hutchinson97}
and the superfluid density is no longer precisely equivalent to the 
condensate density.  The other
local variables appearing in the above equations are the pressure $P$,
entropy density $s$ and thermodynamic chemical potential $\mu$. 
As we shall show, each of
these quantities is given by an explicit expression within our 
microscopic model.

The linearized ZGN$^\prime$ continuity 
equations are given in (\ref{eq49a}) and (\ref{eq50a}). Adding them,
we obtain the continuity equation (\ref{eq57a}).
Combining our two velocity equations in (\ref{eq49b}) and (\ref{eq50b}),
we find
\bea
{\partial\delta{\bf j}\over \partial t} = - \bbox{\nabla}\delta\tilde{P}
&-& 2g(n_{c0} +\tilde{n}_0)\bbox{\nabla}\delta\tilde{n}-g(n_{c0} 
+2\tilde{n}_0) \bbox{\nabla}\delta n_c \nonumber\\
&-&2g\bbox{\nabla} (n_{c0} + \tilde{n}_0)\delta \tilde n
- \bbox{\nabla}U_{ext}\delta \tilde n\,.
\label{eq61}
\eea
One can show that (\ref{eq61}) is equivalent to (\ref{eq57b}) if the 
total pressure is defined as\protect\cite{zaremba98}
\beq
P=\tilde{P} + {1 \over 2} g[n^2 + 2n\tilde{n} -\tilde{n}^2].
\label{eq62}
\eeq
To prove this, we note that (\ref{eq62}) gives
\beq
\delta P = \delta\tilde{P} + 2gn_0\delta\tilde{n}
+ g(n_{c0} + 2\tilde{n}_0)\delta n_c
\label{eq63}
\eeq
and hence,
\bea
{\bbox\nabla}\delta P &=& {\bbox\nabla}\delta\tilde{P} + 
2gn_0{\bbox\nabla}\delta\tilde{n}+g(n_{c0} + 2\tilde{n}_0){\bbox\nabla}
\delta n_c \nonumber \\
&+& 2g{\bbox\nabla}n_0\delta \tilde n +
g({\bbox\nabla}n_{c0}+2{\bbox\nabla} \tilde{n}_0)\delta n_c\,.
\label{eq64}
\eea
Making use of the TF result in (\ref{eq53}) to rewrite the last term as 
$-\delta n_c{\bf\nabla}U_{ext}$, and inserting the resulting expression 
for $\bbox{\nabla}\delta P$ into (\ref{eq57b}), we obtain an equation 
identical to (\ref{eq61}). The results in (\ref{eq61})--(\ref{eq64}) 
were given previously by ZGN~\cite{zaremba98} for a uniform gas, but 
the present derivation is more general in that it is based on the 
ZGN$^\prime$ equations in the context of a nonuniform system.

In order to make contact with the remaining Landau equations 
(\ref{eq57c}) and (\ref{eq57d}), we 
must introduce the appropriate thermodynamic variables corresponding 
to the chemical potential $\mu$, local temperature $T$ and the local 
entropy $s$. Within the HF approximation for the non-condensate, the
total energy density is given by
\bea
\epsilon &=& \tilde{\epsilon} + {1\over 2}g 
\left<\hat{\psi}^{\dagger}\hat{\psi}^{\dagger}
\hat{\psi}\hat{\psi}\right> +nU_{ext} \nonumber\\ 
&=& \tilde{\epsilon} + {1\over 2}g 
\left[n^2+2n\tilde{n}-\tilde{n}^2\right] +nU_{ext}\,,
\label{eq65}
\eea
where $\tilde \epsilon$ is defined by (\ref{eq41}) and (\ref{eq42}).
For the local entropy we use the definition
\bea
s &\equiv& k_{\rm B}\int\frac{d{\bf p}}{(2\pi\hbar)^3} \left [ 
(1+\tilde f)\ln (1+\tilde f)-\tilde f \ln \tilde f \right ]\nonumber \\
&=&{1\over T} \left [ \frac{5}{2}\tilde P - \tilde n(\tilde \mu - U)
\right ]\,.
\label{eq66}
\eea
This result corresponds to the entropy of a uniform ideal gas above 
$T_{BEC}$ with a fugacity $z = e^{\beta(\tilde \mu - U)}$. In static 
equilibrium, (\ref{eq66}) reduces to
\beq
s_0T_0 ={5\over 2}\tilde{P}_0 + gn_{c0}\tilde{n}_0\,,
\label{eq67}
\eeq
where we have noted that
$\tilde \mu_0=\mu_{c0}=U_{ext}+g(2\tilde n_0 +
n_{c0})$ and $U_0 = U_{ext} + 2gn_0$.

From these expressions for $P$, $\epsilon$ and $s$ (all of which
depend on $\br$ and $t$), we obtain the relation
\beq
\epsilon + P - sT = \mu_c n_c + \tilde \mu \tilde n = 
\mu_c n + \mu_{\rm diff} \tilde n\,,
\label{eq68}
\eeq
where $\mu_{\rm diff}$ is the chemical potential difference defined in
(\ref{eq69}). As we shall see, this variable plays a crucial role in 
understanding the approach to complete local equilibrium. If 
$\mu_{\rm diff}$ were zero,
then (\ref{eq68}) would correspond precisely to the usual
thermodynamic relation involving these variables, with $\mu_c$  
playing the role of the equilibrium chemical potential.

Having defined these various thermodynamic functions, we can now
consider their local variations from absolute equilibrium (recall that
the equilibrium properties are still functions of position). 
We begin with the
kinetic pressure given by (\ref{eq43}). Its variation leads to the 
equation
\bea
\delta \tilde P &=& \delta \left ( {1\over \beta \Lambda^3} \right )
g_{5/2}(z_0) + {1\over \beta_0 \Lambda_0^3}  
g'_{5/2}(z_0) \delta z\nonumber \\
&=& s_0 \delta T + \tilde n_0 (\delta \tilde \mu - 2g \delta n)
\label{eq70}
\eea
In obtaining this result, we have made use of (\ref{eq67}) and the Bose
identity $zg_n^{\,\prime}(z) = g_{n-1}(z)$. Combining (\ref{eq63}) and 
(\ref{eq70}), we find
\beq
\delta P = s_0 \delta T + n_0 \delta \mu_c + \tilde n_0 \delta
\mu_{\rm diff}\,.
\label{eq71}
\eeq
The appearance of the last term is once again due to the
non-condensate having a chemical potential different from that of the
condensate. Similarly, from (\ref{eq66}) and (\ref{eq70}), we find 
the variation of the entropy to be given by
\beq
T_0\delta s={3\over 2}\delta\tilde{P} +gn_{c0}\delta\tilde{n}.
\label{eq72}
\eeq
This expression finally allows us to obtain an entropy conservation 
equation. Taking the time derivative of (\ref{eq72}), and making
use of the non-condensate density and pressure equations (\ref{eq49a}) 
and (\ref{eq49c}), we find
\bea
{\partial \delta s \over \partial t} &=& {3\over 2T_0} {\partial \delta
\tilde P \over \partial t} + {gn_{c0} \over T_0} {\partial \delta 
\tilde n \over \partial t}\nonumber \\
&=& -\bbox{\nabla} \cdot (s_0\delta \bv_n)\,.
\label{eq73}
\eea
We note that the source terms on the right hand sides of (\ref{eq49a}) 
and (\ref{eq49c})
cancel out, so that a strict conservation law for the nonequilibrium
local entropy defined in (\ref{eq66}) is obtained.

Finally, a comparison of (\ref{eq50b}) and (\ref{eq57c}) shows that
these two equations are equivalent
if $\delta \bv_s$ is identified with $\delta \bv_c$ and $\delta \mu$ 
with $\delta \mu_c$.  With this identification, we arrive at a set 
of hydrodynamic equations which are precisely of the {\it form} of the 
Landau two-fluid equations. At first sight, 
it might therefore appear that our starting equations are in fact
equivalent to the Landau equations, but this is not the case. The
appearance of $\mu_{\rm diff}$ in (\ref{eq68}) implies that $\mu_c$ is
{\it not} related to the other thermodynamic variables in the same way
that the 
equilibrium chemical potential $\mu$ would be. As a result,
$\delta \mu_c$ and $\delta s$ are not related directly to the
thermodynamic fluctuations $\delta T$ and $\delta P$, as assumed in the
usual derivations of the first and second sound
velocities~\cite{nozieres90}.

To see this more clearly, let us use (\ref{eq67}) to define
the {\it equilibrium} local entropy function, $T s^{eq} \equiv {5\over
2} \tilde P + gn_c \tilde n$, where the notation $s^{eq}$ is used to
distinguish this quantity from the local entropy given by
(\ref{eq66}). It is the variation of the equilibrium entropy $s^{eq}$ 
which can be expressed in terms of the thermodynamic fluctuations 
$\delta T$ and $\delta P$ via equilibrium thermodynamic derivatives.
By considering a quasistatic change in the thermodynamic 
state of the system, (\ref{eq67}) leads to the relation
\beq
T_0 \delta s^{eq} = {5\over 2} \delta \tilde P + gn_{c0} \delta \tilde n
+ g \tilde n_0 \delta n_c - s_0 \delta T,
\label{eq74}
\eeq
and using (\ref{eq70}) to eliminate $\delta T$, we find
\beq
T_0 \delta s^{eq} = {3\over 2} \delta \tilde P + gn_{c0} \delta \tilde n
+ \tilde n_0 \delta \mu_{\rm diff}\,.
\label{eq75}
\eeq
This differs from the result for $\delta s$ given in (\ref{eq72}).
The last term in (\ref{eq75}) indicates that the variation of the
equilibrium entropy is not simply given in the present situation 
by the equilibrium variation.  In terms of $\delta s^{eq}$, the 
entropy equation analogous to (\ref{eq73}) takes the form
\beq
{\partial \delta s^{eq} \over \partial t} =
-\bbox{\nabla} \cdot (s_0\delta \bv_n) + \tilde n_0 {\partial \delta
\mu_{\rm diff} \over \partial t}\,.
\label{eq76}
\eeq
The last term in (\ref{eq76}) can be interpreted as
the production of entropy associated with the equilibration of the
condensate and non-condensate chemical potentials. 
Its appearance implies that the thermodynamic fluctuations
$\delta P$ and $\delta T$ are
coupled to fluctuations in $\delta \mu_{\rm diff}$, which is directly 
related to the condensate-non-condensate collision terms 
$\delta \Gamma_{12}$ appearing on the right hand side of 
Eqs.~(\ref{eq49a}), (\ref{eq49c}) and (\ref{eq50a}). 
To this point we have not made any
assumptions about $\delta \Gamma_{12}$, and as a result,
the equations as they stand differ from the usual Landau
two-fluid theory. To make this final connection we must obtain an
equation of motion for $\delta \mu_{\rm diff}$.

Noting that $\tilde \mu - 2gn = \mu_{\rm diff} - gn_c$,
(\ref{eq70}) can be rewritten in the form
\beq
\delta \tilde P = s_0 \delta T +
\tilde n_0 ( \delta \mu_{\rm diff} - g\delta n_c)\,.
\label{eq77}
\eeq
Similarly, the variation of (\ref{eq46}) yields the equation
\beq
\delta \tilde n = 
\left ( {3\over 2} \tilde n_0 + \tilde \gamma_0 n_{c0} \right ) 
{\delta T \over T_0} + 
{\tilde \gamma_0 \over g}
(\delta \mu_{\rm diff} - g \delta n_c)\,,
\label{eq78}
\eeq
where we have introduced the dimensionless quantity
\beq
\tilde \gamma_0 \equiv {\beta_0 g \over \Lambda_0^3} g_{1/2}(z_0)\,.
\label{eq79}
\eeq
This quantity can be related to thermodynamic derivatives by noting that
(\ref{eq78}) implies the equilibrium variation at constant temperature,
$\delta \tilde n \vert_T = 
(\tilde \gamma_0 /g)\delta(\tilde \mu - 2gn)\vert_T
= -\tilde \gamma_0(\delta n 
- \delta \tilde n)\vert_T$. We thus conclude that
\beq
\tilde \gamma_0 =
{\left ( {\textstyle{\partial \tilde n} \over \textstyle{\partial n}} 
\right )_T \over \left ( {\textstyle{\partial \tilde n} \over 
\textstyle{\partial n} } \right )_T -1}\,.
\label{eq79'}
\eeq
In a similar way, one can show that the coefficient of 
$\delta T$ in (\ref{eq78}) can be expressed as
\beq
\left ( {3\over 2} \tilde n_0 + \tilde\gamma_0 n_{c0} \right ) 
{1\over T_0} = (1-\tilde \gamma_0)\left ( \partial \tilde n \over 
\partial T \right )_n\,.
\label{eq80}
\eeq

Eqs. (\ref{eq77}) and (\ref{eq78}) show that the variations in $\delta
T$ and $\delta \mu_{\rm diff}$ can each be expressed in terms of
$\delta n_c$, $\delta \tilde n$ and $\delta \tilde P$.
Eliminating $\delta \mu_{\rm diff}$ from these equations and making use
of Eqs.~(\ref{eq49a}) and (\ref{eq49c}), we obtain the equation of
motion for $\delta T$,
\beq
{1\over T_0}{\partial \delta T \over \partial t} = - {2\over 3}
\bbox{\nabla} \cdot \delta \bv_n - {g(\tilde n_0 + {2\over 3} \tilde
\gamma_0 n_{c0}) \over {5\over 2} \tilde \gamma_0 \tilde P_0 - 
{3\over 2}g\tilde n_0^2} \delta \Gamma_{12}\,.
\label{eq81}
\eeq
Alternatively, eliminating $\delta T$, we find
\bea
{\partial\delta \mu_{\rm diff} \over \partial t} = \bbox{\nabla} U_0 
\cdot \delta \bv_n &+& {2\over 3} gn_{c0}
\bbox{\nabla}\cdot \delta \bv_n - g\bbox{\nabla}\cdot 
(n_{c0}\delta \bv_c) \nonumber \\
&+& g\left ( {{5\over 2} \tilde P_0 + 2g\tilde n_0 
n_{c0} + {2\over 3} \tilde \gamma_0 g n_{c0}^2 \over 
{5\over 2} \tilde \gamma_0 
\tilde P_0 - {3\over 2} g\tilde n_0^2} -1 \right ) \delta \Gamma_{12}\,.
\label{eq82}
\eea
We next make use of the expression for $\delta\Gamma_{12}$ given in
(\ref{eq83}) to rewrite (\ref{eq82}) as
\beq
{\partial\delta \mu_{\rm diff} \over \partial t}
= \bbox{\nabla} U_0 \cdot \delta \bv_n + {2\over 3} gn_{c0}
\bbox{\nabla}\cdot \delta \bv_n - g\bbox{\nabla}\cdot 
(n_{c0}\delta \bv_c) - {\delta \mu_{\rm diff} \over \tau_\mu}\,,
\label{eq85}
\eeq
where the relaxation time for the chemical potential difference
is defined as
\beq
{1\over \tau_\mu} \equiv {\beta_0 g n_{c0} \over \tau_{12}}
\left ( {{5\over 2} \tilde P_0 + 2g\tilde n_0 
n_{c0} + {2\over 3} \tilde \gamma_0 g n_{c0}^2 \over {5\over 2} 
\tilde \gamma_0 \tilde P_0 - {3\over 2} g\tilde n_0^2} -1 \right )
\equiv {\beta_0 g n_{c0} \over \sigma \tau_{12}}\,.
\label{eq86}
\eeq
Although the relaxation time $\tau_\mu$ is proportional to the 
equilibrium
collision time $\tau_{12}$ defined earlier, it has quite a different 
magnitude and temperature dependence, especially in the neighbourhood of
$T_{BEC}$, as a result of the factors $n_{c0}$ and $\sigma$ which 
appear in its definition (see Fig.~1 in Ref.~\cite{nikuni99}).

The results in Eqs.~(\ref{eq85}) and (\ref{eq86}) were quoted in 
Ref.~\cite{nikuni99} for the special case of a uniform gas. Here we can
see very clearly that $\tau_\mu$ plays a key role in describing the 
equilibration of the condensate and
non-condensate chemical potentials brought about by collisions between 
the two components.  For a uniform gas with spatially 
homogeneous fluctuations, the solution of (\ref{eq85}) is
\beq
\delta \mu_{\rm diff}(t) = \delta \mu_{\rm diff}(0) e^{-t/\tau_\mu}\,,
\label{eq87}
\eeq
which shows that any initial difference in chemical potentials decays 
on a time scale set by $\tau_\mu$. Similar considerations will of course
also apply to an inhomogeneous gas.
Thus, for fluctuations which occur on a time scale longer
than the relaxation time $\tau_\mu$, we expect that $\delta 
\mu_{\rm diff}$ can effectively
be set to zero (i.e., $\tilde \mu = \mu_c$). It then follows from
(\ref{eq68}) that we recover the equilibrium relationships among all the
thermodynamic variables we have defined. As a result, our equations
in the $\tau_\mu \to 0$ limit reproduce {\it precisely} the hydrodynamic
description as provided by the Landau two-fluid equations.

To demonstrate this reduction explicitly it is again 
convenient to consider 
the sound modes in a uniform Bose gas (see also Ref.~\cite{nikuni99}). 
We proceed as in the analysis of the ZGN
equations in the previous Section, but take into account the fact that 
according to (\ref{eq85}), fluctuations in $\delta \mu_{\rm diff}$ are 
coupled to those of the velocities $\delta \bv_c$ and $\delta \bv_n$. 
Introducing velocity potentials according to $\delta
\bv_c = \bbox{\nabla} \phi_c$ and $\delta \bv_n = \bbox{\nabla} \phi_n$,
the equations in (\ref{eq49}) and (\ref{eq50}) can be combined to yield
\begin{mathletters}
\bea
m{\partial^2 \phi_c \over \partial t^2} &=& gn_{c0} \nabla^2 \phi_c +
2g\tilde n_0 \nabla^2 \phi_n +{\sigma\over \tau_\mu} \delta
\mu_{\rm diff} \label{eq88a} \\
m{\partial^2 \phi_n \over \partial t^2} &=& \left ( {5\tilde P_0 \over
3\tilde n_0} + 2g\tilde n_0 \right ) \nabla^2 \phi_n +
2g n_{c0} \nabla^2 \phi_c - {2\sigma \over 3
\tau_\mu} \delta \mu_{\rm diff}\,.
\label{eq88b}
\eea
\label{eq88}
\end{mathletters}
\noindent
These equations in conjunction with (\ref{eq85}) constitute a complete
set of equations for the normal modes of the system. We 
look for solutions having a plane 
wave form
$\phi_{c,n}(\brt) = \phi_{c,n} e^{i(\bk\cdot\br -\omega t)}$. In this
case, (\ref{eq85}) reduces to
\beq
\delta \mu_{\rm diff} = {gn_{c0}\tau_\mu \over 1-i\omega\tau_\mu} 
\left ( \phi_c - {2\over 3}\phi_n \right ) k^2\,.
\label{eq89}
\eeq
Substituting this result into Eq.~(\ref{eq88}), we obtain the two
coupled equations
\beq
m\omega^2 \phi_c = gn_{c0}\Bigg ( 1 - {\sigma
\over 1-i\omega\tau_\mu} \Bigg ) k^2 \phi_c
+ 2g\tilde n_0 \left ( 1 + {\sigma \over 3
(1-i\omega\tau_\mu) } {n_{c0} \over \tilde n_0} 
\right ) k^2 \phi_n 
\label{eq90}
\eeq
and
\bea
m\omega^2 \phi_n = \Bigg ( {5\tilde P_0 \over 3\tilde n_0} &+& 2g\tilde
n_0 \left [ 1- {2\sigma \over 9 
(1-i\omega\tau_\mu) } {n_{c0}^2\over \tilde n_0^2} \right ]
\Bigg ) k^2 \phi_n 
\nonumber \\ &+& 2gn_{c0} \left ( 1 + {\sigma \over 3
(1-i\omega\tau_\mu) } {n_{c0} \over \tilde n_0} 
\right ) k^2 \phi_c\,,
\label{eq91}
\eea
where the thermodynamic factor $\sigma$ is defined in (\ref{eq86}).
In the limit $\omega \tau_\mu \to \infty$, we recover the ZGN 
equations given in (\ref{eq55}). We also see from (\ref{eq89}) that
$\delta \mu_{\rm diff} \to ign_{c0} 
(\phi_c-2\phi_n/3)(k^2/\omega)$, which is proportional to $k$.
In the opposite limit $\omega \tau_\mu \to 0$, we
obtain the equations
\begin{mathletters}
\bea
m\omega^2 \phi_c &=& gn_{c0} ( 1 - \sigma ) k^2 \phi_c
+ 2g\tilde n_0 \left ( 1 + {\sigma n_{c0}
\over 3 \tilde n_0} \right ) k^2 \phi_n\label{eq92a} \\ 
m\omega^2 \phi_n &=& \Bigg ( {5\tilde P_0 \over 3\tilde n_0} + 2g\tilde
n_0\left [ 1 - {2\sigma
n_{c0}^2\over 9\tilde n_0^2} \right ] \Bigg ) k^2 \phi_n 
+ 2gn_{c0} \left ( 1 + {\sigma 
n_{c0} \over 3 \tilde n_0} \right ) k^2 \phi_c\,,
\label{eq92b}
\eea
\label{eq92}
\end{mathletters}
\noindent
with $\delta \mu_{\rm diff} \to gn_{c0} \tau_\mu
(\phi_c-2\phi_n/3)k^2$. We see from this that $\delta \mu_{\rm diff}$ 
vanishes in the $\omega\tau_\mu \to 0$ limit while $\delta \Gamma_{12}$,
which is proportional to $\delta \mu_{\rm diff}/\tau_\mu$, remains
{\it finite} even when $\tau_\mu \to 0$.
Thus, the $\delta \Gamma_{12}$ terms in equations (\ref{eq49})
and (\ref{eq50}) are still present in the Landau limit, reflecting the
fact that the establishment
of complete local equilibrium requires the continued local readjustment
of the number of atoms in the condensate and non-condensate. In other
words, the strong collisional coupling between the two components, which
ensures $\delta \mu_{\rm diff} = 0$, implies that the
condensate and non-condensate densities are {\it not} separately 
conserved as they are in the ZGN ($\omega \tau_\mu \to \infty$) limit.

Although not immediately apparent, the equations in (\ref{eq92}) yield 
first and second sound velocities which are in precise agreement with 
those determined by the Landau two-fluid equations. The latter are
usually expressed in terms of pressure and temperature fluctuations and
yield velocities, $u$, as solutions of an equation of the form
$u^4-Au^2+B=0$, where $A$ and $B$ are expressed in terms of equilibrium
thermodynamic quantities~\cite{griffin97}. If these thermodynamic
quantities are evaluated for our model of the equilibrium properties,
one can show that the Landau $A$ and $B$ coefficients correspond
precisely to those following from (\ref{eq92}). These differ
from the ZGN $A$ and $B$ coefficients given in (\ref{eq56}), and as a
result, the first and second sound speeds found in these two limits 
also differ. However, as shown in Fig.~1, these differences are very
small in the case of a weakly interacting Bose gas. That is, the
transfer of atoms required to equilibrate the condensate and
non-condensate is playing a relatively minor role when it comes to
determining the magnitude of the first and second sound speeds.

The damping of these sound modes is another matter. The modes are 
undamped in either the ZGN ($\omega \tau_\mu \to \infty$) or Landau 
($\omega \tau_\mu \to 0$) limits, but at intermediate values of $\omega
\tau_\mu$, the damping arising from the equilibration of the condensate
and non-condensate can be quite significant. To obtain the complex mode
frequencies, $\omega = \Omega -i\Gamma$, we can solve either the 
hydrodynamic equations in (\ref{eq49}) and (\ref{eq50}) 
[from (\ref{eq83}), we recall that $\delta \Gamma_{12}$ is
proportional to $\delta \mu_{\rm diff}$], or Eqs. (\ref{eq88}) 
together with
the relaxation equation (\ref{eq85}) for the chemical potential
difference. These equations
yield five mode frequencies: four modes corresponding to damped first
and second sound ($\Omega = \pm u_{1,2} k$) and  a purely imaginary 
relaxational mode. The latter is associated with the zero-frequency 
mode mentioned  at the end of Section~\ref{section4} and will
be discussed further at the end of this Section.

In Fig.~2 we show the damping of the second sound mode for a uniform 
gas as a function of $\Omega\tau_\mu$, for
$T/T_{BEC} = 0.9$ and $gn/k_B T_{BEC} = 0.2$. The relative damping 
($\Gamma/\Omega$) of
the first sound mode is barely visible on the scale of Fig.~2 and
therefore is not displayed.  It can be seen that the relative damping 
peaks at 
$\Omega\tau_\mu \simeq 1$, which is the transition between the ZGN and
Landau regimes. The damping of the second sound mode is especially large
since it involves an out-of-phase oscillation of the
condensate and non-condensate density fluctuations. Such an
out-of-phase oscillation would be expected to have the largest 
imbalance of the local chemical
potentials, and thus the largest rate of transfer of atoms between the
condensate and non-condensate. This damping mechanism will arise
whenever the equilibration of the condensate and non-condensate is
incomplete.

The results found here for the uniform gas may be relevant to the MIT 
studies of collective excitations in highly anisotropic
traps~\cite{stamper98}, which are believed (on the basis of $C_{22}$
collisions) to be in the
transition regime between collisionless dynamics and hydrodynamics. The
$m=0$ quadrupolar oscillations of both the condensate and thermal cloud
were observed, as well as an out-of-phase dipole mode. Both the
condensate quadrupolar mode and the dipole mode involve the condensate 
and non-condensate oscillating out-of-phase (see
Section~\ref{section7}). According to the above discussion for a 
uniform gas, this will entail a large chemical potential difference 
between the two components and we would
therefore expect to see a large damping rate associated with the 
equilibration of chemical potentials when the number of non-condensate
atoms is appreciable. Although such large damping rates were indeed 
observed~\cite{stamper98}, they cannot be attributed entirely to this
source since there will also be contributions to the damping from
viscosity and thermal conduction which our present analysis does not
include (see the discussion at the
end of this Section and in Section~\ref{section8}). 
It is nevertheless clear that the relaxation of the local chemical 
potentials of the two components must be taken into account if one 
wants to provide a quantitative description of the damping
processes in condensed Bose gases at finite temperatures.

In addition to the first and second sound modes discussed above, 
the ZGN$^\prime$ equations also yield a nonpropagating relaxational 
mode which, in the $\tau_\mu \to 0$ limit, has a  frequency 
$\omega \simeq -i/\tau_\mu$, consistent with (\ref{eq87}). In the
opposite limit, $\tau_\mu \to \infty$, collisions between condensate
and non-condensate atoms no longer play a role, and the mode becomes 
the zero-frequency mode of the ZGN description. As discussed at the end
of Section~\ref{section4}, this mode is characterized by the fluctuating
variables $\delta \bv_c = \delta \bv_n = 0$, $\delta n_c = -2\delta 
\tilde n$ and $\delta \tilde P = 2g\tilde n_0 \delta \tilde n$. This
mode persists even into the normal phase, in which case the variables
$n_c$ and $\bv_c$ do not exist, and the mode is then characterized by 
amplitudes satisfying $\delta \tilde P = - 2g\tilde n_0 \delta 
\tilde n$ and $\delta \bv_n =0$.

In either case, one can show from (\ref{eq63}) that the zero-frequency 
mode is characterized by $\delta
P=0$. In other words, even though there is a local change in the total 
density, this mode has a zero pressure fluctuation,
and as a result, the fluid remains in mechanical equilibrium. This is
possible since the change in kinetic pressure, $\delta \tilde P$, is
exactly balanced by a corresponding change in the mean-field potential.
It is important to point out that the same situation also arises in a
conventional fluid having some arbitrary equation of state $P(n,T)$. 
Variations in the pressure are then given by $\delta P =
(\partial P/\partial n)_T \delta n + (\partial P/\partial T)_n \delta
T$, and we see that $\delta P$ can be zero if $\delta n/\delta T =
-(\partial P/\partial T)_n / (\partial P/\partial n)_T$.
Our present model is a particular realization of this possibility, 
and we find from (\ref{eq70}) that the fluctuations 
in the local temperature and non-condensate chemical potential
are related by $s_0\delta T = - \tilde n_0 \delta \tilde \mu$. The
zero-frequency mode in the ZGN description therefore has associated 
with it a {\it static} temperature fluctuation. This is consistent from
a thermodynamic point of view, but of course is not expected in a more
complete description of the fluid dynamics.

Extending the ZGN hydrodynamic equations to
include heat conduction will allow the temperature fluctuation to relax
in time, and as a result, the zero-frequency ZGN mode will become the 
usual thermal diffusion mode with a purely imaginary frequency.
However, even in the absence of heat conduction, we have found that
the zero-frequency mode is converted into a relaxational mode within 
the ZGN$^\prime$ description. It is therefore clear that the further
inclusion of heat conduction within the ZGN$^\prime$ description
will simply combine with the collisional relaxation effects due to
$C_{12}$ to produce a hybrid relaxational mode. These two
effects will also contribute to the damping of the sound modes. In
conventional discussions based on the Landau two-fluid equations, the
thermal diffusion mode found above the transition is replaced by
damped second sound modes below the 
transition~\cite{khalatnikov65,gay85}. Since there is one additional
dynamical equation in both the ZGN and ZGN$^\prime$ descriptions as
compared to the Landau theory, there must necessarily be an additional
mode. This is the relaxational mode we have been discussing. However, 
this additional mode is
strongly damped in the $\omega \tau_\mu \to 0$ limit and we then recover
the usual results based on the Landau two-fluid equations. A more
complete discussion of the combined effects of collisional relaxation
and thermal conduction will be given elsewhere.

%% file: kohn.tex
\section {Center of Mass (Kohn) Mode}

In this section, we discuss the center of mass motion of the coupled
condensate and non-condensate systems. For confinement in an arbitrary
harmonic trap potential of the form
\beq
U_{ext}(\br)={1 \over 2}m (\omega_x^2x^2+\omega_y^2y^2+\omega_z^2z^2)\,,
\label{eq6.1}
\eeq
the center of mass degrees of freedom separate from all other internal
degrees of freedom, with the general consequence that there exists a
special normal mode in which the equilibrium density profile oscillates
rigidly without distortion. In the electron gas literature, this is
referred to as the generalized Kohn theorem~\cite{dobson94}, but the 
result is
independent of statistics and applies to any harmonically confined
system at any temperature. The experimental observation of this mode
provides a convenient way of determining the trap frequencies.

In the context of a Bose-condensed trapped gas, 
the theorem implies that the mode densities have the form
\begin{mathletters}
\beq
n_c(\brt) = n_{c0}(\br-\bbox{\eta}(t))\,,
\label{eq6.2a}
\eeq
\beq
\tilde{n}(\brt) =\tilde n_0(\br-\bbox{\eta}(t))\,,
\label{eq6.2b}
\eeq
\label{eq6.2}
\end{mathletters}
\noindent
where the center of mass displacement $\bbox{\eta}(t)$ 
has a time-dependence given by
\beq
{\partial^2\eta_{\alpha}(t)\over \partial t^2} = -\omega_\alpha^2 
\eta_{\alpha}(t)\,.
\label{eq6.3}
\eeq
An approximate theory of the dynamics should, at the very least, be
consistent with a mode of this type. This is indeed the case for the
zero-temperature dynamics of a Bose condensate based on the nonlinear GP
equation~\cite{edwards96}, including the TF limit~\cite{stringari96}. 
However, the simplest extension to
finite temperatures as considered by HZG~\cite{hutchinson97} fails to 
meet this criterion. In fact, 
it was the small deviation of the center of mass frequency from the
trap frequency as a function of temperature that emphasized
the limitation of the static HFP theory. As discussed in Section
\ref{section2}, this approximation does not include the collective
dynamics of the non-condensate.

We now prove that our set of coupled equations 
(\ref{eq27})--(\ref{eq29}) are consistent with the
generalized Kohn theorem. We showed previously\cite{zaremba98} that this
theorem was satisfied for the linearized ZGN equations in (\ref{eq49}) 
and (\ref{eq50}), which are a special case of the more general equations
being considered here. However, it is important to demonstrate that
this conclusion remains valid even when collisions between the atoms are
included. By doing so, we shall be able to clearly see how the effects
of interactions, both in the form of mean fields and collisions, cancel
out in the determination of the center of mass mode frequency.

We start by considering the condensate density. From (\ref{eq6.2a}), 
we obtain
\beq
{\partial n_c(\brt) \over \partial t} = -\dot{\bbox{\eta}}(t) \cdot
\bbox{\nabla} n_{c0}(\br - \bbox{\eta}(t))
=-\bbox{\nabla} \cdot (\dot{\bbox{\eta}}(t) n_c(\brt))\,.
\label{eq6.4}
\eeq
Comparing this with (\ref{eq27a}), we see that (\ref{eq6.2a}) is a 
solution of the continuity equation if the velocity field is given by
\beq
\bv_c(\brt) = {\partial \bbox{\eta}(t) \over \partial t} \equiv 
\dot{\bbox{\eta}}(t) \,,
\label{eq6.5}
\eeq
and {\it provided} the source term $\Gamma_{12}[f]$ is zero. This latter
result will be demonstrated shortly, once the form of the distribution
function $f$ corresponding to (\ref{eq6.2b}) is specified.
Using (\ref{eq6.5}) in (\ref{eq27b}), we have
\beq
m{\partial^2{\eta_\alpha}(t) \over \partial t^2} = -
{\partial \mu_c(\brt) \over \partial x_\alpha} \,.
\label{eq6.6}
\eeq
From the dependence of $\mu_c(\brt)$ on the densities given 
in (\ref{eq28}), we see that (\ref{eq6.2}) implies
\bea
\mu_c(\brt) &=& \mu_{c0}(\br - \bbox{\eta}(t)) +
U_{ext}(\br)-U_{ext}(\br-\bbox{\eta}(t)) \nonumber \\
&=& \mu_{c0} + 
\sum_\alpha m\omega_\alpha^2 x_\alpha \eta_\alpha - {1\over 2}
\sum_\alpha m\omega_\alpha^2 \eta_\alpha^2\,,
\label{eq6.7}
\eea
since as indicated by (\ref{eq48}), $\mu_{c0}$ is a constant,
independent of
position. Inserting (\ref{eq6.7}) into (\ref{eq6.6}) leads immediately 
to (\ref{eq6.3}).
The condensate equations are thus seen to admit a solution consistent
with the assumed form of the densities in (\ref{eq6.2}).

A similar analysis of the non-condensate continuity equation 
(\ref{eq33a}) shows
that (\ref{eq6.2b}) implies a non-condensate velocity
\beq
\bv_n(\brt) = {\partial \bbox{\eta}(t)\over \partial t}\,,
\label{eq6.8}
\eeq
again assuming that $\Gamma_{12}[f] = 0$. This velocity field and the
non-condensate density in (\ref{eq6.2b}) are clearly generated by 
the distribution function
\beq
f_{\eta}(\bp, \br, t) \equiv 
{1 \over e^{\beta_0 [ {1\over 2m}(\bp - m\dot{\bbox\eta}(t))^2 
+ U_0(\br-{\bbox\eta}(t))-\tilde \mu_0]} - 1}\,,
\label{eq6.9}
\eeq
where $\beta_0$ and $\tilde \mu_0$ are equilibrium parameters.
By noting that (\ref{eq6.2}) implies
\beq
U_0(\br -\bbox{\eta}(t))
=U(\brt) + U_{ext}(\br -\bbox{\eta}(t)) -U_{ext}(\br)\,,
\label{eq6.10}
\eeq
we see that (\ref{eq6.9}) is of the local equilibrium form in
(\ref{eq37}), with the non-condensate chemical potential given by
\beq
\tilde \mu(\brt)
=\tilde \mu_0 +U_{ext}(\br) - U_{ext}(\br -\bbox{\eta}(t))\,.
\label{eq6.11}
\eeq
Since $\tilde \mu_0 = \mu_{c0}$ in static equilibrium
[see the discussion after (\ref{eq48})], this local chemical
potential of the non-condensate is equal to the condensate chemical
potential in (\ref{eq6.7}).

We now show that (\ref{eq6.9}) is indeed a solution of our kinetic 
equation in (\ref{eq29}).  
The $C_{22}$ collision integral on the RHS of (\ref{eq29}) is seen to 
vanish as an immediate consequence of (\ref{eq6.9}), namely,
that $f_\eta$ is a local equilibrium Bose distribution.
In addition, from (\ref{eq40}) we see that the $C_{12}$ collision
integral vanishes since $\bv_n = \bv_c$ and $\tilde \mu(\brt) =
\mu_c(\brt)$. This, of course, implies that $\Gamma_{12} = 0$, as we
assumed in the argument leading to (\ref{eq6.5}) and (\ref{eq6.8}).
A straightforward calculation of the LHS of the kinetic equation in 
(\ref{eq29}) using $f_\eta$ in (\ref{eq6.9}) shows that it vanishes if
\beq
m{\partial^2{\bbox \eta}\over\partial t^2} = -{\bbox\nabla}
[U({\bf r}, t) - U_0({\bf r} - {\bbox \eta})]\,.
\label{eq6.12}
\eeq
Using the expression in (\ref{eq6.10}),
we see that (\ref{eq6.12}) reduces to the equation of motion 
in (\ref{eq6.3}).

In summary, we have proven in full generality that the set of dynamical
equations in (\ref{eq27})-(\ref{eq29}) are consistent with the 
generalized Kohn theorem for a harmonic trap described by 
(\ref{eq6.1}). In other words, these equations admit a solution 
consisting of a rigid in-phase oscillation of the 
equilibrium condensate and non-condensate density profiles, the center 
of mass displacement ${\bbox\eta}(t)$ having the trap frequency 
$\omega_\alpha$ for a displacement in the $\alpha$-th direction.
This result is an important check of the internal consistency of our 
treatment of the dynamics. We also find that this particular solution 
is, in fact, independent of collisions and is therefore
valid in the collisionless\protect\cite{bijlsma98} as well as
in the collision-dominated regime. Finally, we
note that no assumptions have been made regarding the
magnitude of the center of mass displacement. These results are
therefore valid for arbitrary amplitude of the center of mass motion.

%% file: varsol.tex
\section{Variational solution of the hydrodynamic equations}
\label{section7}

Our purpose in this Section is to develop a practical means of solving 
hydrodynamic equations of the kind we have derived in this paper. For
simplicity, we shall restrict ourselves to
the ZGN equations given in (\ref{eq49}) and (\ref{eq50}), although a
similar approach could be developed for the more general ZGN$^\prime$
equations. From a practical point of view, the two sets of equations 
give fairly similar results for the frequencies of the first and second 
sound modes in a uniform system, even though they describe different 
regimes. We would expect a similar correspondence in the case of a 
trapped gas. However, it should be noted that the ZGN equations neglect
the damping associated with the relaxation of the chemical potentials,
which is included in the more general ZGN$^\prime$ equations of
Section~\ref{section5}.

A direct solution of the ZGN equations is 
exceedingly difficult, partly for the number of equations
involved and partly for the fact that one is dealing with a
strongly inhomogeneous system. For this reason it is advantageous
to reformulate the problem in terms of a variational 
principle. In doing so, accurate estimates of mode
frequencies can be obtained by using simple trial functions for
the velocity fields whose choice is guided by physical
considerations. Our immediate objective is therefore to
transcribe the ZGN hydrodynamic equations into a variational form.

Although not essential, it is convenient to replace the
velocities in terms of displacement fields. Such a description 
is particularly useful for small amplitude oscillations about 
equilibrium, since in this case each fluid element
makes small excursions from its equilibrium position and the
motion of the fluid is analogous to that of an elastic medium.
The displacement field for the non-condensate, $\bu(\brt)$, and
the condensate, $\bw(\brt)$, are defined by the relations
\beq
{\bf v}_n(\brt) = {\partial \bu(\brt) \over \partial t}\,,\qquad
{\bf v}_c(\brt) = {\partial \bw(\brt) \over \partial t}\,.
\label{eq7.1}
\eeq
Assuming a harmonic time-dependence with frequency $\omega$, these
definitions are equivalent to
\beq
{\bf v}_n(\br) = -i\omega \bu(\br)\,,\qquad {\bf v}_c(\br) =
-i\omega \bw(\br)\,,
\label{eq7.2}
\eeq
and in terms of these variables, the 
continuity equations take the form
\bea
\delta n_c(\br) &=& -\nabla \cdot (n_{c0} \bw) \nonumber \\
\delta \tilde n(\br) &=& -\nabla \cdot (\tilde n_0 \bu) \,.
\label{eq7.3}
\eea

Introducing these definitions into (\ref{eq49}) and eliminating the
variables $\delta n_c$, $\delta \tilde n$ and $\delta \tilde P$, we 
obtain
\bea
-m\tilde n_0 \omega^2 u_i = {\partial \over \partial x_i} \left
( u_j {\partial\tilde P_0 \over \partial x_j} +
{5\over 3}\tilde P_0 {\partial u_j \over \partial x_j} \right )
&+& {\partial U_0 \over \partial x_i} {\partial (\tilde n_0
u_j) \over \partial x_j} \nonumber \\ &+& 2g\tilde n_0 
{\partial^2  \over \partial x_i \partial x_j} \left ( \tilde n_0
u_j + n_{c0} w_j \right )\,.
\label{eq7.4}
\eea
This equation can be simplified considerably by collecting
together terms involving similar derivatives of $u_i$,
and by making use of the
equilibrium condition $\nabla \tilde P_0 + \tilde n_0 \nabla U_0
= 0$. We find
\bea
m\tilde n_0 \omega^2 u_i = -{\partial \over \partial x_i} \left
( \lambda {\partial u_j \over \partial x_j} \right ) - {\partial \over
\partial x_j} \left ( \mu {\partial u_j \over \partial x_i}
\right ) &+& \tilde n_0 u_j {\partial^2 \over \partial x_i
\partial x_j} \left ( U_{ext} + 2gn_{c0} \right ) 
\nonumber \\ &-& 2g \tilde
n_0 {\partial^2 \over \partial x_i \partial x_j} (n_{c0} w_j)\,,
\label{eq7.5}
\eea
where
\beq 
\lambda = {2\over 3} \tilde P_0 + g\tilde n_0^2\,,\qquad
\mu =  \tilde P_0 + g\tilde n_0^2\,.
\label{eq7.6}
\eeq
The first two terms on the right hand side of (\ref{eq7.5}) are 
reminiscent of the divergence of the stress tensor of an anisotropic
elastic medium. To make this correspondence more apparent, we
introduce the symmetric and antisymmetric strain tensors
\bea
u_{ij} &\equiv& {1\over 2}\left ( {\partial u_i \over \partial x_j}
+ {\partial u_j \over \partial x_i} \right )\nonumber \\
\bar u_{ij} &\equiv& {1\over 2}\left ( {\partial u_i \over 
\partial x_j} - {\partial u_j \over \partial x_i} \right )\,.
\label{eq7.7}
\eea
The first two terms on the right hand side of (\ref{eq7.5}) thus become
\begin{displaymath}
-{\partial \over \partial x_i} \left ( \lambda u_{jj} \right )
- {\partial \over \partial x_j} \left ( \mu u_{ij} \right )
+ {\partial \over \partial x_j} \left ( \mu \bar u_{ij} \right )\,.
\end{displaymath}
The part of this expression depending on the symmetric strain
tensor will be recognized as representing an elastic medium, 
with $\lambda$ and $\mu$ playing the role of Lam\'e 
constants~\cite{landau70}.

We are now in a position to define energy functionals for the
non-condensate. First, we introduce the kinetic energy
functional
\beq
K_n[\bu] = {1\over 2} \int d^3r \, m\tilde n_0 u^2\,.
\label{eq7.8}
\eeq
Its variation, $\delta K_n/\delta u_i = m\tilde n_0 u_i$, gives
the coefficient of $\omega^2$ on the left hand side of (\ref{eq7.5}).
Similarly, one can easily check that the variation of the
potential energy functional
\beq
U_n[\bu] = {1 \over 2} \int d^3r \, \left \{ \lambda u_{ii}^2 +
\mu u_{ij}^2 - \mu \bar u_{ij}^2 + u_i u_j \tilde n_0
{\partial^2 \over \partial x_i \partial x_j} (U_{ext} + 2gn_{c0})
\right \}
\label{eq7.9}
\eeq
yields the desired $u_i$-dependent terms on the right hand side
of (\ref{eq7.5}). The two terms proportional to $\mu$ can be written in
several different ways:
\bea
\mu ( u_{ij}^2 - \bar u_{ij}^2) &=& \mu  u_{ij}^2 - {1\over 2}
\mu (\nabla \times \bu)\cdot (\nabla \times \bu) \nonumber \\
&=& \mu {\partial u_i \over \partial x_j} {\partial u_j \over
\partial x_i} \nonumber \\
&=& 2\mu u_{ij}^2 - \mu {\partial u_i \over \partial x_j}
{\partial u_i \over \partial x_j}\,.
\label{eq7.10}
\eea
The first equality in particular shows that the antisymmetric
strain tensor is associated with the solenoidal part of $u_i$.

The last term in (\ref{eq7.5}) accounts for the force exerted on the
non-condensate by fluctuations of the condensate. It can be
represented by the interaction energy functional
\beq
U_{cn}[\bu,\bw] = 2g \int d^3r\, {\partial (\tilde n_0 u_i)
\over \partial x_i} {\partial (n_{c0} w_j) \over \partial
x_j}\,.
\label{eq7.11}
\eeq
We have here indicated explicitly that $U_{cn}$ depends on both
of the displacement fields. 

We next consider the condensate. With the elimination of $\delta n_c$ 
and $\delta \tilde n$ from 
Eqs.~(\ref{eq50}), we arrive at the equation
\beq
m n_{c0} \omega^2 w_i = -n_{c0} {\partial \over \partial x_i} 
\hat K {\partial \over \partial x_j}(n_{c0} w_j) - g n_{c0} 
{\partial^2 \over \partial x_i \partial x_j} (n_{c0} w_j) - 
2g n_{c0} {\partial^2 \over \partial x_i \partial x_j} (\tilde n_0
u_j)\,.
\label{eq7.12}
\eeq
Here we have defined the operator
\beq
\hat K \equiv {1\over 2 \Phi_0(\br)} \hat {\cal L}_0(\br)
{1\over \Phi_0(\br)}\,,
\label{eq7.13}
\eeq
with $\hat {\cal L}_0(\br)$ as given in (\ref{eq52}). We emphasize that
(\ref{eq7.12}) is equivalent to the linearized form of the 
time-dependent GP equation in (\ref{eq4}) (of course with the last two 
terms excluded). In particular, we do {\it not} make the TF 
approximation which neglects the quantum mechanical kinetic energy.

The energy functionals for the condensate analogous to those of
the non-condensate are
\beq
K_c[\bw] =  {1\over 2} \int d^3r \, m n_{c0} w^2
\label{eq7.14}
\eeq
and
\beq
U_c[\bw] = {1\over 2} \int d^3r\, {\partial (n_{c0} w_i)
\over \partial x_i} [ \hat K + g ] {\partial (n_{c0} w_j) 
\over \partial x_j}\,.
\label{eq7.15}
\eeq
The last term in (\ref{eq7.12}) follows from the variation of 
(\ref{eq7.11}) with respect to $w_i$.

We now define the functional
\beq
J[\bu,\bw] \equiv {U[\bu,\bw] \over K[\bu,\bw]}\,,
\label{eq7.16}
\eeq
where
\bea 
U[\bu,\bw] &=& U_n[\bu] + U_c[\bw] + U_{cn}[\bu,\bw]\nonumber \\
K[\bu,\bw] &=& K_n[\bu] + K_c[\bw]\,.
\label{eq7.17}
\eea
The requirement that the functional $J$ be stationary with respect to
variations of $u_i$ and $w_i$  leads to the equations
\bea
{\delta U \over \delta u_i} - \omega^2 {\delta K \over \delta u_i}
&=& 0 \nonumber \\
{\delta U \over \delta w_i} - \omega^2 {\delta K \over \delta w_i}
&=& 0\,,
\label{eq7.18}
\eea
where we have identified $\omega^2$ with the stationary value of
the functional $J$. These equations are identical to (\ref{eq7.5}) 
and (\ref{eq7.12}), which confirms that variation of
the functional $J$ leads to the desired hydrodynamic equations.
More importantly, the functional $J$ offers a convenient variational 
method of estimating the normal mode frequencies of a trapped Bose gas.

One useful approach is the Rayleigh-Ritz method, in which the
velocity components are expanded in some complete set of
functions. This method leads to expressions for the kinetic
and potential energy functionals which are quadratic forms in
the expansion coefficients. Variation of $J$ with respect to
these coefficients results in a set of linear equations which
schematically have the form
\beq
\sum_\beta \left [ U_{\alpha\beta} -\omega^2  K_{\alpha\beta}
\right ] c_\beta = 0\,,
\label{eq7.19}
\eeq
where $U_{\alpha\beta}$ and $K_{\alpha\beta}$ are matrix
elements of the energy functionals. A simplified version of this
method will be used to determined several of the normal modes
of interest.

We now establish some general properties of the solutions to
Eqs.~(\ref{eq7.5}) and (\ref{eq7.12}). For this purpose, it is 
convenient to define the six-component displacement vector
\beq
\bbox{\eta} = \left ( \matrix{\bu\cr \bw \cr} \right )\,.
\label{eq7.20}
\eeq
Eqs.~(\ref{eq7.5}) and (\ref{eq7.12}) can then be combined into one
matrix equation 
\beq
\bbox{L \eta} = \omega^2 \bbox{D \eta}\,,
\label{eq7.21}
\eeq
where the matrix ${\bbox L}$ has the block structure
\beq
{\bbox L} = \left ( \matrix{ \hat L_{11}&\hat L_{12}\cr
                           \hat L_{21}&\hat L_{22}\cr} \right )\,.
\label{eq7.22}
\eeq
The elements of this matrix are three-by-three matrix operators
defined as
\bea
(\hat L_{11}\bu)_i &=& 
-{\partial \over \partial x_i} \left
( \lambda {\partial u_j \over \partial x_j} \right ) - {\partial \over
\partial x_j} \left ( \mu {\partial u_j \over \partial x_i}
\right ) + \tilde n_0 u_j {\partial^2 \over \partial x_i
\partial x_j} \left ( U_{ext} + 2gn_{c0} \right )\nonumber \\
(\hat L_{12}\bw)_i &=& -2g\tilde n_0 {\partial^2\over \partial x_i
\partial x_j}(n_{c0}w_j)\nonumber \\
(\hat L_{21}\bu)_i &=& -2g n_{c0} {\partial^2\over \partial x_i
\partial x_j}(\tilde n_0 u_j)\nonumber \\
(\hat L_{22}\bw)_i &=& 
-n_{c0}{\partial \over \partial x_i} (\hat K + g)
{\partial \over \partial x_j} (n_{c0}w_j)\,.
\label{eq7.23}
\eea
Similarly, the matrix $\bbox{D}$ is block-diagonal ($\hat D_{12} = \hat
D_{21} = 0$) with elements
\bea
\hat D_{11} &=& m\tilde n_0 \hat 1\nonumber \\
\hat D_{22} &=& m n_{c0} \hat 1\,.
\label{eq7.24}
\eea
Here, $\hat 1$ is a three-by-three unit matrix. From the definition of
$\bbox{L}$, it can easily be checked that
\beq
\int d^3r \,\bbox{\zeta}^* \cdot (\bbox{L \eta}) = 
\int d^3r \, (\bbox{L \zeta})^* \cdot \bbox{\eta} \,.
\label{eq7.25}
\eeq
which shows that $\bbox{L}$ is a Hermitian operator. This property can
be used in (\ref{eq7.21}) to show that the eigenvalues $\omega_\alpha^2$
of $\bbox{L}$ are real and that the corresponding eigenvectors
$\bbox{\eta}_\alpha$ satisfy the orthonormality relation
\beq
\int d^3 r\, \bbox{\eta}^*_\alpha \cdot (\bbox{D\eta}_\beta) =
\delta_{\alpha \beta}\,.
\label{eq7.26}
\eeq
 
\subsection{In-phase and out-of-phase dipole oscillations and the 
center of mass mode}

As our first application of the above formalism, we consider the 
center of mass motion of a gas
trapped in a parabolic potential given by (\ref{eq6.1}). According 
to our earlier discussion in Section VI, we expect a solution
whereby the total density oscillates harmonically without distortion.
One can easily show that a uniform displacement $u_i =
w_i = A$ of both components along one of the Cartesian
directions is a solution of (\ref{eq7.5}) and (\ref{eq7.12}), with
frequency $\omega = \omega_i$, $i = x,\,y,\,z$. According to
(\ref{eq7.3}), this implies a density fluctuation $\delta n_c(\br)
= -A{\textstyle{\partial n_{c0}} \over \textstyle{\partial x_i}}$ 
for the condensate and
$\delta \tilde n(\br)= -A{\textstyle{\partial \tilde n_0} \over
\textstyle{\partial x_i}}$ 
for the non-condensate. These forms follow from (\ref{eq6.2}) on 
expanding in the displacement $\bbox{\eta}$.

This center of mass mode has the condensate and non-condensate
oscillating in phase with the same amplitude. One would also
expect a dipole mode in which the two components oscillate out
of phase. In this case, there is no reason to expect the mode
densities to be given by the equilibrium profiles as for the center of
mass mode. However, in obtaining a variational estimate of the mode
frequencies which is accurate to second order in the error of
the mode density, it is reasonable to assume constant
displacements for the two components, but in general with
different amplitudes. Thus we take $u_i = A_n$ and $w_i = A_c$
for a particular direction $i$. For these displacements, the equations
in (\ref{eq7.18}) become
\bea
{\partial U \over \partial A_c} - \omega^2 {\partial K \over
\partial A_c} &=& 0 \nonumber \\
{\partial U \over \partial A_n} - \omega^2 {\partial K \over
\partial A_n} &=& 0 \,.
\label{eq7.27}
\eea
Using the constant displacements to evaluate the energy
functionals, we obtain
\bea
K &=& {1\over 2} M_c A_c^2 + {1\over 2} M_n A_n^2\nonumber \\
U &=& {1\over 2} M_c \omega_i^2 A_c^2 + {1\over 2} M_n \omega_i^2 
A_n^2 + {1\over 2} k_i(A_c-A_n)^2\,,
\label{eq7.28}
\eea
where $M_c = mN_c$, $M_n = m\tilde N$ and 
\beq
k_i \equiv -2g\int d^3r \left ({\partial n_{c0} \over \partial
x_i} \right ) \left ({\partial \tilde n_0 \over \partial x_i} \right
)\,.
\label{eq7.29}
\eeq
(Here, the repeated index is {\it not} summed.) This energy
corresponds to two one-dimensional oscillators confined in an
external parabolic potential of frequency $\omega_i$ and coupled
together by a spring of force constant $k_i$. 

Substituting (\ref{eq7.28}) into (\ref{eq7.27}), 
we obtain the pair of coupled equations
\beq
\pmatrix{\omega_i^2+\omega_1^2-\omega^2 & -\omega_1^2 \cr
         -\omega_2^2 & \omega_i^2+\omega_2^2-\omega^2 \cr}
\pmatrix{A_c\cr A_n} = 0
\label{eq7.30}
\eeq
where $\omega_1^2 = k_i/M_c$ and $\omega_2^2 = k_i/M_n$. One solution 
of (\ref{eq7.30}) is $\omega = \omega_i$ with $A_c = A_n$, which is
the center of mass mode discussed more generally in the previous 
section. However, the present analysis gives another solution
\bea
\omega^2 &=& \omega_i^2+\omega_1^2+\omega_2^2\nonumber \\
&=& \omega_i^2 - 2g{M_c + M_n \over M_c  M_n} 
\int d^3r \left ({\partial n_{c0} \over \partial x_i} \right
) \left ({\partial \tilde n_0 \over \partial x_i} \right)\,,
\label{eq7.31}
\eea
with the corresponding amplitudes satisfying $M_c A_c + M_n A_n = 0$. 
This implies that this mode is essentially an {\it out-of-phase}
oscillation of the condensate and non-condensate and that
the center of mass remains stationary.
To the extent that the displacement fields for this
mode are indeed uniform, one finds that $N_c/\tilde N = - A_n/A_c$ and 
hence a measurement of the relative amplitudes would provide a direct
estimate of the relative number of atoms in the condensate and
non-condensate. 

These results for the in-phase and out-of-phase
oscillations of the condensate and non-condensate were previously 
reported in ZGN\cite{zaremba98}. The first observation of such an
out-of-phase dipole oscillation was made by the MIT 
group~\cite{stamper98} for an 
anisotropic cigar-shaped trap. They find that this mode has a frequency
slightly below the axial frequency of the trap, and is strongly damped
at higher temperatures. Our prediction 
would give a frequency slightly {\it above} the trap frequency
since the slopes of the condensate and non-condensate tend to have
opposite signs (see Fig.~3) in the region of space contributing to the
integral in (\ref{eq7.31}). It is possible that the damping of the mode
not included in the ZGN equations
would have to be taken into account in order to resolve this
discrepancy.

\subsection{Monopole and Quadrupole Modes in Isotropic Traps}

We next consider some of the other low-lying modes of the trapped gas.
For simplicity, we shall work with an isotropic trap with a common 
oscillator frequency $\omega_0$ in all directions. The particular modes
we shall consider here are the monopole and quadrupole modes, which
respectively have $l=0$ and $l=2$ angular character. The lowest monopole
mode is expected to be a breathing mode whose mode density would have
the approximate form $\lambda^{-3} n_0(\br/\lambda)$ where $\lambda$ is
a time-dependent, but spatially independent, scaling parameter. The
equilibrium situation corresponds to $\lambda = 1$ and small deviations
from this value imply a displacement field of the form $\bu \propto
\br$. The quadrupole mode can be viewed as an anisotropic scaling
solution with $l=2$ symmetry. This requires a displacement field with
components $(x,y,-2z)$. Both the $l=0$ and $l=2$ modes can be 
represented by displacement fields of the kind
\begin{mathletters}
\bea
\bu &=& (ax,ay,bz)\\
\bw &=& (cx,cy,dz)\,,
\eea
\label{eq7.32}
\end{mathletters}
\noindent
where $a$, $b$, $c$ and $d$ are constant variational parameters.
For the monopole mode, we have $a=b$ and $c=d$, while for the quadrupole
mode, $b= -2a$ and $d=-2c$. The various terms required in the
evaluation of the energy functional $J$ for these modes are given in 
Appendix C, where the expressions are reduced to
the form of radial integrals over equilibrium quantities.

One advantage of a variational approach is that the calculated mode
frequencies can be improved systematically by introducing trial
displacements which are more general than those given,
for  example, by (\ref{eq7.32}).
In the case of the monopole modes, displacement fields of the form
$\bu = \br\sum_{\nu=0}^n a_\nu r^{2\nu}$ were examined~\cite{king98},
where the number of terms $n$ in the expansion was varied from 1 to 5. 
The expansion coefficients $a_\nu$ (together with a similar set for 
the condensate) play the role of the variational parameters,
and the functional $J$ can be minimized with respect to this set. 
The total number of collective modes
generated in this way is $2n$, with the lowest pair corresponding to the
modes found using (\ref{eq7.32}). One can examine the convergence of the
mode frequencies with $n$ and one finds that the results for the lowest
monopole modes are already quite good with $n=1$. We therefore have
confidence in the accuracy of the present results obtained using
the simplest variational form given in (\ref{eq7.32}).

In Fig.~4, we present our results for the lowest $l=0$ and $l=2$ 
modes in an isotropic trap with radial frequency $\nu_0 = 200$ Hz. 
As for the case of the dipole modes discussed earlier, the restricted 
form of the variational displacement
fields gives rise to two modes for each symmetry.  The calculations
were performed for 5000 $^{87}$Rb atoms having an s-wave scattering 
length of $a = 110$ a$_B$. In the lower part of Fig.~4  we show the
condensate fraction as a function of temperature.
The transition temperature is found to
be $T_{BEC} \approx 149$ nK. In Fig.~3, we show the condensate and
non-condensate radial densities at $T = 100$ nK, where the number of
condensate and non-condensate atoms is approximately equal. 
The depletion in the non-condensate density at the centre of the trap
due to the repulsive interaction with the
condensate in the overlap region is clearly evident. These results are 
very similar to those obtained within the static HFP 
approximation~\cite{hutchinson97}, and confirmed by other
calculations~\cite{giorgini97}.

In the upper part of Fig.~4 we show the mode frequencies as a function
of temperature. The results below $T_{BEC}$ can be viewed in a first 
approximation as a superposition of the
condensate modes given by  the HFP scheme~\cite{hutchinson97}
and the non-condensate modes found above $T_{BEC}$~\cite{griffin97b},
extended to lower temperatures. At $T=0$, the {\it condensate} modes 
start at $\omega \approx 2.25 \omega_0$ for $l=0$
and  $\omega \approx 1.5 \omega_0$ for $l=2$, which are close to the TF
limits ($\sqrt{5}\omega_0$ and $\sqrt{2}\omega_0$, 
respectively~\cite{stringari96}). Apart from some hybridization effects
to be discussed, these two modes basically follow the HFP 
behaviour with
increasing temperature, merging together as $T$ approaches $T_{BEC}$.
The two {\it non-condensate} modes are analogous to the $l=0$ mode at 
$\omega = 2\omega_0$ and the $l=2$ mode at $\omega = \sqrt{2}\omega_0$,
as found by Griffin {\it et al.}~\cite{griffin97b} above $T_{BEC}$. 
However, unlike this earlier work, the present results include the
effect of interactions, and even above $T_{BEC}$, the mode frequencies 
are shifted slightly from their non-interacting values, although it 
turns out that the $l=2$ mode frequency ($2^-$) remains at 
$\sqrt{2}\omega_0$ within the present variational approximation.
For temperatures
below $T_{BEC}$, the non-condensate density fluctuations couple with
the condensate fluctuations and these modes show an increasing frequency
shift with decreasing $T$. In addition, hybridization of the modes is 
observed at points were the uncoupled modes cross. For example, the
$\omega=2\omega_0$ mode is seen in Fig.~4 to hybridize with 
the condensate mode just below
$T_{BEC}$, and once again at lower temperatures.

This hybridization is most clearly revealed by looking at the amplitudes
of the condensate and non-condensate fluctuations associated with
each of the
modes. In Fig.~5 we show the amplitudes for the two $l=0$ modes, where
the upper $l=0$ mode is labelled $0^+$ and the lower, $0^-$. Starting at
low temperatures, the $0^+$ mode is essentially a non-condensate
oscillation ($|A_{n+}| \approx 1$) with a very small condensate
component. After passing through the hybridization point near 
$T \simeq 35$ nK, the
upper $l=0$ mode changes over to a condensate oscillation, with a small
out-of-phase non-condensate amplitude. At the second hybridization point
near 136 nK, the amplitude of the non-condensate grows dramatically as
the mode switches over to the $2\omega_0$ non-condensate mode.
In the temperature range between this
hybridization point and $T_{BEC} \simeq$ 149 nK, the
condensate amplitude is still appreciable; however, the condensate
has a minimal effect on the non-condensate oscillation since the
condensate fraction is quite small here. One can simply view the
condensate oscillation as being driven by the much more massive
non-condensate component. The amplitudes for the lower frequency
$l=0$ mode ($0^-$)
are shown in the lower part of Fig.~5. Above the $T \simeq 136$ nK
hybridization point, this mode is seen to be essentially a condensate 
oscillation ($|A_{c-}| \simeq 1$),
with a very small non-condensate amplitude. A mode
of this type is to be expected since the non-condensate is relatively
massive and the condensate simply oscillates in the presence of a static
non-condensate density. In this respect, this mode is clearly equivalent
to the mode found in the HZG approximation~\cite{hutchinson97},
which ignored the collective dynamics of the non-condensate completely.

The behaviour of the $l=2$ amplitudes shown in Fig.~6 follows a 
similar pattern. The high frequency $l=2$ mode ($2^+$) 
starts off as a non-condensate oscillation at low $T$
and then switches over to a condensate oscillation all the way up to
$T_{BEC}$, since in this case there is no second hybridization point. 
The lower $l=2$ mode ($2^-$) is mainly a non-condensate oscillation for
all temperatures, although a small condensate amplitude is
evident as well. Interestingly, the condensate amplitude goes to zero at
$T \approx 125$ nK, which implies that at this temperature, the 
condensate is effectively stationary
in the presence of the oscillating non-condensate. A similar
zero-crossing can be seen in Fig.~5 in the case of the lower $l=0$ mode.
This behaviour can perhaps be understood most easily in the case of the
breathing mode, which corresponds to a dilation of the density. By 
referring to the non-condensate density in Fig.~3, one can see that the
corresponding density fluctuation, $\delta \tilde n \propto 3\tilde n_0
+ r(\partial
\tilde n_0 / \partial r)$, will be both positive and negative in
different regions of space. Thus the zero-amplitude condensate
oscillation arises when the radial force on the condensate due to the
breathing non-condensate averages to zero. Although the situation for
the $l=2$ mode is not as easy to interpret, presumably the force on the
condensate is being averaged to zero in a similar way. 
We conclude that a 
simultaneous measurement of the condensate and non-condensate amplitudes
as a function of temperature should reveal a particular temperature at
which the condensate amplitude passes through a minimum. This would be
an especially interesting experimental signature of the
modes we have been discussing.

We have checked that the
results shown here for 5000 atoms are in fact qualitatively
similar to those for other values of $N$, from a lower limit of 2000 
to an upper limit of 20,000 atoms. There is no reason why
this qualitative behaviour should not persist to arbitrarily large $N$,
where a hydrodynamic description would be expected to be valid. Thus the
numerical calculations we have performed can be viewed as 
representative of the modes in the hydrodynamic regime. 

We are
not claiming that a trapped gas containing 5000 atoms, approximately the
number of atoms in the experiments of Jin {\it et al.}~\cite{jin97},
is actually in the hydrodynamic regime. The crossover into this regime
occurs at $\omega\tau_c \approx 1$, where $\tau_c$ is the mean time
between collisions. If one uses the classical gas expression
$\tau_c^{-1} = \sigma \bar n \bar v$ (where $\sigma=8\pi a^2$ is the 
s-wave cross-section, $\bar n$ is the maximum non-condensate 
density, and $\bar v$ is a characteristic thermal velocity), one obtains
for the JILA experiment~\cite{jin97} the estimate $\omega \tau_c 
\simeq 4$, which places this experiment in the collisionless regime.
However, this estimate based on the classical collision time is not
entirely secure. Our recent work~\cite{nikuni99} has shown that
collision times defined by the $C_{12}$ and $C_{22}$ collision integrals
can in fact be much smaller than $\tau_c$ in the region around
$T_{BEC}$. Thus there is the possibility that the JILA data near
$T_{BEC}$ is actually closer to the hydrodynamic regime than would be
expected on the basis of the classical gas expression for $\tau_c$. 
In any case, it would be very useful to have further experimental
studies of collective modes at finite temperatures in traps containing a
much larger number of atoms. Performing such
experiments in isotropic traps would further facilitate a comparison
with our theoretical predictions, not only with regard to mode 
frequencies, but also with respect to the relative amplitude of the 
condensate and non-condensate oscillations.

Finally, we comment on the results of Shenoy and
Ho~\cite{shenoy98}, which were obtained using the Landau two-fluid
hydrodynamic equations. Above $T_{BEC}$, they find hydrodynamic modes in
agreement with our results, namely, an $l=0$ mode at $2\omega_0$, the
center of mass mode at $\omega_0$ and an $l=2$ mode at
$\sqrt{2}\omega_0$. Below $T_{BEC}$, these modes continue as
non-condensate-like modes (i.e., the non-condensate has the larger
amplitude as indicated in Figs.~5 and 6) and their results are again in
agreement with ours. However, there are large differences in the
condensate-like oscillations which, in both approaches, are essentially
out-of-phase oscillations of the two components. Referring to Fig.~4, we
see that our $l=0$ and $l=2$ condensate modes are essentially the modes
obtained~\cite{hutchinson97} from the $T=0$ time-dependent GP equation,
but with the condensate number given by the finite-temperature value
$N_c(T)$.  These modes then become degenerate as $T$
approaches $T_{BEC}$, with a common value close to $2\omega_0$. In the
work of Shenoy and Ho, however, the lowest $l=0$ condensate mode
frequency falls steadily from $2\omega_0$ at $T_{BEC}$ with decreasing
temperature and gives no indication of approaching the expected GP
frequency of about $2.25\omega_0$ at low temperatures. Their lowest
$l=2$ condensate mode starts out at $\sqrt{2}\omega_0$ at $T_{BEC}$ and
stays close to this value with decreasing temperature. 

As a partial
explanation of these differences, we recall that our results are based 
on the ZGN equations which assume only partial local equilibrium and 
therefore are not in the Landau regime studied by Shenoy and Ho.
However, as shown in Fig.~1,  the results for the first and second
sound mode frequencies of a uniform gas are in fact very similar in the
two regimes, which would lead one to expect the differences also to be 
small in the case of a trapped Bose gas.
Perhaps the major difference can be accounted for by the different
equilibrium properties assumed in the two calculations. As discussed in
Appendix B, Shenoy and Ho based their calculations on the Lee-Yang
equation of state~\cite{lee58}, which is only correct to first order in
the interactions and, moreover, leads to a uniform non-condensate
density in the region of overlap with the condensate. In contrast, our
calculations make use of condensate and non-condensate densities
determined self-consistently as shown in Fig.~3. It would be useful to
recalculate the hydrodynamic modes in the Landau limit with
these equilibrium densities.

%% file: conclusions.tex
\section {Conclusions}
\label{section8}

In this paper, we have derived a set of equations which describe the
dynamics of a trapped Bose-condensed gas at finite temperatures. These
equations consist of a generalized Gross-Pitaevskii equation for the
condensate order parameter $\Phi(\brt)$ and a semiclassical kinetic
equation for the excited atom (non-condensate) distribution function 
$f(\bp,\br,t)$. By limiting ourselves to higher temperatures, we arrive
at a simple and intuitive picture in which the excited atoms   
behave as particles moving in a dynamic Hartree-Fock field. Collisions
between all the atoms are included and in particular, allow for the
transfer of atoms between the two components.

Most of our discussion is devoted to the hydrodynamic regime and the
derivation of a closed set of generalized hydrodynamic equations for the
two components. These equations are based on the assumption that 
collisions between excited atoms are sufficiently rapid to drive the 
distribution 
function $f(\bp,\brt)$ toward the local equilibrium Bose-Einstein 
distribution $\tilde f(\bp,\brt)$, with the consequence that the 
$C_{22}[\tilde f]$ collision integral for the excited atoms vanishes. 
However, the $C_{12}[\tilde f]$ collision integral describing collisions
between the condensate and non-condensate atoms remains finite, and as 
a result, our equations can be used to describe the 
situation in which the condensate is not in diffusive local equilibrium
with the non-condensate. As noted in Section~\ref{section1}, this
situation was discussed many years ago in the context of superfluid
helium near the $\lambda$ point, but using a more phenomenological 
approach~\cite{pitaevskii59,miyake76}. Previous microscopic
studies~\cite{bogoliubov70,kirkpatrick} of the hydrodynamics of Bose
superfluids have generally ignored this possibility and simply 
assumed that the superfluid and normal fluid were in local equilibrium 
with each other.

We find that the equilibration process associated with the transfer of 
atoms between the two components leads to the existence of a new 
relaxational mode characterized by the relaxation time $\tau_\mu$. 
If we consider the dynamics of the system on time scales set by the
frequency $\omega$, we find that the linearized form of our theory is
equivalent to the Landau two-fluid equations in the limit
$\omega\tau_\mu \ll 1$. In the opposite limit, $\omega\tau_\mu \gg 1$,
the $C_{12}$ collisions can be neglected and we recover the ZGN limit 
recently considered by the authors~\cite{zaremba98}. The transition
between these two limits can be analyzed straightforwardly for the case 
of a uniform Bose gas. As shown in Section~\ref{section5}, we find that
the first and second sound modes are coupled to the new relaxational 
mode and that this coupling gives rise to anomalous sound absorption
near $\omega\tau_\mu \approx 1$. Of course, a complete theory of sound
absorption must also take into account other dissipative mechanisms such
as viscosity and thermal conduction~\cite{nikuni98}.
We briefly discussed the
qualitative effect of a finite thermal conductivity at the end of
Section~\ref{section5}.

One outcome of our study has been the realization that the various
collision and relaxation times which appear are strongly temperature 
dependent. In Ref.~\cite{nikuni99}, these were evaluated for a uniform 
Bose gas, where it was found that $\tau_{12}$ and $\tau_{22}$ become 
very small as $ T \to T_{BEC}$, but that $\tau_\mu$ becomes very long. 
It is therefore possible to have a situation in which $\omega\tau_{22} 
\ll 1$, as required by our local equilibrium assumption, but with
$\omega\tau_\mu\gg 1$. These conditions will always occur close enough
to $T_{BEC}$ and define the regime in which the ZGN equations can be
used. The Landau regime will only be valid at intermediate temperatures
where $\tau_\mu \simeq \tau_{22}$ and $\omega\tau_\mu\ll 1$. In a
separate paper, we shall give a detailed discussion of how these
relaxation times depend on position within a trapped Bose gas. Since the
the edge of the condensate is effectively in the critical regime, we 
anticipate anomalous collision and relaxation times in this region.

In Section~\ref{section7}, we developed a powerful variational approach
to solve our generalized hydrodynamic equations. Although we restricted
our discussion to the collective modes in isotropic traps, the 
method can also be applied to anisotropic traps, where a direct 
solution of the hydrodynamic equations (as in Ref.~\cite{shenoy98})
would be extremely difficult. Using this variational approach to obtain
solutions of the ZGN equations, we found significant differences in 
some of the low-lying hydrodynamic oscillations from the results
of Shenoy and Ho~\cite{shenoy98} based on the Landau two-fluid 
equations. At
present, we believe these differences arise largely from the 
use in Ref.~\cite{shenoy98} of the
TF approximation in conjunction with the Yang-Lee~\cite{lee58} free 
energy approximation for the equilibrium thermodynamic properties (see 
Appendix B). To confirm this, however, will require detailed 
calculations based on the full ZGN$^\prime$ equations. Such calculations
will have the added benefit of allowing us to explore the full range of
behaviour from the ZGN limit to the Landau two-fluid regime.

There are other possible generalizations and extensions of the work 
presented in this paper. We briefly discuss some of these, partly for
the purpose of pointing out some of the limitations of the present work.
Perhaps the most obvious limitation of our simple model is the use of 
the Hartree-Fock approximation, as given in (\ref{eq24}), to describe
the dynamics of the thermally excited atoms. As noted at the end of
Section~\ref{section3}, this corresponds to thermal excitations having 
an energy $E_p(\brt) = p^2/2m + gn_c(\brt)$, which is the correct high 
momentum limit of the Popov approximation for the excitation
spectrum~\cite{giorgini97}. This emphasizes the fact that our kinetic 
equation (\ref{eq29}) can only be used to describe low energy collective
modes at {\it finite} temperatures, where (\ref{eq24}) gives a
good description of the dominant elementary excitations determining the
thermodynamic properties.

In the region of very low temperatures, the HF spectrum (\ref{eq24}) 
we have used in this paper is clearly inadequate. A first step in
improving our treatment would be to use the full Popov approximation for
the excitation spectrum. In this approximation, the excitations are
identical to the Bogoliubov spectrum, with the exception that the
condensate density is now temperature dependent. Indeed, 
KD~\cite{kirkpatrick} have used this Popov spectrum in the case of a 
uniform gas to derive a kinetic equation valid at very low
temperatures. The main difference this leads to in the $C_{12}$
collision integral is the appearance of Bose coherence factors  which
result from the transformation from a description in terms of excited
atoms to one in terms of Popov-Bogoliubov quasiparticles.

In this connection, we note that generalized kinetic equations can be
derived using the well-known Kadanoff-Baym method~\cite{kadanoff62},
which starts from a specific self-energy approximation for
non-equilibrium single-particle Green's functions. The full,
self-consistent Hartree-Fock-Bogoliubov (HFB) approximation for the 
thermal excitations includes the off-diagonal self-energy of the 
non-condensate ($\tilde m$) neglected in the Popov 
approximation~\cite{griffin96} used here. (See Section 6 of
Ref.~\cite{griffin99} for a recent attempt to classify the various
approximations used in the treatment of static and dynamic properties of
Bose-condensed gases.)
As is well-known in the Bose 
gas literature~\cite{hohenberg65,griffin96}, the full first order HFB 
does not give a consistent ``gapless" approximation of the thermal 
excitations, since the off-diagonal self-energies are second order in 
the interaction $g$. Nevertheless, the kinetic equation based on the 
HFB spectrum does allow one to discuss collective modes which are 
consistent with conservation laws.
The HFB approximation was recently used to 
derive~\cite{imamovic98} coupled kinetic equations for a Bose-condensed
gas in the collisionless regime. Work to include collisions within this
scheme is currently in progress.

Turning to another basic limitation of our present analysis, we recall
that the derivation of the ZGN$^\prime$ hydrodynamic equations 
depended on the assumption that the excited atoms were in local 
equilibrium with each other, i.e., $f(\bp,\brt)\simeq \tilde
f(\bp,\brt)$, where by definition $C_{22}[\tilde f] = 0$. The next step
would be to look at the effect of small deviations from this local Bose
distribution, $f \simeq \tilde f +\delta f$. This procedure parallels
the Chapman-Enskog treatment of transport processes in classical gases.
It was implemented recently for trapped Bose gases~\cite{nikuni98} 
to generate corrections to the ZGN hydrodynamic
equations due to shear viscosity and thermal conductivity. This gives
explicit expressions for hydrodynamic damping associated with the
Uehling-Uhlenbeck collision integral $C_{22}$ in (\ref{eq23b}). 
Similar corrections to the ZGN$^\prime$ equations
will be discussed elsewhere. 

Finally, it should be noted that the situation in a trapped Bose gas is
quite different from that in a uniform gas. Due to the decreasing 
density in the tail of a trapped thermal cloud, the assumption of local
equilibrium enforced by collisions must always break down.
This leads to the interesting result that
the damping of certain collective modes above $T_{BEC}$ is governed by 
the low-density tail of the thermal cloud. It was shown in recent
work~\cite{kavoulakis98,nikuni98} that a good estimate of this damping
could be obtained by introducing a spatial cutoff in the hydrodynamic
equations. Although this type of calculation can also be extended to 
below $T_{BEC}$, it is clear that a more rigorous
solution of the kinetic equation in (\ref{eq29}) is needed in order to
deal with the kinetics in the low-density tail of a trapped Bose gas.

\acknowledgements
E.Z. and A.G. would like to thank both the Institute for Theoretical 
Physics at
Santa Barbara and JILA at the University of Colorado, for a stimulating
environment and financial support during various stages of this work. In
addition, E.Z. would like to thank the Institute for Theoretical Physics
at Utrecht University for its hospitality and support. We also
acknowlege stimulating discussions with E. Cornell, T.L. Ho and H.T.C. 
Stoof.  T.N. is supported
by a JSPS fellowship and A.G. and E.Z. are supported by research grants
from NSERC of Canada.

%% file: append.a.tex
\section{Derivation of Collision Integrals}
\label{appendixa}

In this Appendix, we give a more detailed derivation of
the expressions for the collision integrals given by Eqs.~(\ref{eq23b}) 
and (\ref{eq23a}), as well as the three-field correlation function given
by (\ref{eq25}).  We closely follow the approach of Kirkpatrick and
Dorfman~\cite{kirkpatrick} in their calculation of $C_{12}$ and 
$C_{22}$, apart from the fact that we work in the lab
frame as opposed to the local rest frame. 
In order to obtain a closed kinetic equation
for the distribution function $f(\bp,\br,t)$, we must express
higher order correlation functions in terms of $f(\bp,\br,t)$.
For this purpose,
we treat $\hat H'(t)$ as a perturbation to the zeroth-order Hamiltonian
$\hat H_0(t)$, as defined in (\ref{eq14}). We shall effectively
calculate collision integrals to second order in $g$, while explicitly 
keeping interaction effects in excitation energies and chemical 
potentials only to first order in $g$.

The formal solution to (\ref{eq16}) can be written as (with $\hbar =1$)
\begin{equation}
\tilde\rho(t,t_0)=\hat S_0(t,t_0)\hat\rho(t_0)\hat S^{\dagger}_0(t,t_0)
-i\int_{t_0}^{t}dt' \hat S_0(t,t')[\hat H'(t'),\tilde\rho(t',t_0)]
\hat S^{\dagger}_0(t,t'),
\label{A1}
\end{equation}
where the unperturbed evolution operator $S_0(t,t_0)$
satisfies the equation of motion
\begin{equation}
i\frac{d\hat S_0(t,t_0)}{dt}=\hat H_0(t)\hat S_0(t,t_0)\,.
\label{A2}
\end{equation}
The solution of this equation is
\begin{equation}
\hat S_0(t,t_0)\equiv T\exp \left[-i\int_{t_0}^tdt'\hat H_0(t')\right]
\label{A3}
\end{equation}
where $T$ is the time-ordering operator.
Iterating (\ref{A1}) to first order in $\hat H'$, one has
\begin{equation}
\tilde\rho(t,t_0)\simeq \hat S_0(t,t_0)\hat\rho(t_0)\hat 
S^{\dagger}_0(t,t_0) -i\int_{t_0}^{t}dt' \hat S_0(t,t')[\hat H'(t'),
\hat S_0(t',t_0)\hat\rho(t_0)\hat S^{\dagger}_0(t',t_0)]
\hat S^{\dagger}_0(t,t').
\label{A4}
\end{equation}
Using this in (\ref{eq15}), the expectation value of an arbitrary 
operator $\hat O(t)$ can be expressed as
\begin{eqnarray}
\langle \hat O \rangle_t =
\langle \hat O(t) \rangle&=&{\rm Tr}\hat\rho(t_0)
\Bigl\{\hat S^{\dagger}_0(t,t_0)\hat O(t_0)\hat S_0(t,t_0)  \cr
&&\qquad\qquad -i\int_{t_0}^{t}dt'
\hat S^{\dagger}_0(t',t_0)[\hat S^{\dagger}_0(t,t')
\hat O(t_0) \hat S_0(t,t'), \hat H'(t')]\hat S_0(t',t_0)\Bigr\}.
\label{A5}
\end{eqnarray}

We first consider the three-field correlation function
$\langle \tilde\psi^{\dagger}\tilde\psi\tilde\psi\rangle$, which appears
in (\ref{eq5a}). Referring to (\ref{A5}), we can see that there are two
ways in which such an anomalous average can arise. 
The first term on the RHS of (\ref{A5}) gives the initial
correlation, while the term proportional to $\hat H'$ defines those
correlations (to lowest order in $g$) which build up in time as a result
of the Bose broken symmetry. It is the latter which we identify with
collision processes. Thus, for the purpose of evaluating higher order
correlation functions, we assume that the initial density matrix is such
that the initial correlations vanish. This is the basic assumption we 
make in the following~\cite{kirkpatrick}. Since the evolution according
to $\hat S_0(t,t_0)$ is number conserving, this assumption implies that
in evaluating various correlation functions we need retain only those
terms which preserve the total number of particles.
Thus, to lowest order in the interaction $g$, 
the three-field correlation function is given by
\bea
&&\langle \tilde\psi^{\dagger}(\brt)\tilde\psi(\brt)\tilde\psi(\brt)
\rangle \nonumber \\ &&\hskip .5truein =-i {\rm Tr}\hat\rho(t_0) 
\int_{t_0}^{t}dt'
\hat S^{\dagger}_0(t',t_0)\Big [\hat S^{\dagger}_0(t,t')
\tilde\psi^{\dagger}(\brt_0)\tilde\psi(\brt_0)\tilde\psi(\brt_0)
\hat S_0(t,t'), \hat H_1'(t')+H_3'(t')\Big ] 
\nonumber \\ &&\hskip 5truein \times \hat S_0(t',t_0)\,,
\label{A6}
\eea
where it is understood that only those parts of $\hat H_1'$ and 
$\hat H_3'$ are kept which yield an equal number of creation and 
annihilation operators.

Let us first consider the contribution coming from $\hat H_1'(t')$,
which is defined in (\ref{eq14}). The evaluation of (\ref{A6}) and
similar correlation functions is facilitated by two key
assumptions~\cite{kirkpatrick}: (i) that the effect of $\hat H_1'(t')$ 
in the interval $t_0 \le t' \le t$ is essentially a collision process,
which occurs on a time scale much shorter than all other time scales in
the problem, and (ii) that the hydrodynamic variables vary slowly in 
space and time. One therefore expects the dominant contribution from
the commutator of field operators $\tilde \psi(\brt)$ and terms 
contained in the effective Hamiltonian to come from values of $\br'$ 
and $t'$ close to $\br$ and $t$.  In this situation, we can expand the 
quantities $n_c$, $\tilde n$, $\theta$ and $U$ that 
appear in $\hat H_{\rm eff}$ about this point.
It is sufficient to use
\begin{eqnarray}
n_c({\bf r'},t')&\simeq& n_c(\brt), \ 
\tilde n({\bf r'},t')\simeq \tilde n(\brt), \ 
U({\bf r'},t')\simeq U(\br,t), \nonumber \\
\theta({\bf r'},t')&\simeq& \theta(\brt)+{\partial \theta \over \partial
t} (t'-t) + \bbox{\nabla}\theta\cdot(\br'-\br) \nonumber \\
&=& \theta(\brt)-\varepsilon_c(\brt)(t'-t)
+m{\bf v}_c(\brt)\cdot(\br'-\br) \,,
\label{A7}
\end{eqnarray}
and
\beq
\hat S_0(t,t') \simeq e^{-i\hat H_0(t)(t-t')}\,.
\label{A8}
\eeq

Introducing the Fourier transform of the non-condensate field operators 
according to ($V$ is the volume of the system)
\begin{equation}
\tilde\psi({\bf r},t_0)=\frac{1}{\sqrt{V}}\sum_{\bf p}a_{\bf p}
e^{i{\bf p}\cdot {\bf r}},\ \ 
\tilde\psi^{\dagger}({\bf r},t_0)=\frac{1}{\sqrt{V}}\sum_{\bf p}
a^{\dagger}_{\bf p}e^{-i{\bf p}\cdot{\bf r}},
\label{A9}
\end{equation}
and using (\ref{A7}) and (\ref{A8}), one contribution to the commutator
in (\ref{A6}) is
\bea
&&[\hat S^{\dagger}_0(t,t')
\tilde\psi^{\dagger}(\brt_0)\tilde\psi(\brt_0)\tilde\psi(\brt_0)
\hat S_0(t,t'),\hat H_1'(t')]\nonumber \\
&&= -{2g\over V^2}\sum_{\bp_1,\bp_2,\bp_3,\bp_4}
e^{-i(\bp_1-\bp_2-\bp_3)\cdot \br} 
e^{i(\tilde \varepsilon_1-\tilde \varepsilon_2
-\tilde \varepsilon_3)(t-t')} \int d\br' \tilde n(\br',t')\Phi(\br',t')
e^{-i\bp_4\cdot \br'}[a_{\bp_1}^\dagger a_{\bp_2} a_{\bp_3}, 
a_{\bp_4}^\dagger]\nonumber \\
&&\simeq -2g \tilde n(\brt) \sqrt{n_c(\brt)} e^{i\theta(\brt)}
{1\over V}\sum_{\bp_1,\bp_2,\bp_3,\bp_4}
e^{-i(\bp_c+\bp_1-\bp_2-\bp_3)\cdot \br} e^{i(\varepsilon_c+
\tilde \varepsilon_1 -\tilde \varepsilon_2 -\tilde \varepsilon_3)(t-t')}
\nonumber \\ &&\hskip 3.25truein
\times \delta_{\bp_4,\bp_c}\left [
a_{\bp_1}^\dagger a_{\bp_2} \delta_{\bp_3,\bp_4} +
a_{\bp_1}^\dagger a_{\bp_3} \delta_{\bp_2,\bp_4} \right ]\,,
\label{A10}
\eea
where we have defined the condensate momentum $\bp_c \equiv m\bv_c$.
Substituting this expression into (\ref{A6}), we obtain
\bea
&&\langle \tilde\psi^{\dagger}(\brt)\tilde\psi(\brt)\tilde\psi(\brt)
\rangle_{(1)} \nonumber \\ 
&&= 2ig \tilde n(\brt) \sqrt{n_c(\brt)} e^{i\theta(\brt)}
{1\over V}\sum_{\bp_1,\bp_2,\bp_3,\bp_4}
e^{-i(\bp_c+\bp_1-\bp_2-\bp_3)\cdot \br} \delta_{\bp_4,\bp_c}
\nonumber \\ &&\hskip .5truein \times \int_{t_0}^t dt' 
e^{i(\varepsilon_c+ \tilde \varepsilon_1 -\tilde \varepsilon_2 
-\tilde \varepsilon_3)(t-t')} \left [ \delta_{\bp_3,\bp_4} 
\langle a_{\bp_1}^\dagger a_{\bp_2} \rangle_{t'}
+ \delta_{\bp_2,\bp_4} \langle a_{\bp_1}^\dagger a_{\bp_3} 
\rangle_{t'} \right ]\,,
\label{A11}
\eea
where
\beq
\langle a_{\bp_1}^\dagger a_{\bp_2} \rangle_{t'}
={\rm Tr}\hat \rho(t_0)\hat S_0^\dagger(t',t_0)
a_{\bp_1}^\dagger a_{\bp_2} S_0(t',t_0)
\approx e^{i(\tilde \varepsilon_1 - \tilde \varepsilon_2)(t'-t_0)}
\langle a_{\bp_1}^\dagger a_{\bp_2} \rangle_{t_0}\,.
\label{A12}
\eeq
Following KD~\cite{kirkpatrick}, we assume that the initial 
statistical density matrix $\hat \rho(t_0)$ has the form appropriate 
for the HF Hamiltonian $\hat H_0(t)$, in which case
\beq
\langle a_{\bp_1}^\dagger a_{\bp_2} \rangle_{t_0}\simeq
\delta_{\bp_1,\bp_2}f(\bp_1,\br,t)\,.
\label{A13}
\eeq
This identification of the expectation value with the distribution
function at $\br$ and $t$ is based on the picture that the collision
processes are short-ranged in both space and time.
Finally, the remaining time integral in (\ref{A11})
is performed by setting $t_0 \to -\infty$ and introducing the
convergence factor $e^{-\delta(t-t')}$, with the result
\beq
\int_{-\infty}^t dt' 
e^{i(\varepsilon_c+ \tilde \varepsilon_1 -\tilde \varepsilon_2 
-\tilde \varepsilon_3+i\delta)(t-t')}
\simeq \pi \delta( \varepsilon_c+ \tilde \varepsilon_1 
-\tilde \varepsilon_2 - \tilde \varepsilon_3)
+i P \left ({1\over 
\varepsilon_c+ \tilde \varepsilon_1 -\tilde \varepsilon_2 
-\tilde \varepsilon_3}\right ) \,.
\label{A14}
\eeq

It is useful to note that the approximations made in the above 
calculations can be reproduced simply by replacing $\hat H_1'(t')$ by 
\beq
H'_1(t')\approx 
-2gV^{1/2}\tilde n n_c^{1/2}
\sum_\bp\delta_{\bp,\bp_c}[e^{-i\theta}e^{i\varepsilon_c(t'-t)}
e^{i\bp_c\cdot{\bf r}}a_\bp+{\rm h.c.}]\,,
\label{A15}
\eeq
where the variables $\tilde n$, $n_c$, $\bv_c$, $\varepsilon_c$ and
$\theta$ are all evaluated at $\br$ and $t$. In the subsequent
calculations it is convenient to make similar replacements for the other
parts of the interaction Hamiltonian. In particular, we have
\begin{mathletters}
\begin{eqnarray}
H'_2(t')&\approx& {gn_c\over 2}\sum_{\bp_1,\bp_2} 
\delta_{\bp_1+\bp_2,2\bp_c} [e^{-i2\theta} e^{i2\varepsilon_c(t'-t)}
e^{i2\bp_c\cdot\br}a_{\bp_1}a_{\bp_2}+{\rm h.c.}], \label{A16a}\\
H'_3(t')&\approx&
\frac{gn_c^{1/2}}{V^{1/2}}
\sum_{\bp_1,\bp_2,\bp_3}\delta_{\bp_c+\bp_1,\bp_2+\bp_3}
[e^{-i\theta}e^{i\varepsilon_c(t'-t)}e^{i\bp_c\cdot\br}
a^{\dagger}_{\bp_1}
a_{\bp_2}a_{\bp_3}+{\rm h.c.}], \label{A16b}\\
H'_4(t')&\approx&
\frac{g}{2V}\sum_{\bp_1,\bp_2,\bp_3,\bp_4}
\delta_{\bp_1+\bp_2,\bp_3+\bp_4}a^{\dagger}_{\bp_1}
a^{\dagger}_{\bp_2}a_{\bp_3}a_{\bp_4}-2g\tilde n\sum_\bp 
a^{\dagger}_\bp a_\bp\,.
\label{A16c}
\end{eqnarray}
\label{A16}
\end{mathletters}
One thus ends up with expressions for $\hat H_i'(t')$ which now depend
explicitly on $(\brt)$. These forms are of course restricted to the 
evaluation of correlation functions such as (\ref{A6}) involving the 
field operators $\tilde \psi(\brt)$.

The contribution of $\hat H_3'(t')$ to the three-field correlation
function in (\ref{A6}) can now be obtained in a similar way using 
(\ref{A16b}). 
In the course of the calculation we encounter higher order correlation
functions of the type $\langle a^\dagger a^\dagger aa \rangle$, which we
express in terms of the normal products $\langle a^\dagger a \rangle$
by means of Wick's theorem. We thus find
\begin{eqnarray}
\langle \tilde\psi^{\dagger}(\brt)\tilde\psi(\brt)
\tilde\psi(\brt) \rangle_{(3)} &=&
-i2 g n_c^{1/2} e^{i\theta} {1\over V^2} \sum_{\bp_1,\bp_2,\bp_3}
e^{i(\bp_2+\bp_3-\bp_1-\bp_c)\cdot\br} \nonumber \\
&&\times \int_{t_0}^t dt' e^{i(\varepsilon_c +\tilde\varepsilon_1
-\tilde\varepsilon_2-\tilde\varepsilon_3)(t-t')}
\sum_{\bp'_1,\bp'_2,\bp'_3}
\delta_{\bp_c+\bp'_1,\bp'_2+\bp'_3} \nonumber \\
&&\times
[\delta_{\bp_2,\bp'_2}\delta_{\bp_3,\bp'_3}
\langle a^{\dagger}_{\bp_1}a_{\bp'_1}\rangle_{t'}
+\delta_{\bp_2,\bp'_2}\langle a^{\dagger}_{\bp_1}a_{\bp'_1}\rangle_{t'}
\langle a^{\dagger}_{\bp'_3}a_{\bp_3} \rangle_{t'} \nonumber \\
&&+\delta_{\bp_3,\bp'_3}\langle a^{\dagger}_{\bp_1}a_{\bp'_1} 
\rangle_{t'} \langle a^{\dagger}_{\bp'_2}a_{\bp_2} \rangle_{t'}
-\delta_{\bp_1,\bp'_1}\langle a^{\dagger}_{\bp'_2}a_{\bp_2} \rangle_{t'}
\langle a^{\dagger}_{\bp'_3}a_{\bp_3} \rangle_{t'} \nonumber \\
&&+\delta_{\bp_2,\bp'_2}\langle a^{\dagger}_{\bp_1}a_{\bp_3} 
\rangle_{t'} \langle a^{\dagger}_{\bp'_3}a_{\bp_1'} \rangle_{t'}
+\delta_{\bp_3,\bp'_3}\langle a^{\dagger}_{\bp_1}a_{\bp_2} \rangle_{t'}
\langle a^{\dagger}_{\bp'_2}a_{\bp_1'} \rangle_{t'} ]\,,
\label{A17}
\end{eqnarray}
where $\langle a^{\dagger}_{\bp_1}a_{\bp_2}\rangle_{t'}$ is given by 
(\ref{A12}). The last two terms in (\ref{A17}) have a special form.
Isolating this pair of terms, we have the sum
\bea
&&\sum_{\bp'_1,\bp'_2,\bp'_3} 
\delta_{\bp_c+\bp'_1,\bp'_2+\bp'_3}
\left [ \delta_{\bp_2,\bp'_2}\langle a^{\dagger}_{\bp_1}a_{\bp_3} 
\rangle_{t'} \langle a^{\dagger}_{\bp'_3}a_{\bp_1'} \rangle_{t'}
+\delta_{\bp_3,\bp'_3}\langle a^{\dagger}_{\bp_1}a_{\bp_2} \rangle_{t'}
\langle a^{\dagger}_{\bp'_2}a_{\bp_1'} \rangle_{t'} \right ]\nonumber \\
&=&\sum_{\bp'_1,\bp'_2,\bp'_3} 
\delta_{\bp_c+\bp'_1,\bp'_2+\bp'_3} \left [
\delta_{\bp_2,\bp'_2}\langle a^{\dagger}_{\bp_1}a_{\bp_3} \rangle_{t'}
+\delta_{\bp_3,\bp'_2}\langle a^{\dagger}_{\bp_1}a_{\bp_2} \rangle_{t'}
\right ] \langle a^{\dagger}_{\bp'_3}a_{\bp_1'} \rangle_{t'}\nonumber\\
&=&\tilde n V \sum_{\bp'_2} \delta_{\bp_c,\bp'_2} \left [
\delta_{\bp_2,\bp'_2}\langle a^{\dagger}_{\bp_1}a_{\bp_3} \rangle_{t'}
+\delta_{\bp_3,\bp'_2}\langle a^{\dagger}_{\bp_1}a_{\bp_2} \rangle_{t'}
\right ]\,,
\nonumber
\eea
where in going from the first to the second line we have interchanged
the $\bp_2'$ and $\bp_3'$ summation variables, and have used (\ref{A13})
to obtain the last line. When this expression is inserted back into
(\ref{A17}), we see that the last two terms in this equation exactly
cancel the contribution from $\hat H_1'$ in (\ref{A11}).
We thus obtain the following 
explicit expression for the three-field correlation function
\begin{eqnarray}
\langle \tilde\psi^{\dagger}(\brt)\tilde\psi(\brt)
\tilde\psi(\brt)\rangle &=&
-i2\pi gn_c^{1/2}e^{i\theta}{1\over V^2}\sum_{\bp_1,\bp_2,\bp_3} \cr
&&\times \left [ \delta(\varepsilon_c+\tilde\varepsilon_1
-\tilde\varepsilon_2-\tilde\varepsilon_3) + {i\over \pi}P
{1\over \varepsilon_c+\tilde\varepsilon_1
-\tilde\varepsilon_2-\tilde\varepsilon_3} \right ] \cr
&&\times\delta_{\bp_c+\bp_1,\bp_2+\bp_3}
[f_1(1+f_2)(1+f_3)-(1+f_1)f_2f_3]\,,
\label{A18}
\end{eqnarray}
with $f_i\equiv f(\bp_i,\br,t)$.
We emphasize that in deriving (\ref{A18}), we have assumed that initial
values of anomalous correlations, such as
$\langle a_{\bp_1}a_{\bp_2}\rangle_{t_0}$, vanish.
In addition, we have treated the system as locally homogeneous, with the
consequence that the local HF single-particle energies,
$\tilde \varepsilon_p(\brt) = p^2/2m+U(\brt)$, appear. We finally note 
that the three-field correlation function in (\ref{A18}) 
is explicitly proportional to $\sqrt{n_c}$ and the interaction
strength $g$. Thus it vanishes above $T_{\rm BEC}$ as well as in a
non-interacting Bose gas.

Following the same procedure, one can also calculate the anomalous pair
correlation function
\bea
\tilde m(\brt) &\equiv& \langle \tilde \psi(\brt) \tilde \psi(\brt)
\rangle \cr
&=& {1\over V} \sum_{\bp_1,\bp_2} \langle a_{\bp_1}a_{\bp_2} \rangle_t
e^{i(\bp_1 + \bp_2)\cdot \br}\,.
\label{A1'}
\eea
The final result analogous to (\ref{A18}) is
\bea
\tilde m(\brt) &=& -i\pi g \Phi^2 {1\over V} \sum_{\bp_1,\bp_2} 
\delta_{\bp_1+\bp_2,2\bp_c}[1+f_1+f_2] \cr && \hskip 1truein \times
\left [ \delta(\tilde \varepsilon_1+\tilde\varepsilon_2
-2\varepsilon_c) + {i\over \pi}P
{1\over \tilde \varepsilon_1+\tilde\varepsilon_2
-2\varepsilon_c} \right ] \,.
\label{A2'}
\eea
We note that if (\ref{A2'}) is evaluated for a static equilibrium
situation ($\bv_c =0$, $\varepsilon_c = \mu_{c0}$ and  $f_i$ equilibrium
Bose distributions), the imaginary part vanishes and the real part
simplifies to
\beq
\tilde m_0(\br) = -gn_{c0}(\br) \int {d\bp \over (2\pi)^3}
{1+2f^0_p \over 2E_p(\br)}\,,
\label{A3'}
\eeq
where $E_p(\br) = \tilde \varepsilon_p(\br)-\mu_{c0}$ is the local HF
excitation energy. This result for the anomalous pair correlation is 
consistent with the well-known HFB expression~\cite{shi98}
for a uniform Bose gas (in the HF limit).

The real parts of (\ref{A18}) and (\ref{A2'}) contribute to the local
condensate chemical potential in (\ref{eq8}). As is well known (see,
for example, Section 5 of Ref.~\cite{griffin99} and
Ref.~\cite{proukakis98a}), the anomalous
pair correlation in (\ref{A3'}) is ultraviolet divergent. This
divergence is removed in the more complete Beliaev
approximation~\cite{shi98}, which includes all self-energy contributions
to second order in $g$. Since we do not include renormalization effects
consistently to order $g^2$ in the present treatment,  we will drop
these contributions to the chemical potential. The imaginary parts
of (\ref{A18}) and (\ref{A2'})
contribute to the source term on the right hand side of (\ref{eq5a}).
However, the contribution from $\tilde m$ involves the energy-conserving
delta function $\delta(\tilde \varepsilon_1 +\tilde \varepsilon_2 -
2\varepsilon_c)$. Taking the momentum conservation condition in
(\ref{A2'}) into account, one finds that the argument of the delta
function in the TF limit is equal to $(\bp_1-\bp_2)^2/4m+2gn_c$. Since 
this expression is positive definite, the imaginary part of $\tilde
m$ makes no contribution to the source term in (\ref{eq5a}).

Using similar techniques and approximations, we next evaluate
the collision terms given in the right hand side of (\ref{eq21}).
The $C_{12}$ collision integral is defined as the contribution from
the $\hat H'_3$ perturbation in (\ref{A16b}),
\begin{eqnarray}
C_{12}[f]&\equiv&
-i{\rm Tr}\tilde\rho(t,t_0)[\hat f(\bp,\brt_0),\hat H'_3(t)] \cr
&\simeq&
-ign_c^{1/2}V^{-1/2}\sum_{\bf q}
\sum_{\bp_1,\bp_2,\bp_3} e^{i{\bf q}\cdot\br}\delta_{\bp_c+\bp_1
,\bp_2+\bp_3} e^{-i\theta}e^{i\bp_c\cdot\br} \cr
&&\qquad\times \Bigl[\delta_{\bp_1,\bp+{\bf q}/2}
\langle a^{\dagger}_{\bp-{\bf q}/2}a_{\bp_2}a_{\bp_3} \rangle_t
-\delta_{\bp_2,\bp-{\bf q}/2} 
\langle a^{\dagger}_{\bp_1}a_{\bp+{\bf q}/2}a_{\bp_3} \rangle_t \cr
&& \hskip 1.25truein  -\delta_{\bp_3,\bp-{\bf q}/2}
\langle a^{\dagger}_{\bp_1}a_{\bp_2}a_{\bp+{\bf q}/2} \rangle_t -
{\rm h.c.}\Bigr].
\label{A19}
\end{eqnarray}
Eq.~(\ref{A18}) can be used to identify the
three-field correlation function $\langle a_{\bp_1}^\dagger
a_{\bp_2} a_{\bp_3} \rangle_t$. Inserting the result in (\ref{A19}), 
we find
\begin{eqnarray}
C_{12}[f]&=&4\pi g^2n_cV^{-1}\sum_{\bp_1,\bp_2,\bp_3}
\delta(\varepsilon_c+\tilde\varepsilon_1-\tilde\varepsilon_2
-\tilde\varepsilon_3)
\delta_{\bp_c+\bp_1,\bp_2+\bp_3} \cr
&&\times[\delta_{\bp_1,\bp}-\delta_{\bp_2,\bp}
-\delta_{\bp_3,\bp}] [(1+f_1)f_2f_3-f_1(1+f_2)(1+f_3)].
\label{A20}
\end{eqnarray}

The $C_{22}$ collision term is the contribution from the $\hat H'_4$
perturbation defined in (\ref{A16c}). We note that the second term in
(\ref{A16c}) is proportional to the number operator which commutes with
any number-conserving operator.  We thus have
\begin{eqnarray}
C_{22}[f]&\equiv&
-i{\rm Tr}\tilde\rho(t,t_0)[\hat f(\bp,\brt_0),\hat H'_4(t)] \cr
&\simeq&
-\frac{ig}{2V}\sum_{\bf q}\sum_{\bp_1,\bp_2,\bp_3,p_4}
e^{i{\bf q}\cdot\br}
\delta_{\bp_1+\bp_2,\bp_3+\bp_4} \cr
&&\ \ \ \times \Bigl[
\delta_{\bp_1,\bp+{\bf q}/2}
\langle a^{\dagger}_{\bp-{\bf q}/2}a^{\dagger}_{\bp_2}a_{\bp_3}
a_{\bp_4}\rangle_t 
+ \delta_{\bp_2,\bp+{\bf q}/2}
\langle a^{\dagger}_{\bp-{\bf q}/2}a^{\dagger}_{\bp_1}a_{\bp_3}
a_{\bp_4}\rangle_t \cr
&&\ \ \ \ \ \ - \delta_{\bp_3,\bp-{\bf q}/2} \langle 
a^{\dagger}_{\bp_1}a^{\dagger}_{\bp_2}a_{\bp_4}a_{\bp+{\bf q}/2}
\rangle_t -
\delta_{\bp_4,\bp-{\bf q}/2}
\langle a^{\dagger}_{\bp_1}a^{\dagger}_{\bp_2}a_{\bp_3}a_{\bp+{\bf q}/2}
\rangle_t \Bigr ]\,.
\label{A21}
\end{eqnarray}
According to (\ref{A5}), 
\begin{eqnarray}
&&\langle a^{\dagger}_{\bp_1}a^{\dagger}_{\bp_2}a_{\bp_3}a_{\bp_4}
\rangle_t ={\rm Tr}\hat\rho(t_0)\Bigl\{\hat S^{\dagger}(t,t_0)
a^{\dagger}_{\bp_1} a^{\dagger}_{\bp_2}a_{\bp_3}a_{\bp_4}\hat S(t,t_0) 
\cr
&&\hskip 1truein -i\int_{t_0}^t dt' e^{i(\tilde\varepsilon_1
+\tilde\varepsilon_2 -\tilde\varepsilon_3-\tilde\varepsilon_4)(t-t')}
\hat S^{\dagger}(t',t_0) \left [
a^{\dagger}_{\bp_1}a^{\dagger}_{\bp_2}a_{\bp_3}a_{\bp_4},\hat H'_4(t')
\right ] \hat S(t',t_0)\Bigr\}\,.
\label{A22}
\end{eqnarray}
We find that the first term of order $g^0$ in (\ref{A22}), while finite,
makes no contribution to (\ref{A21}). Thus the
$C_{22}[f]$ collision integral is explicitly of second order in $g$.
Using Wick's theorem to evaluate the second term in 
(\ref{A22}), we obtain the relevant contribution
\begin{eqnarray}
\langle a^{\dagger}_{\bp_1}a^{\dagger}_{\bp_2}a_{\bp_3}a_{\bp_4}
\rangle_t&\simeq&
-\frac{2\pi ig}{V}
\delta(\tilde\varepsilon_1+\tilde\varepsilon_2
-\tilde\varepsilon_3-\tilde\varepsilon_4)
\delta_{\bp_1+\bp_2,\bp_3+\bp_4} \cr
&&\times\Big [ f_1f_2(1+f_3)(1+f_4)-(1+f_1)(1+f_2)f_3f_4\Big ]\,.
\label{A23}
\end{eqnarray}
Inserting (\ref{A23}) into (\ref{A21}), we finally obtain the
expression
\begin{eqnarray}
C_{22}[f]&=&\frac{\pi g^2}{V^2}\sum_{\bp_1,\bp_2,\bp_3,\bp_4}
\delta(\tilde\varepsilon_1+\tilde\varepsilon_2
-\tilde\varepsilon_3-\tilde\varepsilon_4)
\delta_{\bp_1+\bp_2,\bp_3+\bp_4} \cr
&&\times[\delta_{\bp,\bp_1}+\delta_{\bp,\bp_2}
-\delta_{\bp,\bp_3}-\delta_{\bp,\bp_4}] \cr
&&\times\Big [f_1f_2(1+f_3)(1+f_4)-(1+f_1)(1+f_2)f_3f_4\Big ]\,.
\label{A24}
\end{eqnarray}

Replacing the momentum sum $(1/V)\sum_\bp$ by the integral
$\int d\bp/(2\pi)^3$, and the Kronecker delta function
$V\delta_{\bp,{\bf 0}}$ by the Dirac delta function 
$(2\pi)^3\delta(\bp)$ in (\ref{A18}), (\ref{A20}) and (\ref{A24}),
we recover the expressions written down in Eqs.~(\ref{eq25}), 
(\ref{eq23a}) and (\ref{eq23b}), respectively. As discussed in the body
of the paper, there is a close relation between the expressions in
(\ref{A18}) and (\ref{A20}). This ensures that the separate 
continuity equations for the condensate and the non-condensate 
[see (\ref{eq27a}) and (\ref{eq46'a})] are consistent with the 
requirement that the total number of atoms is conserved. (We note that
KD~\cite{kirkpatrick} did not obtain the equivalent of these
equations since they did not explicitly evaluate $\langle \tilde
\psi^\dagger \tilde \psi \tilde \psi \rangle$.)

%% file: append.b.tex
\section{Equilibrium Properties}
\label{appendixb}
In this Appendix, we  briefly discuss the difference between the static 
equilibrium functions we use for a weakly-interacting Bose gas 
and the well-known results obtained by Lee and Yang \cite{lee58}
at finite temperatures ($T\lesssim T_{BEC}$). The Lee-Yang expressions 
for the thermodynamic functions were derived as {\it first-order}
corrections in the interaction $g$ to the ideal Bose-condensed gas 
results.  We list them here for convenience:
\begin{eqnarray}
P&=&\tilde P_{\rm cr}+\frac{1}{2}g(n^2+\tilde n^2_{\rm cr})\,, 
\nonumber \\
\epsilon&=&\frac{3}{2}\tilde P_{\rm cr}+\frac{1}{2}g(n^2-n\tilde 
n_{\rm cr} +2\tilde n_{\rm cr}^2)\,, \label{B1} \\
Ts&=&\frac{5}{2}\tilde P_{\rm cr}+\frac{3}{2}g(\tilde n_{\rm cr}^2
-n\tilde n_{\rm cr})\,. \nonumber
\end{eqnarray}
The associated free energy density is
\begin{equation}
f\equiv \epsilon-Ts=-\tilde P_{\rm cr}+\frac{1}{2}g(n^2+2n\tilde 
n_{\rm cr} -\tilde n_{\rm cr}^2)\,. 
\label{B2}
\end{equation}
Here
\begin{eqnarray}
\tilde P_{\rm cr}&\equiv&\frac{1}{\beta\Lambda^3}g_{5/2}(z=1) \cr
\tilde n_{\rm cr}&\equiv&\frac{1}{\beta\Lambda^3}g_{3/2}(z=1),
\label{B2'}
\end{eqnarray}
are the ideal-gas values of the pressure and non-condensate density at
$T_{\rm BEC}$. One can see that (\ref{B1}) and (\ref{B2})
explicitly involve only first order 
corrections in $g$. To this order, these expressions are entirely
consistent with our thermodynamic expressions given by (\ref{eq62}),
(\ref{eq65}) and (\ref{eq66}). For example, we have
\begin{eqnarray}
\tilde P&=&\frac{1}{\beta\Lambda^3}g_{5/2}(z=
e^{-\beta gn_{c}}) \cr &\simeq&\frac{1}{\beta\Lambda^3}
g_{5/2}(z=1)-gn_{c}\frac{1}{\Lambda^3} g_{3/2}(1)+\cdots \cr
&=&\tilde P_{\rm cr}-g\tilde n_{\rm cr}(n-\tilde n_{\rm cr})\,.
\label{B3}
\end{eqnarray}
Substituting this into (\ref{eq62}) with $\tilde n = \tilde n_{\rm cr}$,
we obtain the expression for the pressure in (\ref{B1}). However, it 
is clear that our description
of the equilibrium properties (based on the HF approximation) goes
beyond the Lee-Yang description, which is only correct to first order in
$g$.

The implications of this become apparent when these results are
considered in the context of a trapped Bose-condensed gas. Within our
semiclassical HF description, the non-condensate density is given by
\beq
\tilde n = {1\over \Lambda^3}g_{3/2}(z)\,,
\label{B4}
\eeq
with the fugacity given by
\beq
z=e^{\beta(\mu_c - U)}\,.
\label{B5}
\eeq
Here, $\mu_c$ is the condensate eigenvalue as obtained from (\ref{eq47})
and $U(\br) = U_{\rm ext}(\br)+ 2gn(\br)$ is the effective potential as
seen by the non-condensate atoms. For (\ref{B4}) to be well-defined,
$\mu_c$ must lie {\it below} the minimum of the effective potential
$U(\br)$. Because of the localized condensate, the non-condensate atoms
experience a localized repulsive interaction $2gn_c(\br)$ which has the
effect of depleting the non-condensate density at the center of the trap
where the non-condensate density overlaps with the condensate. 
This is evident in Fig.~3 (see
also Refs.~\cite{hutchinson97} and \cite{giorgini97}), and gives rise to
corresponding spatial variations in the other thermodynamic functions.

In contrast, in the Lee-Yang description, the non-condensate takes on
the critical value $\tilde n_{\rm cr}$ wherever the condensate density
is finite.  This is the approximation used by 
Shenoy and Ho~\cite{shenoy98}
in their calculation of the normal modes of a trapped Bose gas. Working
within the Thomas-Fermi approximation, their condensate density is
given by
\beq
n_c(\br) = {1\over g} (\mu - U_{\rm ext}(\br) - 2g\tilde n_{\rm cr})\,,
\eeq
which goes to zero at a radius $R_{\rm TF}$ 
defined by $U_{\rm ext}(R_{\rm TF}) = \mu - 2g\tilde n_{\rm cr}$.
For $r<R_{\rm TF}$, all the thermodynamic coefficients required in the 
Landau equations were evaluated using the Lee-Yang free energy density 
in (\ref{B2}).

Above $T_{\rm BEC}$, Lee and Yang \cite{lee58} obtain the 
non-perturbative result for the free energy density (generalized to 
include a trap)
\begin{equation}
f=-\frac{1}{\beta\Lambda^3}g_{5/2}(z_0)-gn^2({\bf r})+
n\mu_0-nU_{ext}({\bf r}),
\end{equation}
where the density $n({\bf r})$ is obtained from  (\ref{B4}) with the
fugacity given by
\begin{equation}
z_0= e^{\beta(\mu_0-U_{ext}({\bf r}) -2gn({\bf r}))}\,.
\end{equation}
The chemical potential $\mu_0$ is chosen to give the desired total
number of atoms, $N$. The other thermodynamic functions are given in
this temperature range by 
\begin{eqnarray}
P&=&\frac{1}{\beta\Lambda^3}g_{5/2}(z_0)+gn^2, \cr
\epsilon&=&\frac{3}{2}\frac{1}{\beta\Lambda^3}g_{3/2}(z_0)+
gn^2+nU_{ext}({\bf r}), \cr
sT&=&\frac{5}{2}\frac{1}{\beta\Lambda^3}g_{5/2}(z_0)-nk_{\rm B}T
\ln z_0\,.
\end{eqnarray}
These are the results used by Shenoy and Ho~\cite{shenoy98} for the
thermal cloud in those regions where $\tilde n_0 < \tilde n_{cr}$, that
is, for $r > R_{\rm TF}$ when $T < T_{BEC}$. We note that these
expressions take interactions into account in a self-consistent fashion,
unlike the situation for $r<R_{\rm TF}$ where the non-condensate density
takes on the ideal gas value of $\tilde n_{cr}$. 

%% file: append.c.tex
\section{}
\label{appendc}

We list here the various integrals required for the evaluation of the
energy functional $J$ in Section~\ref{section7}. 
The form of the non-condensate displacement is
$\bu(\br) = (ax,ay,bz)$ and that of the condensate is
$\bw(\br) = (cx,cy,dz)$. For the kinetic energy terms, we have
\bea
K_n[\bu] &=& {1\over 6}(2a^2+b^2) m \int d^3 r \, r^2 \tilde n_0(r)
\nonumber \\ 
K_c[\bw] &=& {1\over 6}(2c^2+d^2) m \int d^3 r \, r^2 n_{c0}(r)\,.
\nonumber
\eea
In obtaining these results, we have made use of the spherical 
symmetry of the equilibrium
densities. The various potential energy contributions are the terms
appearing in (\ref{eq7.9}), (\ref{eq7.11}) and (\ref{eq7.15}):
\bea
{1\over 2} \int d^3 r\,u_{ii}^2 &=& {1\over 2} (2a+b)^2 \int d^3 r\,
\lambda(r) \nonumber \\
{1\over 2} \int d^3 r \mu ( u_{ij}^2-\bar u_{ij}^2)
&=& {1\over 2} (2a^2 + b^2) \int d^3 r\, \mu(r) \nonumber \\
{1\over 2} \int d^3 r 
u_i u_j \tilde n_0 {\partial^2 \over \partial x_i x_j} U_{ext} &=&
{1\over 6}(2a^2+b^2)m\omega_0^2 \int d^3 r\, r^2\tilde n_0\nonumber\\
{1\over 2} \int d^3 r 
u_i u_j \tilde n_0 {\partial^2 \over \partial x_i x_j} 2g n_{c0} &=&
{2\over 15} (a-b)^2 g \int d^3 r\, r \tilde n_0 
{\partial n_{c0} \over \partial r} \nonumber \\ &&
+ {1\over 15}(8a^2+4ab+3b^2)g \int
d^3 r\, r^2 \tilde n_0  {\partial^2 n_{c0} \over
\partial r^2} \nonumber \\
2g \int d^3 r\, 
{\partial (\tilde n_0 u_i) \over \partial x_i}
{\partial (n_{c0} w_j) \over \partial x_j} &=&
{2\over 15} (8ac+2ad+2bc+3bd)g\int d^3 r\, r^2 {\partial \tilde n_0
\over \partial r}
{\partial n_{c0}\over \partial r}\nonumber \\
{1\over 2}g \int d^3 r\, 
{\partial (n_{c0} w_i) \over \partial x_i}
{\partial (n_{c0} w_j) \over \partial x_j} &=&
{1\over 30} (8c^2+4cd+3d^2)g\int d^3 r\, r^2 \left ( {\partial n_{c0}
\over \partial r}\right )^2 \nonumber \\
{1\over 2} \int d^3 r\, 
{\partial (n_{c0} w_i) \over \partial x_i} \hat K
{\partial (n_{c0} w_j) \over \partial x_j} &=&
{1\over 15}(8c^2+4cd+3d^2)\int d^3 r\, 
\left ( r {\partial \Phi_0 \over \partial r}\right ) \hat {\cal L}_0 
\left ( r {\partial \Phi_0 \over \partial r}\right ) \nonumber \\&&+
{4\over 15}(c-d)^2{\hbar^2\over m} \int d^3 r\, 
\left ( {\partial \Phi_0 \over \partial r}\right )^2\,. \nonumber 
\eea